\documentclass[aps,10pt,a4paper,amsfonts,nofootinbib]{revtex4}\usepackage{graphicx}
\newcommand{\be}{\begin{equation}}
\newcommand{\ee}{\end{equation}}
\newcommand{\ba}{\begin{eqnarray}}
\newcommand{\ea}{\end{eqnarray}}
\newcommand{\bra}[1]{\left(#1\right)}
\newcommand{\bras}[1]{\left[#1\right]}
\newcommand{\brac}[1]{\left\{#1\right\}}

\renewcommand{\:}[2]{{\textstyle\frac{#1}{#2}}}
\renewcommand{\;}[2]{{\frac{#1}{#2}}}

\newcommand{\forget}[1]{\iffalse#1\fi}
\newcommand{\forgetmenot}[1]{\iftrue#1\fi}

\newcommand{\del}{\nabla}

\renewcommand{\div}{{\mathsf{div}\,}}
\newcommand{\curl}{{\mathsf{curl}\,}}
\newcommand{\sdel}{{\mathrm{D}}}

\newcommand{\<}{\langle}
\renewcommand{\>}{\rangle}
\newcommand{\sech}{{\mathrm{sech}}\,}
\newcommand{\cosech}{{\mathrm{cosech}}\,}

\newcommand{\subsubsubsection}[1]{\vskip 4mm \paragraph{#1}}

\newcommand{\udot}{{\cal A}}
\newcommand{\uudot}{\dot{u}}
\newcommand{\n}{n}
\newcommand{\N}{N}
\newcommand{\E}{{\cal E}}
\renewcommand{\H}{{\cal H}}
\newcommand{\lc}{\varepsilon}
\newcommand{\hatn}{a}%{{\hat{\n}}}
\newcommand{\dotn}{\alpha}%{{\dot{\n}}}
\newcommand{\lb}{\{}%{\lceil}
\newcommand{\rb}{\}}%{\rfloor}

\renewcommand{\S}{_{\mathsf{S}}}
\newcommand{\V}{_{\mathsf{V}}}
\newcommand{\T}{_{\mathsf{T}}}

\renewcommand{\ll}{L}%l(l+1)
\newcommand{\lll}{l}%(l-1)(l+2)

\newcommand{\vo}{{\mathbf{V_{\mathsf{O}}}}}
\newcommand{\ve}{{\mathbf{V_{\mathsf{E}}}}}
\newcommand{\odd}{{_{\mathbf{\mathsf{O}}}}}
\newcommand{\even}{{_{\mathbf{\mathsf{E}}}}}

\newcommand{\gen}{^{\mathsf{g}}}
\newcommand{\cross}{^{\mathsf{c}}}

\begin{document}

\title{Covariant Perturbations of Schwarzschild Black Holes}
\author{Chris A. Clarkson}
\affiliation{Relativity and Cosmology Group,
Department of Mathematics and Applied Mathematics, University of Cape Town,
Rondebosch 7701, Cape Town, South Africa}
\email{clarkson@maths.uct.ac.za}
\author{Richard K. Barrett}
\affiliation{Astronomy and Astrophysics group, Department of Physics and
Astronomy, University of Glasgow, University Avenue, Glasgow, G12 8QQ, UK}
\email{richard@astro.gla.ac.uk}

\date{\today}

%\maketitle

\begin{abstract}
We present a new \emph{covariant} and \emph{gauge-invariant} perturbation
formalism for dealing with spacetimes having spherical symmetry (or some
preferred spatial direction) in the background, and apply it to the case of
gravitational wave propagation in a Schwarzschild black hole spacetime. The 1+3
covariant approach is extended to a `1+1+2 covariant sheet' formalism by
introducing a radial unit vector in addition to the timelike congruence, and
decomposing all covariant quantities with respect to this. The background
Schwarzschild solution is discussed and a covariant characterisation is given.
We give the full first-order system of linearised 1+1+2 covariant equations,
and we show how, by introducing (time and spherical) harmonic functions, these
may be reduced to a system of \emph{first-order} ordinary differential
equations and algebraic constraints for the 1+1+2 variables which may be solved
straightforwardly. We show how both the odd and even parity perturbations may
be unified by the discovery of a covariant, frame- and gauge-invariant,
transverse-traceless tensor describing gravitational waves, which satisfies a
covariant wave equation equivalent to the Regge-Wheeler equation for both even
and odd parity perturbations. We show how the Zerilli equation may be derived
from this tensor, and derive a similar transverse traceless tensor equivalent
to this equation. The so-called `special' quasinormal modes with purely
imaginary frequency emerge naturally. The significance of the degrees of
freedom in the choice of the two frame vectors is discussed, and we demonstrate
that, for a certain frame choice, the underlying dynamics is governed purely by
the Regge-Wheeler tensor. The two transverse-traceless Weyl tensors which carry
the curvature of gravitational waves are discussed, and we give the closed
system of four first-order ordinary differential equations describing their
propagation. Finally, we consider the extension of this work to the study of
gravitational waves in other astrophysical situations.

\end{abstract}

\maketitle

\section{Introduction}

The 1+3 covariant approach has proven to be an extremely useful technique for
developing a detailed understanding of many aspects of relativistic cosmology,
both in terms of fully nonlinear GR effects and through the application of the
gauge-invariant, covariant perturbation formalism (see~\cite{HvE}, for example)
to the formation and evolution of density perturbations~\cite{BE} in the
universe and to the physics of the cosmic microwave background~\cite{CL,MGE},
amongst other things (see~\cite{HvE} for a comprehensive review). Its strength
in cosmological applications lies in the fact that it is well adapted to the
system it is describing: all essential information can be captured in a set of
(1+3) covariant variables (defined with respect to a preferred timelike
observer congruence), that have an immediate physical and geometrical
significance. They satisfy a set of \emph{evolution} and \emph{constraint}
equations, derived from Einstein's field equations, and the Bianchi and Ricci
identities, which form a closed system of equations when an equation of state
for the matter is chosen. The covariant and gauge-invariant linearisation
procedure is easy and transparent: it consists of deciding which variables are
`first order' (or `of order $\epsilon$') and those which are `zeroth order'~--
i.e., those which do not vanish in the background, which is usually a
Friedmann-Lema\^\i tre-Robertson-Walker (FLRW) model. Products of first-order
quantities can then be ignored in the equations. Whenever the background is a
homogeneous and isotropic FLRW model, all projected vectors and tensors are
first-order, so there is no vector-tensor and tensor-tensor coupling in the
equations. Harmonic functions can then be introduced which re-write the
equations in scalar form; the resulting system is then in the form of algebraic
constraints and some \emph{first-order} ordinary differential equations; the
solution is then straightforward. The key point of the approach is that it
deals with physically or geometrically relevant quantities, such as the
fractional density gradient, ${\cal D}_a$, or the electric and magnetic parts
of the Weyl tensor, $E_{ab}$ and $H_{ab}$, respectively, which represent the
non-local parts of the gravitational field, and which describe, amongst other
things, the propagation of gravitational waves (GW). The variables are also
(coordinate) gauge invariant (although there is a frame-gauge freedom in the
choice of~$u^a$~-- see below).

The aim of this paper is to extend the gauge-invariant, covariant perturbation
formalism to an astrophysical setting. The 1+3 approach is not appropriate for
many situations where such techniques would seem highly desirable: when the
spacetime in inhomogeneous, for example, the 1+3 equations usually become
intractable. However, by introducing an additional frame vector, assuming that
the background spacetime has some preferred spatial direction (such as in
spherically symmetric, or more general \emph{locally rotationally symmetric}
spacetimes, or $G_2$ spacetimes) we can in many cases recover all of the
advantages of the 1+3 equations, but in a \emph{1+1+2 covariant}
framework~\cite{henk}. In this paper we introduce the 1+1+2 formalism, and
apply it to linear perturbations of a Schwarzschild spacetime. Not only is this
a first step in applying the procedure to more general astrophysical situations
(such as perturbations of the interior of compact objects, collapsing and
exploding stars, etc.), it also represents an important field of study in
itself: with the development of large gravitational wave detectors
(e.g.,~\cite{LIGO}) an improved understanding of the problem of GW propagation
around compact objects is certainly timely. The power of the 1+1+2 technique is
clearly shown by the significant results we are able to obtain, relatively
simply. For example, we show here that the full description of gravitational
waves around a Schwarzschild black hole is governed by closed covariant wave
equation, unifying both parities in a single covariantly defined gauge- and
frame-invariant transverse-traceless 2-tensor, $W_{ab}$.

Linear perturbations of Schwarzschild black holes have been conventionally
studied through perturbations of the metric tensor (or via the Newman-Penrose
formalism~\cite{chandra}). In the metric approach, fluctuations of the
spacetime geometry are characterised by perturbations in the metric tensor;
these fluctuations are determined by closed wave equations~-- the Regge-Wheeler
equation for odd parity perturbations~\cite{RW} and the Zerilli equation in the
even parity regime~\cite{zerilli}. These wave equations act on linear
combinations of the functions (and their derivatives) appearing in the
perturbed metric, but these functions do not determine directly the
gravitational waves which they represent; a general coordinate transformation
would preserve neither the perturbation functions themselves, nor the wave
equations which they satisfy. The approach we develop here is completely
covariant, so such issues do not arise. Instead, corresponding wave equations
we derive here are formed from covariant and gauge-invariant variables which
have a physical significance; furthermore, they do not require a harmonic
splitting for their derivation.

The formalism we develop here relies on a further splitting of the spacetime
using a radial vector $\n^a$, in addition to the usual splitting with the
timelike vector $u^a$ used in the 1+3 approach. We split the Ricci and Bianchi
identities using $u^a$ and $\n^a$ into a coupled set of \emph{first order}
differential equations, plus some constraints. The differential operators we
use are along the two vector fields which give us a set of \emph{evolution} and
\emph{propagation} (along the `radial' direction) equations, while a derivative
formed from a projection orthogonal to $u^a$ and $\n^a$ gives a small number of
constraints. The differential equations involve the covariant variables derived
from splitting the Weyl tensor (and more generally the Ricci tensor, but this
is zero here as we only consider vacuum perturbations), and the kinematics of
$u^a$ and $\n^a$. As our background is static and spherically symmetric, we may
use harmonic functions for our evolution and projected derivatives, putting our
equations into the form of a first order system of ordinary DE's and
constraints, which can then be tackled relatively easily.

Previously,~\cite{GS,MGG,ST,J} (and references therein) developed approaches to
stellar and black hole perturbations  similar to the method presented here in
the sense that they use two orthogonal vectors to form their time and space
derivatives. These approaches are fundamentally different from our approach,
however, in that they formulate their differential equations as \emph{second
order} PDE's derived from Einstein's field equations (EFE), the solutions of
which give the metric functions (as in all metric perturbation approaches). On
the other hand, our system of DE's is manifestly first order, as it is derived
from the first order Ricci and Bianchi identities~\footnote{We use the Ricci
identities applied to our vectors $u^a$ and $\n^a$, so they are second order
DE's of $u^a$ and $\n^a$; however, our covariant objects are different
projections of $\del_au_b$ and $\del_a\n_b$, so that the Ricci identities are
first order DE's in these covariant objects.}, as it involves physical or
geometric quantities, and not the metric functions. Second order wave equations
may be derived if desired. This is one of the key properties of covariant or
tetrad methods. This change in derivative level is conceptually analogous to
the change in going from the Lagrangian to Hamiltonian formulations of
classical mechanics, from configuration space to phase space.

%%%%%%%%%%%%%%%%%%%%%%%%%%%%%%%%%%%%%%%%%%%%%%%%%%%%%%%%

%%%%%%%%%%%%%%%%%%%%%%%%%%%%%%%%%%%%%%%%%%%%%%%%%%%%%%%%%

The layout of the paper is as follows: in the following section we discuss the
merits of a 1+1+2 decomposition of the field equations and set out the 1+1+2
covariant formalism. Then, in section~\ref{the-equations} we present the full
set of 1+1+2 covariant, gauge-invariant, first-order equations, linearised
about a Schwarzschild background and introduce the (spherical and temporal)
harmonics on the `sheet', which enable the equations to be reduced to a set of
coupled ODEs for the 1+1+2 covariant variables. In section~\ref{RWsec} we prove
the existence of a transverse-traceless (TT) tensor that satisfies a closed
wave equation equivalent to the Regge-Wheeler equation, valid for harmonics of
either parity. Following this we discuss the even parity variable which
satisfies a wave equation equivalent to the Zerilli equation; we demonstrate
here the existence of an odd parity counterpart, but defer the derivation of
the `Zerilli tensor' until later, Sec.~\ref{zertens}. Then, in
section~\ref{solution-section} we describe in detail the (matrix-based) method
of solution of the linear, first-order system of ODEs for the harmonic
components, emphasising the significance of the freedom to choose the frame
vectors and showing that with an appropriate choice we can reduce the whole
solution for both parities to a single 2-dimensional ODE. The wave equations
for the TT electric and magnetic Weyl tensors are also presented, and the
closed four-dimensional ODE which they also satisfy is discussed. Finally, we
discuss the results we have obtained.

We follow the notation and conventions of~\cite{HvE}.

\section{The 1+1+2 covariant sheet approach}

Before setting out the principles and equations of the 1+1+2 covariant
formalism it will be illuminating to examine the 1+3 approach to see where
its strengths and weaknesses lie, and why it is so successful in a cosmological
setting, but is less useful in other situations, such as in the study of
gravitational radiation in a Schwarzschild spacetime which we consider here.
In the process we hope to indicate that the 1+1+2 formalism neatly fills a gap
between the 1+3 and tetrad approaches.

\subsection{1+3 covariant perturbation theory: why it works in cosmology, but
not for black holes}
\label{why1+3}

In a nutshell, the 1+3 formalism is successful in cosmological applications
because the assumed spatial homogeneity and isotropy of the background
spacetime means that the only essential coordinate is time: the introduction of
an appropriate timelike observer congruence~$u^a$ allows the full structure of
the spacetime to be described solely in terms of ordinary differential
equations involving (1+3) scalar quantities because all spatial derivatives,
spatial projections of vectors, and projected, symmetric, trace-free parts of
tensors must vanish by symmetry.
%Ordinary
%differential equations are easy to solve (see section~\ref{solution-section}).

\forget{To illustrate this in more detail, consider an FLRW spacetime, and choose
$u^a$ orthogonal to surfaces of homogeneity and isotropy. Then, the 1+3
formalism reduces the \emph{tensor} equations:
\ba
\begin{array}{rcll}
G_{ab}&=&T_{ab}&\mbox{Einstein field equations}\\
2\del_{[a}\del_{b]}u_c&=&R_{abcd}u^d&\mbox{Ricci identities}\\
\del_{[a}R_{bc]de}&=&0&\mbox{Bianchi identities}
\end{array}
\ea
to much simpler \emph{scalar} equations:
\ba
\dot\theta+\:13\theta^2+\:12\bra{\mu+p}&=&0,\\
\dot\mu+\theta\bra{\mu+p}&=&0,\\
{\cal R}-2\mu+\:23\theta^2&=&0.
\ea
($\theta$, $\mu$ and $p$ are expansion, energy density and pressure, as usual.)
All non-zero covariant quantities are scalars, as there is no preferred
direction in the surfaces of homogeneity. In addition to the background
equations being simple, under any first-order perturbation all vectors and
tensors are \emph{first-order}, which means there is no tensor-vector coupling
etc. And because there is no coupling between tensors and vectors and tensors
and tensors (in contrast to the example below), we can introduce harmonic
functions and expand all first-order functions in terms of these. These
harmonic functions allow one to effectively divide out the vector or tensor
nature of the equations in favour of scalar equations of fixed harmonic index.
Furthermore, these harmonic functions, if chosen correctly, remove all spatial
derivatives and replace them with a harmonic index. One is left then with a
first-order system of ordinary  differential equations, plus some algebraic
constraints, which can be solved or manipulated by standard techniques.}

On the other hand, when the spacetime is not homogeneous and isotropic the
resulting equations are not simple ODEs. Consider, for example, a family of
\emph{static} observers around a Schwarzschild black hole, that is, observers
on the congruence $u^a$ parallel to the timelike conformal Killing vector. Then
$u^a$ has zero rotation, shear and expansion, but has non-zero
acceleration~$\uudot^a$ (reflecting the fact that a force must be applied to
prevent infall), and the electric Weyl curvature, $E_{ab}$, is non-zero, while
all other covariant quantities are zero. Relative to these observers, $E_{ab}$
measures non-local gravitational effects: in this case the (time-independent)
radial tidal forces only, which can be described by a single function of
distance from the hole~-- thanks to the spherical symmetry we can think of it
is a \emph{tensor} describing an essentially \emph{scalar} phenomenon. The 1+3
covariant equations describing the spacetime are~\footnote{We use the standard
notation whereby a dot represents differentiation along the observers'
four-velocity $u^a$, $\dot
\psi _{a\cdots b}\equiv u^c\del_c \psi _{a\cdots b}$, and $\sdel_a$ is a
derivative in the rest space of the observers; $\sdel_c
\psi _{a\cdots b}\equiv h_{c}^{~d} h_a^{~e}\cdots h_b^{~f}\del_d \psi _{e\cdots f}$,
where $h_{ab}\equiv g_{ab}+u_au_b$ is the usual projection tensor orthogonal to
$u^a$. We use angled brackets on indices to donate the projected, symmetric and
trace-free part of a tensor.}
\ba
\dot E_{\< ab\>}&=&0,\\
\div\uudot+\uudot^2&=&0,\\
\curl\uudot_a&=&0,\\
E_{ab}-\sdel_{\<a}\uudot_{b\>}-\uudot_{\<a}\uudot_{b\>}&=&0,\\
\curl E_{ab}+2\lc_{cd\<a}\uudot^cE_{b\>}^{~~d}&=&0,\\
\div E_a&=&0.
%\\^3{\cal R}_{ab}-E_{ab}&=&0.
\ea
This is already a formidable set of tensor equations; indeed, their solution is
basically an intractable problem unless we introduce a full tetrad or
revert to a metric based approach. These problems become even more severe
when we consider the perturbed spacetime.

In the 1+3 approach this is achieved by assuming that in general all of the 1+3
covariant quantities are non-zero, but that any that vanish in the background
are small~-- small enough that we can neglect products of such terms. To get an
idea of the horrendous nature of the equations, consider the (gauge-dependent)
wave equation for $E_{ab}$, which gives information about gravitational waves
(or linear dynamical tidal forces):
\ba
\ddot
E_{\<ab\>}-\sdel^2E_{ab}&=&-\sdel^c\sdel_{(a}E_{b)c}+5\uudot^c\sdel_cE_{ab}
-2\uudot^c\sdel_{\<a}E_{b\>c}%+\cdots+\mbox{more awful stuff...}\nonumber
-3E_{c\< a}\sdel^c\uudot_{b\>}-E_{c\< a}\sdel_{b\>}\uudot^c\nonumber\\
&+&\:12\lc_{cd\< a}
\curl\uudot^c E_{b\>}^{~~d}+2\bra{\div\uudot+\uudot^2}E_{ab}+3E^c_{~\<a}E_{b\>c}
-6\uudot^c\uudot_{\<a}E_{b\>c}.
\ea
This also contains information about \emph{non-linear} tidal forces, which
don't propagate at the speed of light, by virtue of the presence of the parts
of $E_{ab}$ that do not vanish in the background. But there is no way to
separate the two physical effects. In addition, how could we solve this wave
equation? In contrast to FLRW models, which have only scalars describing the
model after a 1+3 decomposition, it is the presence of non-zero vectors and
tensors in the background spacetime which makes a black hole impossible to
deal with in the 1+3 approach: all the equations have vector-tensor and
tensor-tensor coupling in them, rendering them intractable.

The key, then, for the covariant perturbation approach lies in the fact that,
in the background, the congruence $u^a$ is orthogonal to 3-surfaces of
homogeneity and isotropy. \forget{Projection of all tensors using
$h_{a}^{~b}=g_{a}^{~b}+u_au^b$ basically removes the tensor character of the
equations, as all 4-tensors are `parallel' to $u^a$.} In the case of
inhomogeneous, spherically symmetric systems, projection via $u^a$ is simply
not enough. Another vector field is required that is orthogonal to homogeneous
and isotropic surfaces. Clearly, after an appropriate projection with $u^a$,
such surfaces are provided by spheres surrounding the centre of symmetry, and
the vector orthogonal to these is a radial vector. We turn now to developing
such a formalism.

\subsection{The 1+1+2 formalism}

In the 1+3 approach, a timelike threading vector field $u^a$ ($u^au_a=-1$) is
introduced, representing the observers' congruence. Given this vector field,
the projection tensor $h_{a}^{~b}=g_{a}^{~b}+u_au^b$ is introduced, which
projects all vectors and tensors orthogonal to~$u^a$. Using $h_{ab}$,
any 4-vector may be split into a (1+3 scalar) part parallel
to $u^a$ and a (3-vector) part orthogonal to $u^a$. Any second rank
tensor may be covariantly and irreducibly split into scalar, vector and
\emph{projected, symmetric, trace-free (PSTF)} 3-tensor parts, which requires
the alternating tensor $\lc_{abc}=u^d\eta_{dabc}$~\cite{HvE}. Tensors of higher
rank may be similarly split, but are rarely used (an important exception being
CMB physics~\cite{CL,MGE}). These are the fundamental quantities describing the
spacetime, after the introduction of~$u^a$.

We now introduce another vector field and perform \emph{another} split, but
this time of the 1+3 equations. The `1+1+2' decomposition we develop here has
been partially studied before, mostly in the context of symmetries of solutions
of the EFE~\cite{zafiris,TM2,henk}. It was introduced by~\cite{greenberg} and
further developed in~\cite{TM1,henk}. However, there are importances
differences with the work presented here. \emph{In the following we assume the
1+3 covariant split of the equations} (as given in~\cite{HvE}, for example),
\emph{with all tensors split into scalars, vectors and PSTF tensors with
respect to~$u^a$.}

Take a unit vector~$\n^a$ orthogonal to $u^a$: $\n^a\n_a=1,~u^a\n_a=0$, and
define the projection tensor
\be
\N_a^{~b}\equiv h_a^{~b}-\n_a\n^b=g_{a}^{~b}+u_au^b-\n_a\n^b,
\ee
which projects vectors orthogonal to $\n^a$ (and $u^a$):
$\n^a\N_{ab}=0=u^a\N_{ab}$, onto 2-surfaces ($\N_a^{~a}=2$) which we refer to
as the `sheet' (to carry the sewing analogy of the threading approach into the
realm of the ridiculous).

Any 3-vector $\psi^a$ can now be irreducibly split into a scalar, $\Psi$, which
is the part of the vector parallel to $\n^a$, and a vector, $\Psi^a$, lying in
the sheet orthogonal to $\n^a$;
\be
\psi^a=\Psi\n^a+\Psi^{a},~~~\mbox{where}~~~\Psi\equiv \psi_a\n^a,~~~\mbox{and}~~~\Psi^{a}\equiv
\N^{ab}\psi_b\equiv \psi^{\bar a},
\label{vector-decomp}
\ee
where we use a bar over an index to denote projection with $\N_{ab}$.
\forget{See Fig.~(\ref{sheet-cylinderplot.ps}).
\begin{figure}[ht]
\includegraphics[width=10cm]{{sheet-cylinderplot.ps}}
\caption{\small This shows the steps involved in splitting the 4-vector $\psi^a$
into the 3-vector $\psi^{\< a\>}$, using the projection tensor $h_a^{~b}$, and
the 2-vector $\Psi^a$ on the sheet, which is a projection of $\psi^{\< a\>}$
with $N_a^{~b}$ onto the sheet. This is specialised to the case of a
spherically symmetric black hole, with $u^a$ chosen for static observers, and
$\n^a$ pointing radially outwards. The sheet in this case is actually a
2-sphere surrounding the centre of symmetry.
\label{sheet-cylinderplot.ps}}
\end{figure}}
Similarly, any PSTF tensor, $\psi_{ab}$, can now be split into scalar, vector
and tensor (which are PSTF with respect to $\n^a$) parts:
\be
\psi_{ab}=\psi_{\<ab\>}=\Psi\bra{\n_a\n_b-\:12\N_{ab}}+2\Psi_{(a}\n_{b)}+\Psi_{{ab}},
\label{tensor-decomp}
\ee
where
\ba
\Psi&\equiv &\n^a\n^b\psi_{ab}=-\N^{ab}\psi_{ab},\nonumber\\
\Psi_a&\equiv &\N_a^{~b}\n^c\psi_{bc}=\Psi_{\bar a},\nonumber\\
\Psi_{ ab}&\equiv &\psi_{\lb ab\rb}\equiv
\bra{\N_{(a}^{~~c}\N_{b)}^{~~d}-\:12\N_{ab}\N^{cd}}\psi_{cd}\label{PSTF-TT}.
\ea
We use curly brackets to denote the PSTF with respect to $\n^a$ part of a
tensor. \emph{Note that for 2nd-rank tensors in the 1+1+2 formalism `PSTF' is
precisely equivalent to `transverse-traceless'}.\footnote{Our use of the term
`transverse' only refers to the fact that the tensor is orthogonal to $\n^a$;
this does not imply it is divergence free, which is an additional property of
tensors in the commonly used TT gauge of the plane wave approximation.} Note
also that $h_{\lb
ab\rb}=0$,~$\N_{\<ab\>}=-\n_{\<a}\n_{b\>}=\N_{ab}-\:23h_{ab}$.

We also define the alternating Levi-Civita 2-tensor
\be
\lc_{ab}\equiv\lc_{abc}\n^c = u^d\eta_{dabc}n^c,
\ee
so that $\lc_{ab}\n^b=0=\lc_{(ab)},~
\lc_{abc}= \n_a\lc_{bc}-\n_b\lc_{ac}+\n_c\lc_{ab},~
\lc_{ab}\lc^{cd}=\N_a^{~c}\N_b^{~d}-\N_a^{~d}\N_b^{~c},~\lc_a^{~c}\lc_{bc}=\N_{ab},$
and $\lc^{ab}\lc_{ab}=2.$ Note that for a 2-vector
$\Psi^a$, $\lc_{ab}$ may be used to form a vector orthogonal to $\Psi^a$ but
of the same length.

With these definitions, then, we may split any object into \emph{scalars,
2-vectors in the sheet, and transverse-traceless 2-tensors, also defined in the
sheet.} These three types of objects are the only objects which appear, after a
complete splitting. Hereafter, we will assume such a split has been made, and
`vector' will generally refer to a vector projected orthogonal to $u^a$ and
$\n^a$, and `tensor' will generally mean transverse-traceless tensor, defined
by Eq.~(\ref{PSTF-TT}).

There are two new derivatives of interest now, which $\n^a$ defines, for any
object $\psi_{\cdots}^{~~\cdots}$:
\ba
\hat \psi_{a\cdots b}^{~~~~~c\cdots d}&\equiv &
\n^e \sdel_e\psi_{a\cdots b}^{~~~~~c\cdots d},\label{hatdef}\\
%\delta_a \psi_{\cdots}^{~~\cdots}&\equiv & \N_a^{~b}\N_\cdot^{~\cdot}\cdots
%\N_\cdot^{~\cdot}\sdel_b \psi_{\cdots}^{~~\cdots}.\label{deltadef}
\delta_e \psi_{a\cdots b}^{~~~~~c\cdots d}&\equiv & \N_e^{~j}\N_a^{~f}\cdots
\N_b^{~g}\N_h^{~c}\cdots\N_i^{~d}\sdel_j \psi_{f\cdots g}^{~~~~~h\cdots i}.\label{deltadef}
\ea
The hat-derivative is the derivative along the vector field $\n^a$ in the
surfaces orthogonal to $u^a$. This definition represents a conceptual
divergence from the tetrad approach, in which the basis vectors appear on an
equal footing [i.e.,~with $\del_a$ rather than $\sdel_a$ in
Eq.~(\ref{hatdef})]. As a result, the congruence~$u^a$ retains the primary
importance it has in the 1+3 covariant approach.
%(Actually this really only determines notation:
(We choose to think of ${\cal A}\equiv u^an^b\del_a u_b= -u^au^b\del_a n_b$ as
the radial component of the acceleration of~$u^a$, rather than the time
component of $\dot{n}^a$.) The $\delta$-derivative, defined by
Eq.~(\ref{deltadef}) is a projected derivative on the sheet, with projection on
every free index.

With these definitions we may now decompose the covariant derivative of $\n^a$
orthogonal to $u^a$:
\be
\sdel_a\n_b=\n_a\hatn_b+\:12\phi \N_{ab}+\xi\lc_{ab}+\zeta_{ab},
\ee
where
\ba
\hatn_a &\equiv &\n^c\sdel_c\n_a=\hat \n_a,\\
\phi &\equiv &\delta_a \n^a,\\
\xi &\equiv &\:12\lc^{ab}\delta_a\n_b,\\
\zeta_{ab} &\equiv &\delta_{\lb a}\n_{b\rb}.
\ea
We may interpret these as follows: travelling along $\n^a$, $\phi$ represents
the sheet expansion, $\zeta_{ab}$ is the shear of $\n^a$ (distortion of the
sheet), and $\hatn^a$ its acceleration, while $\xi$ represents a `twisting' of
the sheet~-- the rotation of $\n^a$~\cite{TM1}. The other derivative of $\n^a$
is its change along $u^a$,
\be
\dot n_{a}=\udot u_a+\dotn_a~~~\mbox{where}~~~\dotn_a\equiv\dot n_{\bar a}
~~~\mbox{and}~~~\udot=\n^a\uudot_a.
\ee
The new variables $\hatn_a$, $\phi$, $\xi$, $\zeta_{ab}$ and $\dotn_a$ are
fundamental objects in the spacetime, and their
dynamics gives us information about the spacetime geometry. They are
treated on the same footing as the kinematical variables of $u^a$ in the 1+3
approach (which also appear here).

Note that for a scalar, we have $\sdel_a\Psi=\hat\Psi\n_a+\delta_a\Psi$, while
for any vector $\Psi^a$ orthogonal to $\n^a$ and $u^a$
(i.e.,~$\Psi^a=\Psi^{\bar a}$), we may decompose the different parts of its
spatial derivative:
\be
\sdel_a \Psi_b=-\n_a\n_b\Psi_c\hatn^c+\n_a\hat\Psi_{\bar b}-\n_b\bras{\:12\phi\Psi_a+
\bra{\xi\lc_{ac}+\zeta_{ac}}\Psi^c}+\delta_a\Psi_b.
\ee
Similarly, for a tensor $\Psi_{ab}$: $\Psi_{ab}=\Psi_{\{ab\}}$, we have
\be
\sdel_a\Psi_{bc}=-2\n_a \n_{(b}\Psi_{c)d}\hatn^d
 +\n_a\hat \Psi_{bc}-2\n_{(b}\bras{
\:12\phi \Psi_{c)a}+\Psi_{c)}^{~~d}\bra{\xi \lc_{ad}+\zeta_{ad}}}
 +\delta_a\Psi_{bc}.
\ee
Note that $
\dot\N_{ab}=2u_{(a}\uudot_{b)}-2\n_{(a}\dot\n_{b)},~
\hat\N_{ab}=-2\n_{(a}\hatn_{b)},~
\delta_c\N_{ab}=0;$ while
$\dot\lc_{ab}=-2u_{[a}\lc_{b]c}\udot^c+2\n_{[a}\lc_{b]c}\dotn^c,~
\hat\lc_{ab}=2\n_{[a}\lc_{b]c}\hatn^c,~
\delta_c\lc_{ab}=0.$

%\subsection{Splitting the kinematical, Weyl and matter tensors}

We take $\n^a$ to be arbitrary at this point, and then split the usual 1+3
kinematical and Weyl quantities into the irreducible
set~$\{\theta,\udot,\Omega,\Sigma,{\cal E},{\cal H},\udot^a,\Sigma^a,{\cal
E}^a,{\cal H}^a,\Sigma_{ab},{\cal E}_{ab},{\cal H}_{ab}\}$ using
(\ref{vector-decomp}) and~(\ref{tensor-decomp}) as follows:
\ba
\uudot^a&=&\udot \n^a+\udot^a,\\
\omega^a&=&\Omega \n^a+\Omega^a,\\
\sigma_{ab}&=&\Sigma\bra{\n_a\n_b-\:12\N_{ab}}+2\Sigma_{(a}\n_{b)}+\Sigma_{ab},\\
E_{ab}&=&{\cal E}\bra{\n_a\n_b-\:12\N_{ab}}+2{\cal E}_{(a}\n_{b)}+{\cal E}_{ab},\\
H_{ab}&=&{\cal H}\bra{\n_a\n_b-\:12\N_{ab}}+2{\cal H}_{(a}\n_{b)}+{\cal
H}_{ab}.
\ea
\forget{Similarly we may split the fluid variables $q^a$ and $\pi_{ab}$,
\ba
q^a&=&Q \n^a+Q^a,\\
\pi_{ab}&=&\Pi\bra{\n_a\n_b-\:12\N_{ab}}+2\Pi_{(a}\n_{b)}+\Pi_{ab},
\ea
although we won't use these here.}

Having described the splitting of the 1+3 variables to obtain their 1+1+2 parts,
and the introduction of the new 1+1+2 variables corresponding to the irreducible
parts of~$\del_a\n_b$, it only remains to apply this splitting procedure to
the 1+3 equations themselves. We give these equations in
section~\ref{the-equations}, linearised about a Schwarzschild background.

\subsection{The Ricci identities}

Once the vector $\n^a$ has been introduced it is possible, and necessary,
to augment the 1+3 equations with the Ricci identities for~$\n^a$;
without these we do not have enough equations to determine the new
1+1+2 variables. The Ricci identities for $\n^a$ are
\be
R_{abc}\equiv2\del_{[a}\del_{b]}\n_c-R_{abcd}\n^d=0,\label{ricci}
\ee
where $R_{abcd}$ is the Riemann curvature tensor. This 3-index tensor may be
covariantly split using the two vector fields $u^a$ and $n^a$, and
gives dynamical equations for the covariant parts of the derivative of $\n^a$
(namely $\dotn_a$, $\hatn_a$, $\phi$, $\xi$ and~$\zeta_{ab}$) in the form
of \emph{evolution} equations, involving dot-derivatives of these variables,
and \emph{propagation} equations, involving hat-derivatives. In order to
facilitate the calculation of these Ricci identities, which appear
in the following section, we give here the expression for the full
covariant derivative of $\n^a$ in terms of the relevant 1+1+2 variables:
\be
\del_a\n_b=-\udot u_au_b-u_a\dotn_b +\bra{\Sigma+\:13\theta}\n_a u_b
+ \bras{\Sigma_a-\lc_{ac}\Omega^c}u_b +\n_a\hatn_b
+\:12\phi\N_{ab}+\xi\lc_{ab}+\zeta_{ab},
\ee
which may be inserted into Eq.~(\ref{ricci}).

\subsection{Commutation relations}

In general the three derivatives we now have defined,
$`\dot{\phantom{x}}$',~$`\hat{\phantom{x}}$' and $`\delta_a$' do not commute.
Instead, when acting on a scalar $\psi$, they satisfy:
\ba
\hat{\dot \psi}-\dot{\hat \psi}&=&-\dot\psi\udot+\bra{\:13\theta+\Sigma}\hat\psi
+\bra{\Sigma_a+\lc_{ab}\Omega^b-\dotn_a}\delta^a\psi,\label{comm-un}
\\
\delta_a\dot\psi-\N_a^{~b}\bra{\delta_b\psi}^\cdot&=&-\udot_{a}\dot\psi+
\bra{\dotn_{a}+\Sigma_a-\lc_{ab}\Omega^b}\hat\psi
+\bra{\:13\theta-\:12\Sigma}\delta_a\psi +
\bra{\Sigma_{ab}+\Omega\lc_{ab}}\delta^b\psi,
\\
\delta_a\hat\psi-\N_a^{~b}\widehat{\bra{\delta_b\psi}}
&=&-2\lc_{ab}\Omega^b\dot\psi+\hatn_a\hat\psi+\:12\phi\delta_a\psi+
\bra{\zeta_{ab}+\xi\lc_{ab}}\delta^b\psi,
\\
\delta_a\delta_b\psi-\delta_b\delta_a\psi&=& 2\lc_{ab}\bra{\Omega\dot\psi-\xi\hat\psi}
+2\hatn_{[a}\delta_{b]}\psi.\label{comm1}
\ea
These last two equations are the decomposition of the 1+3 commutation relation
\be
\curl\sdel_a\psi=2\dot\psi\omega_a.
\ee
These relations are considerably more complicated for vectors and tensors.

From Eq.~(\ref{comm1}), we see that our sheet will be a genuine 2-surface in
the spacetime (and, in particular, that the derivative~$\delta_a$ will be a
true covariant derivative on this surface) if and only if
$\xi=\Omega=\hatn^a=0$. (Recall that the 1+3 spatial metric $h_{ab}$
corresponds to a genuine 3-surface when $\omega^a=0$.) Otherwise, the sheet is
really just a collection of tangent planes. In addition, the two vectors $u^a$
and $\n^a$ are 2-surface forming if and only if the commutator $[u,n]$
in~(\ref{comm-un}) has no component in the sheet: that is, when Greenberg's
vector, $\Sigma^a+\lc^{ab}\Omega_b-\dotn^a$, vanishes~\cite{zafiris}~-- see
Eq.~(\ref{comm-un}).

\section{Perturbations around a Schwarzschild black hole}
\label{the-equations}

For an exact Schwarzschild black hole it turns out the the only non-zero 1+1+2
variables are the scalars $\{\udot,{\cal E},\phi\}$ (and their derivatives
$\{\hat{\udot},\hat{\cal E},\hat\phi \}$), a covariant characterisation of the
Schwarzschild solution. (We saw in section~\ref{why1+3} that $\udot$ and ${\cal
E}$ are the only non-zero parts of the 1+3 variables, and it is clear by
considering Gauss' theorem that $\phi$, the divergence of~$\n^a$, must also be
non-zero. We will consider the background solution in more detail in
section~\ref{background-sol}.) \emph{Because the background solution involves
only scalars,  under any perturbation all vectors and tensors are first-order},
which greatly simplifies things, as we discussed in section~\ref{why1+3}.

The usual 1+3 evolution and constraint equations may be further split with the
vector $\n^a$, into a set of \emph{evolution} (along $u^a$) and
\emph{propagation} (along $\n^a$) equations. Together with the Ricci identities
for $\n^a$, we find a complete set of propagation, evolution and constraint
equations~-- the constraints being those equations with no hat- or
dot-derivatives in them. We will give the complete nonlinear equations
elsewhere, as they are large and unpleasant. Here, however, we will give the
vacuum equations linearised around a Schwarzschild black hole background.
Our linearisation procedure is straightforward: as in the 1+3 approach we
neglect all products of first-order quantities; first-order quantities being
those which vanish in the background:
\be
\{\theta,\Omega,\Sigma,\xi,{\cal H},\dotn^a,\hatn^a,\udot^a,\Omega^a,\Sigma^a,{\cal E}^a,{\cal
H}^a,\zeta_{ab},\Sigma_{ab},{\cal E}_{ab},{\cal H}_{ab}\}={\cal O}(\epsilon)
\ee
(along with their derivatives, and dot- and $\delta$-derivatives of
$\{\udot,{\cal E},\phi\}$), where $\epsilon$ is a `smallness' parameter, which
measures departures from an exact black hole. So, for example, one could define
$\epsilon\simeq\Omega/\phi$, or $\epsilon\simeq\sqrt{\E_{ab}\E^{ab}}/\E$, and
so on. From now on all equations are linearised about a Schwarzschild black
hole, and equations of the form $A=B$ generally mean $A-B={\cal O}(\epsilon^2)$
(in keeping with usual practice, we will not distinguish between this and real
equality).

{When studying cosmological perturbations using the 1+3 approach, the evolution
equations are of prime importance, since time is the only remaining essential
parameter: the goal is to find the evolution of seed perturbations
corresponding to the various spatial harmonics. In contrast, for the black hole
perturbations analysed here the time invariance and spherical symmetry of the
background mean that radius is the interesting parameter, and so the
propagation equations are the key: we want to find the variation with radius of
the various (time and spherical) harmonic components. Thus, we present the
propagation (hat) equations first, and relegate the evolution equations to a
secondary position. This will be helpful when we come to solve the equations in
Section~\ref{solution-section}.}

\vskip\baselineskip

Propagation\footnote{These equations are derived as follows:
    Eq.~(\ref{phihatBH}) from $\n^a\N^{bc}R_{abc}$;
    Eq.~(\ref{xihat}) from $\n^a\lc^{bc}R_{abc}$;
    Eq.~(\ref{hatndotnl}) from $u^a\n^bR_{ab\bar c}=\n^au^bR_{ab\bar c}$;
    Eqs.~(\ref{hattheta-sigma}) and (\ref{hatSig_anl}) from the shear divergence equation, $(C_1)^a$;
    Eq.~(\ref{thetadotfull}) from the Raychaudhuri equation;
    Eq.~(\ref{hatOmega}) from the rotation divergence equation, $C_2$;
    Eqs.~(\ref{EhatBH}) and (\ref{hatE_a}) from the electric Weyl divergence equation, $(C_4)^a$;
    Eqs.~(\ref{HhatBH}) and (\ref{hatH_a}) from the magnetic Weyl divergence equation, $(C_5)^a$;
    Eq.~(\ref{hatA_a}) from the shear evolution equation;
    Eq.~(\ref{hatSigma_ab}) from the $H_{ab}$-equation, $(C_3)_{\lb ab\rb}$;
    Eq.~(\ref{zetahat}) from $\n^aR_{a\lb bc\rb}$;
    Eq.~(\ref{Eabdot}) from the electric Weyl evolution equation;
    and Eq.~(\ref{Habdot}) from the magnetic Weyl evolution equation.
    }:
\ba
\hat\phi &=&-\:12\phi^2-{\cal E}+\delta_a\hatn^a,\label{phihatBH}\\
\hat\xi &=&-\phi\xi+\:12\lc_{ab}\delta^a\hatn^b,\label{xihat}\\
\hat\dotn_{\bar a}-\dot\hatn_{\bar a} &=& -\bra{\:12\phi+\udot}\dotn_{a}
+\bra{\:12\phi-\udot}\bra{\Sigma_a+\lc_{ab}\Omega^b}-\lc_{ab}\H^b,\label{hatndotnl}\\
\:23\hat\theta-\hat\Sigma
&=&\:32\phi\Sigma+\delta_a\Sigma^a+\lc_{ab}\delta^a\Omega^b,\label{hattheta-sigma}\\
\hat{\udot}-\dot\theta &=&-\udot\bra{\phi+\udot}-\delta_a\udot^a,\label{thetadotfull}\\
\hat\Omega &=&\bra{\udot-\phi}\Omega-\delta_a\Omega^a,\label{hatOmega}\\
   % =-2\udot\Omega+\:12\delta_a\Omega^a-\:32{\cal H}+\:32\lc_{ab}\delta^a\Sigma^b,\\
\hat{\cal E}&=&-\:32\phi{\cal E}-\delta_a{\cal E}^a,\label{EhatBH}\\
\hat{\cal H}&=& -\:32\phi{\cal H}-\delta_a{\cal H}^a-3{\cal E}\Omega,\label{HhatBH}\\
%\hat\Omega_{\bar a}&=&-\bra{\:12\phi-2\udot}\Omega_a-2{\cal H}_a-\delta_a\Omega
%-\lc_{ab}\bra{\hat\Sigma^b-\:12\Sigma^b-\:32\delta^b\Sigma}
%+\lc_{bc}\delta^b\Sigma_a^{~c},\\
\hat\Sigma_{\bar a}-\lc_{ab}\hat\Omega^b &=& -\:32\phi\Sigma_a
+{\bra{2\udot+\:12\phi}\lc_{ab}\Omega^b}+\:23\delta_a\theta+
\:12\delta_a\Sigma-\lc_{ab}\delta^b\Omega-\delta^b\Sigma_{ab}
,\label{hatSig_anl}\\
\hat\udot_{\bar a}-2\dot\Sigma_{\bar a}&=&2\E_a-\udot\hatn_a-\delta_a\udot
-2\bra{\udot-\:14\phi}\udot_a,\label{hatA_a}\\
%\hat\Sigma_{\bar a}&=&-\:32\phi\Sigma_a+\:12\delta_a\Sigma-\delta^b\Sigma_{ab}
%+\:23\delta_a\theta+\lc_{ab}\bra{\hat\Omega^b+\bra{\:12\phi+2\udot}\Omega^b
%+\delta^b\Omega},\\
\hat{\cal E}_{\bar a}&=&-\:32\phi{\cal E}_a+\:12\delta_a{\cal E}-\:32{\cal
E}\hatn_a-\delta^b{\cal E}_{ab},\label{hatE_a}\\
\hat{\cal H}_{\bar a}&=&-\:32\phi{\cal H}_a+\:12\delta_a{\cal H}-\delta^b
{\cal H}_{ab}
+\:32{\cal E}\bra{\Omega_a-\lc_{ab}\Sigma^b},\label{hatH_a}\\
%\lc_{c(a}\hat\Sigma_{b)}^{~~c}&=&-\lc_{c(a}\bra{\:12\phi\Sigma_{b)}^{~~c}
%+\delta^c\Sigma_{b)}+\delta_{b)}\Sigma^c}+\delta_{\lb a}\Omega_{b\rb }+{\cal
%H}_{ab}.
\hat\Sigma_{\lb ab\rb}&=&-\:12\phi\Sigma_{ab}
+\delta_{\lb a}\Sigma_{b\rb}-\lc_{c\lb a}\delta^{c}\Omega_{b\rb }-\lc_{c\lb
a}{\cal H}_{b\rb}^{~~c},\label{hatSigma_ab}\\
\hat\zeta_{\lb ab\rb }&=&-\phi\zeta_{ab}-{\cal E}_{ab}+\delta_{\lb a}\hatn_{b\rb }
,\label{zetahat}\\
\dot{\cal E}_{\lb ab\rb }-\lc_{c\lb a}\hat\H_{b\rb}^{~~c}&=&-\:32{\cal E}\Sigma_{ab}
+\lc_{c\lb a}\bra{-\delta^c{\cal H}_{b\rb} +\bra{\:12\phi+2\udot}{\cal H}_{b\rb}^{~~c}},\label{Eabdot}\\
\dot{\cal H}_{\lb ab\rb }+\lc_{c\lb a}\hat\E_{b\rb}^{~~c}&=&\lc_{c\lb a}\bra{\delta^c{\cal E}_{b\rb }
+\:32{\cal E}\zeta^{~~c}_{b\rb}-\bra{\:12\phi+2\udot}{\cal E}_{b\rb
}^{~~c}}\label{Habdot}.
\ea

Evolution\footnote{These are derived as follows:
    Eq.~(\ref{phidot}) from $u^a\N^{bc}R_{abc}$;
    Eq.~(\ref{xidot}) from $u^a\lc^{bc}R_{abc}$;
    Eq.~(\ref{dotOmega}) from the rotation evolution equation;
    Eq.~(\ref{sigmadotfull}) from $u^a\n^bu^cR_{abc}=-\n^au^bu^cR_{abc}$ and
    the Raychaudhuri equation;
    Eqs.~(\ref{Edot}) and~(\ref{Eadot}) from the electric Weyl evolution
    equation [(\ref{Eadot}) also uses~(\ref{hatH_a})];
    Eqs.~(\ref{Hdot}) and~(\ref{Hadot}) from the magnetic Weyl evolution
    equation [(\ref{Hadot}) also uses~(\ref{hatE_a})];
    Eq.~(\ref{dotsigom}) from the rotation and shear evolution equations;
    Eq.~(\ref{Sigmaabdot}) from the shear evolution equation;
    and Eq.~(\ref{dotzeta}) from $u^cR_{c\lb ab\rb}$.
    }:
\ba
\dot\phi &=&\bra{\:12\phi-\udot}\bra{\Sigma-\:23\theta}+\delta_a\dotn^a,\label{phidot}\\
\dot\xi &=&-\bra{\:12\phi-\udot}\Omega+\:12{\cal H}+\:12\lc_{ab}\delta^a\dotn^b,\label{xidot}\\
\dot\Omega &=&\xi\udot+\:12\lc_{ab}\delta^a\udot^b,\label{dotOmega}\\
\:23\dot\theta-\dot\Sigma &=&{\cal E}+{\phi{\udot}}+\delta_a\udot^a,\label{sigmadotfull}\\
\dot{\cal E}&=&\bra{\:32\Sigma-\theta}{\cal E}+\lc_{ab}\delta^a{\cal H}^b,\label{Edot}\\
\dot{\cal H}&=&-3{\cal E}\xi-\lc_{ab}\delta^a{\cal E}^b,\label{Hdot}\\
\dot\Sigma_{\bar a}-\lc_{ab}\dot\Omega^b &=&-{\cal E}_a+\delta_a\udot
+\bra{\udot-\:12\phi}\udot_a,\label{dotsigom}\\
\dot{\cal E}_{\bar a}&=&-\:32\E\dotn_{a}
+\:12\lc_{ab}\bra{\delta^b{\cal H} +\bra{\phi-2\udot}{\cal H}^b}-
\lc_{c\lb d}\delta^d{\cal H}_{a\rb}^{~~c},\label{Eadot}
\\
\dot{\cal H}_{\bar a}&=&
-\:32\E\lc_{ab}{\udot^b}-\:12\lc_{ab}\bra{\delta^b{\cal E}
+\bra{\phi-2\udot}{\cal E}^b}+\lc_{c\lb d}\delta^d{\cal
E}_{a\rb}^{~~c},\label{Hadot}
\\
\dot\Sigma_{\lb ab\rb }&=&\udot\zeta_{ab}-{\cal E}_{ab}+\delta_{\lb a}\udot_{b\rb},\label{Sigmaabdot}\\
\dot\zeta_{\lb ab\rb }&=& -\bra{\:12\phi-\udot}\Sigma_{ab}-\lc_{c\lb a}{\cal
H}_{b\rb }^{~~c}+\delta_{\lb a}\dotn_{b\rb }.\label{dotzeta}
\ea

Constraint\footnote{These are derived as follows:
    Eq.~(\ref{conom_anl}) from $(C_3)_{ab}\n^b$ and $(C_1)_{\bar a}$, or $\n^a u^cR_{a\bar
    bc}$;
    Eq.~(\ref{consig_anl}) from $\lc^{ab}u^cR_{abc}$;
    and Eq.~(\ref{newcons}) from $\N^{bc}R_{\bar abc}$. Note that
    the equation formed from $(C_3)_{ab}\n^a\n^b$ is equivalent to
    Eqs~(\ref{hatOmega}) and~(\ref{consig_anl}).
    }:
\ba
\delta_a\Sigma-\:23\delta_a\theta +2\lc_{ab}\delta^b\Omega+2\delta^b\Sigma_{ab}
&=& -\phi\Sigma_a+\phi\lc_{ab}\Omega^b- 2\lc_{ab}\H^b
,\label{conom_anl}\\
%%%%%%%%%%%%%%%%%%%%%%%%%%%%%%%%%%%%%%%%%%%%%%%%%%%%%%%%%%%%%%%%%%%%%%%%%%%%%%%%%%%%
\delta_a\Omega^a+\lc^{ab}\delta_a\Sigma_b &=& \bra{2\udot-\phi}\Omega
 +
{\cal H},\label{consig_anl}\\
% 0 &=& \lc_{ab} (\E^b + (1/2)\delta^b\phi) + \delta_a\xi -
%\lc^{bc}\delta_b\zeta_{ca}~~~TIDY UP
\:12\delta_a\phi-\lc_{ab}\delta^b\xi-\delta^b\zeta_{ab}&=&-\E_a.\label{newcons}\label{barEV}
\ea

These equations must of course be consistent with one another: the
constraints must evolve and propagate consistently, and each first-order
variable must satisfy the commutation relation
\be
{\cal C}[\Psi]\equiv\dot{\hat \Psi}-\hat{\dot \Psi}-\udot\dot
\Psi=0,\label{integrab}
\ee
while the background scalars must satisfy the commutation
relation~(\ref{comm-un}). It turns out after an arduous calculation that all
the equations are consistent.

\subsection{Frame choice}
\label{frame-choice}

In this work we are presenting a \emph{gauge invariant}, covariant approach to
perturbations of spherically symmetric spacetimes, based on the introduction of
a \emph{partial frame}, that is, of two basis vectors\forget{, one representing
a timelike observer congruence, and the other the radial direction at each
point}. \forget{It will be helpful here to consider precisely the extent to
which such partial-frame (and, indeed, full-frame) techniques are independent
of the degrees of freedom available in the description of the perturbations
(see e.g.,~\cite{BS}).} In GR there are two `gauge' freedoms: the choice of
coordinates and the freedom to choose a frame basis in the tangent space at
each point. These have their direct analogues in perturbation theory, where we
imagine that the true spacetime we are studying is `close to' some given,
idealised background spacetime. Although not formally the same thing, choosing
a coordinate system in the true spacetime is in practice equivalent to fixing
the mapping between the true and background spacetimes allowing the direct
comparison of scalar, vector and tensor objects in the two spacetimes at
corresponding points (see~\cite{BS}). In metric-based (non-covariant)
approaches to the perturbation problem the first step is to find a nice
coordinate system in the true spacetime, corresponding to that in the
background, and to write equations for the derivatives of the perturbations of
scalars, vectors and tensors with respect to these coordinates.

In covariant (partial-)frame approaches, on the other hand, one tries to
avoid explicit reference to the background, using it merely to determine
which quantities are zeroth order (i.e.,~which do not vanish in the
background). Given the frame vectors, a set of covariant (that is, coordinate
invariant) equations describing the true spacetime are written down.
(Coordinate-)gauge invariance will hold, according to the Stewart-Walker
lemma~\cite{SW}, if we can find a complete set of variables all of which
vanish in the background. However, since the covariant variables are the
projections of tensors with respect to the frame vectors and the projected
parts of the of frame-vector covariant derivatives, the equations and their
solutions will, in general, depend on the particular choice of frame vectors.
\forget{For example, when it is said that the 1+3 covariant perturbation
formalism is gauge invariant, what is meant is that the equations and solutions
are the same regardless of the mapping of the true spacetime to the background,
by virtue of the Stewart-Walker lemma.} However, since the true spacetime lacks
the symmetry of the background, there is, in general, no unique covariant
definition of the frame vectors, and one can always perform a first-order
rotation of these. This freedom can easily be seen in the preceding set of
equations: there are no evolution equations for $\udot,~\udot_a$, and
$\alpha_a$, while there is no propagation equation for $a_a$. Indeed, this is
true in any spacetime, as one can \emph{choose} the frame vectors at any point
freely, the motion of which must be put into the equations by hand (GR can't
predict this!).

\forget{timelike congruence~$u^a$: given any choice of this frame vector it is possible
to obtain many other valid choices by making a (small, first-order) boost of
the observer's velocity at each point\forget{ (although note that in
cosmological applications, a unique, covariant choice of $u^a$ may be made when
the matter is assumed to be perfect fluid, since there will be only one frame
for which this is true)}. The equations and solutions will in general depend on
the $u^a$ chosen. This is doubly so in the 1+1+2 formalism, where we must
choose two frame vectors\forget{ (see section~\ref{solution-frame}, in which we
discuss these issues further)}.}

In what follows we will reserve the term `gauge invariant' to refer to the
invariance of the equations under the mapping between the true and background
spacetimes (in the sense of the Stewart-Walker lemma), and will use `frame
invariant' to describe invariance under the choice of frame vectors. Our
1+1+2 formalism, in common with the 1+3 and Newman-Penrose approaches, is
gauge invariant, but not frame invariant. \forget{In general, frame and
partial-frame methods may be made \emph{gauge invariant} through a judicious
choice of variables (variables that vanish in the background), but they will
not be frame invariant.}

The equations as they are presented above are completely general, involving
no particular choice of either frame vector. Given a timelike
congruence~$u^a$, obvious choices are possible: for example,  we could take
$\n^a$ to be parallel to the acceleration~$\dot{u}^a$, leading to the frame
condition~$\udot^a=0$, a physically plausible choice for observers hovering
above the black hole as one could always dexterously align ones rocket to
make this so (objects dropped onto the floor of the rocket would fall
directly `down'); or we could demand that $\n^a$ be parallel to the `radial'
eigenvector of the electric Weyl tensor~$E_{ab}$, which leads to the frame
choice~$\E^a=0$. Indeed, we can set any 2-vector to zero by these
considerations. We shall not impose a frame condition at the moment, however,
as there is no need.

\forget{
However, the system of equations is already ferociously complicated and it
will simplify matters somewhat if we eliminate some of our freedom to choose
a frame immediately by making a specific choice for the radial vector~$\n^a$.
What we require is a covariant definition of~$\n^a$ that results in $\n^a$
being purely radial in the background. Given the timelike congruence~$u^a$,
obvious choices are possible: for example,  we could take $\n^a$ to be
parallel to the acceleration~$\dot{u}^a$, leading to the frame
condition~$\udot^a=0$; or we could demand that $\n^a$ be parallel to the
`radial' eigenvector of the electric Weyl tensor~$E_{ab}$, which leads to the
frame choice~$\E^a=0$. Although this latter option seems appealing in many
ways we will choose $\n^a$ to lie parallel to the acceleration vector, so
that $\udot^a=0$ always. This fixes completely our choice of $\n^a$ in the
perturbed spacetime; we have not, however, fully determined our choice of
observer. This is our remaining frame freedom. \forget{We will discuss in
section~\ref{solution-frame} the significance of this remaining frame
freedom, and the possible effect of dropping the frame condition on~$\n^a$.}
}

\subsection{Background Solutions}
\label{background-sol}

In the background we have, by setting all vectors, tensors and time derivatives
to zero and retaining only the zeroth-order scalars:
\ba
\hat\phi &=&-\:12\phi^2-{\cal E},\label{phihatrad}\\
\hat{\cal E}&=&-\:32\phi{\cal E}\label{ehatrad};
\ea
together with
\be
{\cal E}+\udot\phi=0,\label{udotbackground}
\ee
which is Eq.~(\ref{sigmadotfull}). Note that these three equations are
sufficient: Eq.~(\ref{thetadotfull}) is satisfied.
% and~(\ref{sigmadotfull}) are satisfied.

If we associate our hat-derivative with an affine parameter $\rho$, i.e.,
$\hat{\phantom{x}} =d/d\rho$, then we may solve these equations. The parametric
solutions to our background equations are, in terms of either a parameter $x$
or $r$,
\ba
{\cal E}&=&-\;{1}{(2m)^2}\sech^6 x=-\;{2m}{r^3},\\
\phi &=&\;{1}{m}\sinh x \,\sech^3 x=\;2r\sqrt{1-\;{2m}{r}},\\
\udot&=&\;{1}{4m}\cosech x\, \sech^3 x=\;{m}{r^2}\bra{1-\;{2m}{r}}^{-1/2};
\label{backsols(x)}
\ea
where
\be
\rho = 2m\brac{x+\sinh x\, \cosh x},\label{rho(x)}
\ee
and the usual Schwarzschild coordinate
\be
r=2m\cosh^2 x.\label{rsch}
\ee
These form a one-parameter family of solutions, parameterised by the constant
$m$, which is just the Schwarzschild mass. The Schwarzschild solution is given
for $2m<r<\infty$ for $0<x<\infty$; the interior solution may be found by this
approach, but we will not require it here.

\begin{figure}[ht]
\includegraphics[width=10cm]{{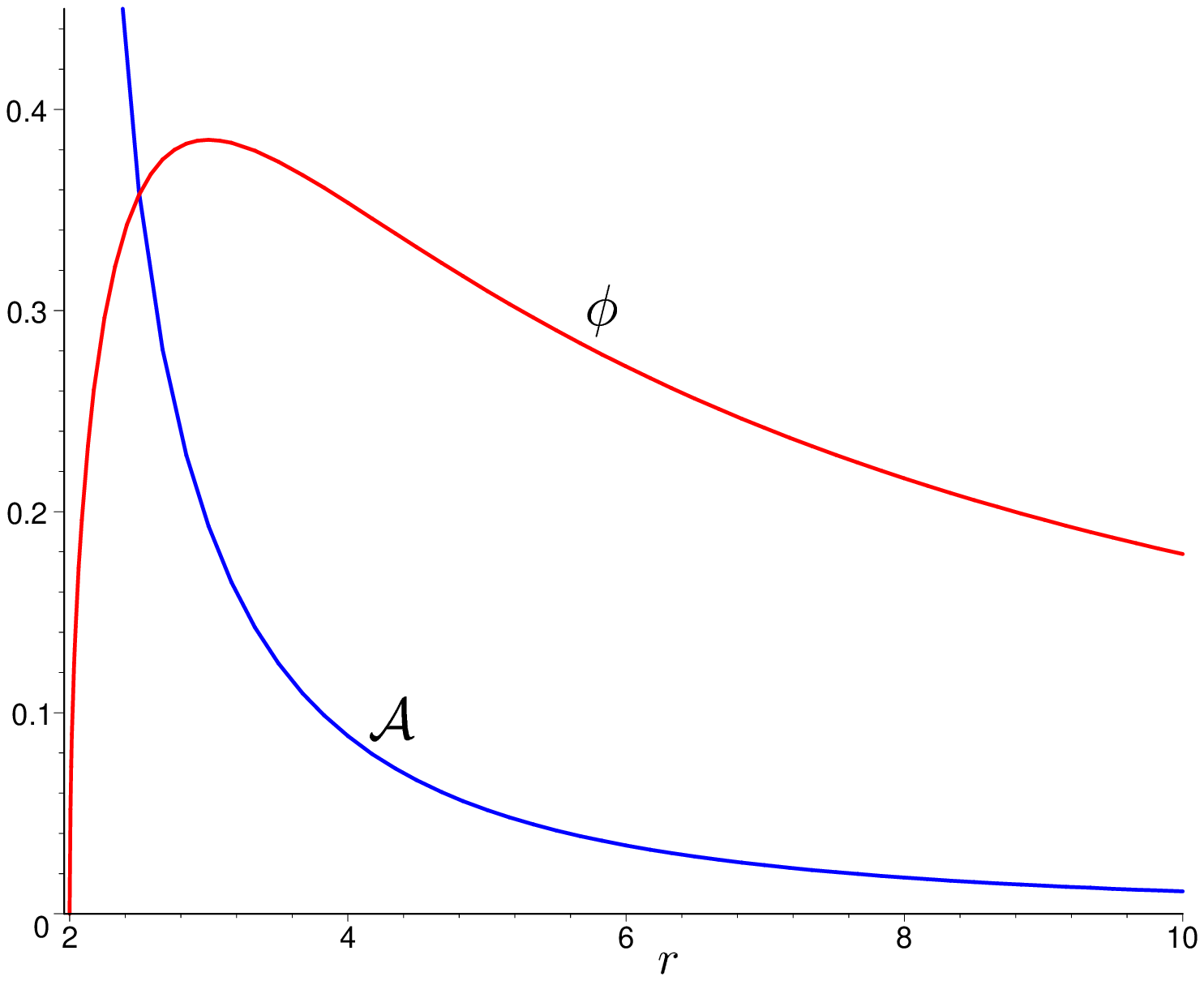}}
\caption{\small A plot of $\phi$ and $\udot$ with $r$, showing the maxima of $\phi$
at the photon sphere, $r=3m$. In contrast, $\udot$ falls from $\infty$ at the
horizon.
\label{phi-and-udot-vs-r.ps}}
\end{figure}

We show a plot of $\phi$ and $\udot$ with $r$ in
Fig.~\ref{phi-and-udot-vs-r.ps}. This shows how the expansion of $\n^a$ starts
from zero at the horizon, is largest at the photon sphere, before dropping to
zero again as $r\rightarrow\infty$.

The solutions given by~(\ref{backsols(x)}) represent the general solution to
the system of equations~(\ref{phihatrad}),~(\ref{ehatrad})
and~(\ref{udotbackground}), which are the covariant 1+1+2 equations under the
conditions of the spacetime being static, spherically symmetric and vacuum;
hence, we may covariantly characterise the Schwarzschild solution by
$\{\udot,{\cal E},\phi\}\neq0$, with all other covariant quantities being
zero.\footnote{We would like to thank an anonymous referee for drawing this
to our attention.}

\forget{It is interesting to note that upon making the replacements in our background
equations
\be
{\cal E}\rightarrow\:13\kappa^2\mu_{RW},~~~
\phi\rightarrow\:23\kappa\theta_{RW},~~~
\rho\rightarrow\;1\kappa t_{RW},
\ee
($\kappa$ is a constant) we have the evolution equations for a dust FLRW model;
Eq.~(\ref{udotbackground}) becomes the Friedman equation, with the acceleration
taking the role of the curvature, and the radial tidal force the energy
density. Such a substitution works for a $\gamma=-1$ equation of state RW
model, now with the alternative replacements
\be
{\cal E}\rightarrow-\:53\kappa^2\mu_{RW},~~~
\phi\rightarrow-\:23\kappa\theta_{RW},~~~
\rho\rightarrow\;1\kappa t_{RW}.
\ee
}

\subsection{Gauge invariant variables}\label{GIvars}

We have developed a set of covariant equations describing a perturbed black
hole, with all variables defined with respect to a family of observers~$u^a$
with a preferred radial vector field~$\n^a$, which these observers can choose;
first-order variables have a clear physical or geometrical meaning. However,
not all variables appearing in the equations are gauge invariant, because they
are not first order. Recall the Stewart-Walker lemma~\cite{SW}, which states
that if a variable vanishes in the background then it is gauge invariant in the
perturbed spacetime.

Equations~(\ref{thetadotfull}),~(\ref{sigmadotfull}),~(\ref{phihatBH}),
and~(\ref{EhatBH}) are not gauge invariant because zeroth-order terms
appear in these equations, i.e.,~isolated terms involving
$\{\udot,{\cal E},\phi\}$ not multiplied by a first-order quantity. We therefore
introduce the set of gauge-invariant variables
\ba
X_a &=& \delta_a {\cal E},\\
Y_a &=& \delta_a \phi, \\
Z_a &=&\delta_a\udot,
\ea
which do vanish in the background ($\{\udot,{\cal E},\phi\}$ do not vary over a
sphere in the background), and so are gauge invariant, by the Stewart-Walker
lemma~\cite{SW}.
It is important to notice that we lose no true degrees of freedom in the
solutions to the equations by introducing these variables, since we
only eliminate possible spherically symmetric perturbations (for which $X_a$,
$Y_a$ and $Z_a$ are automatically zero), but we know from Birkhoff's theorem
that all spherically symmetric vacuum spacetimes are Schwarzschild,
and therefore any such nontrivial solution must be purely a result of the
freedom to chose the frame vectors in the Schwarzchild background.

We may directly obtain the following evolution and propagation
equations for these new gauge-invariant variables:
\ba
\dot X_{ a} &=& \:32\phi\E\bra{\dotn_{a}+\Sigma_a-\lc_{ab}\Omega^b}
+\:32\E\bra{\delta_a\Sigma-\:23\delta_a\theta} +\lc_{bc}\delta_a\delta^b\H^c,\label{Xdot}\\
\dot Y_{a} &=& \bra{\:12\phi^2+\E}\bra{\dotn_{a}+\Sigma_a-\lc_{ab}\Omega^b}
+\bra{\:12\phi-\udot}\bra{\delta_a\Sigma-\:23\delta_a\theta}
+\delta_a\delta_c\dotn^c,\label{Ydot}\\
\hat X_{a} &=& -2\phi X_a-\:32\E Y_a +\:32\phi\E\hatn_a-\delta_a\delta_b\E^b,\label{Xhat}\\
\hat Y_a &=& -X_a -\:32\phi Y_a +\bra{\:12\phi^2+\E}\hatn_a
+\delta_a\delta_b\hatn^b,\label{Yhat}\\
\hat Z_a &=& -\bra{\:32\phi+2\udot}Z_a-\udot
Y_a+\udot\bra{\phi+\udot}\hatn_a+\delta_a\dot\theta-\delta_a\delta_b\udot^b.\label{Zhat}
\ea
Note that there is not an equation for $\dot Z_{ a}$ because there is no
equation for $\dot\udot$.

We also find the following constraints by applying the commutator~(\ref{comm1})
to our new variables:
\ba
\lc_{ab}\delta^aX^b&=&3\phi\E\xi,\label{barX}\\
\lc_{ab}\delta^aY^b&=&\bra{\phi^2+2\E}\xi,\label{barY}\\
\lc_{ab}\delta^aZ^b&=&2\udot\bra{\phi+\udot}\xi.\label{Znewcons}\label{barZ}
\ea

Equations~(\ref{Xdot}) and~(\ref{Ydot}) replace our equations for~(\ref{Edot})
and~(\ref{phidot}) respectively. Similarly Eq.~(\ref{Xhat}) replaces
Eq.~(\ref{EhatBH}), Eq.~(\ref{Yhat}) replaces~(\ref{phihatBH}),
and~(\ref{Zhat}) replaces~(\ref{thetadotfull}). The constraint~(\ref{Znewcons})
allows us to derive a pseudo-evolution equation for $Z^a$ (but we shall not
require it). We may also replace our $\dot\Sigma$ equation~(\ref{sigmadotfull})
with
\be
\delta_a\dot\Sigma-\:23\delta_a\dot\theta=-X_a-\udot Y_a-\phi Z_a-\delta_a\delta_b\udot^b.
\label{XYZ}
\ee
\emph{All equations are now gauge invariant.} This means that when we have
chosen our frame $u^a,\n^a$ in a unique way, then all quantities appearing in
the equations are uniquely defined covariant and gauge-invariant quantities,
with a physical or geometrical meaning.

\forget{We could go further than this, and define dimensionless variables, such as
\be
r\;{\delta_a {\cal E}}{\E},
\ee
etc., but we shall not pursue this here.}

\subsection{Commutation relations for first-order variables}

Having discussed which variables are zeroth order and which are first order,
it will be useful to present the following commutation relations for the
derivatives of first-order scalars, vectors and tensors. For any scalar,
vector or tensor,~$\Psi$:
\ba
\dot{\hat \Psi}-\hat{\dot \Psi}&=&\udot\dot\Psi,\\
\delta_a\dot\Psi-\bra{\delta_a\Psi}^\cdot &=&0,\\
\delta_a\hat\Psi-\widehat{\bra{\delta_a\Psi}} &=&\:12\phi\delta_a\Psi;
\ea
while for a scalar
\be
\delta_{[a}\delta_{b]}\Psi=0,
\ee
and for a vector
\be
\delta_{[a}\delta_{b]}\Psi_c=\bra{\:14\phi^2-\E}\N_{c[a}\Psi_{b]},\label{vectcomm}
\ee
and a tensor
\be
\delta_{[a}\delta_{b]}\Psi_{cd}=\:12\bra{\:12\phi^2-\E}\bras{\N_{c[a}\Psi_{b]d}+\N_{d[a}\Psi_{b]c}}.
\ee
\forget{We can define the two curvature Riemann tensor $^2{\cal R}_{abcd}$ using
Eq.~(\ref{vectcomm}):
\be
2\delta_{[a}\delta_{b]}\Psi_c={^2{\cal R}_{abcd}}\Psi^d,
\ee
but we won't pursue this further here.}

\subsection{Spherical harmonics}

As the equations stand we can't find solutions because of the appearance of
angular derivatives `$\delta_a$'. An appropriate choice of basis functions will
allow us to write all first-order variables as an infinite sum over these basis
functions, and allow us to replace angular derivatives by a harmonic
coefficient. Clearly the spherical symmetry of the background begs us to use
spherical harmonics as our basis functions, so here we will develop these
appropriately for our formalism. We define our harmonics by analogy with the
FLRW case~\cite{HvE}, and we refer to~\cite{thorne} for details of other
approaches.

Note that all functions and relations below are defined in the background only;
we only expand first-order variables, so zeroth-order equations are sufficient.

We introduce spherical harmonic functions $Q=Q^{(\ell,m)}$, with
$m=-\ell,\cdots,\ell$, defined on the background, such that
\be
\delta^2 Q = -\ell\bra{\ell+1}r^{-2} Q,~~~\hat Q=0=\dot Q.\label{SH}
\ee
The function $r$ is covariantly defined by
\be
\phi=2\;{\hat r}{r},~~~\dot r=0=\delta_a r.\label{rdefc}
\ee
This factor is included in our definition~(\ref{SH}) so that the equation
propagates; it is trivial to show that it evolves also. We have defined $r$ so
far up to an arbitrary constant, which reflects our freedom in choosing a
particular normalisation of the spherical harmonic functions; we will find it
most useful for our purposes to fix this freedom by covariantly defining
\be
r\equiv\bra{\:14\phi^2-\E}^{-1/2},\label{rdef}
\ee
i.e., we identify $r$ defined here with the parameter defined by
Eq.~(\ref{rsch}).
%In fact, if Eq.~(\ref{rdef}) is calculated in Schwarzschild
%coordinates, $r$ corresponds simply to the Schwarzschild coordinate defined by
%equation~(\ref{rsch}), and $Q$ are simply the usual spherical harmonics
%$Y_\ell^m$ in Schwarzschild coordinates.
We can now expand any first order
scalar ${\Psi}$ in terms of these functions as
\be
{\Psi}=\sum_{\ell=0}^{\infty}\sum_{m=-\ell}^{m=\ell} {\Psi}\S^{(\ell,m)}
Q^{(\ell,m)} = {\Psi}\S Q,
\ee
where the sum over $\ell$ and $m$ is implicit in the last equality. The
$\mathsf{S}$ subscript reminds us that ${\Psi}$ is a scalar, and that a
spherical harmonic expansion has been made. Due to the spherical symmetry of
the background, $m$ never appears in any equation; we will just ignore it from
now on.

We also need to expand vectors and tensors in spherical harmonics. We therefore
define the \emph{even} (electric) parity vector spherical harmonics for
$\ell\geq1$ as
\be
Q_a^{(\ell)}=r\delta_a Q^{(\ell)} ~~~\Rightarrow ~~~\hat Q_a=0=\dot
Q_a,~~~\delta^2Q_a=\bra{1-\ell\bra{\ell+1}}r^{-2}Q_a;
\ee
where the $(\ell)$ superscript is implicit, and we define \emph{odd} (magnetic)
parity vector spherical harmonics as
\be
\bar Q_a^{(\ell)}=r\lc_{ab}\delta^b Q^{(\ell)}~~~\Rightarrow
~~~\hat{\bar{Q}}_a=0=\dot{\bar{Q}}_a,~~~\delta^2\bar
Q_a=\bra{1-\ell\bra{\ell+1}}r^{-2}\bar Q_a.
\ee
Note that $\bar Q_a=\lc_{ab}Q^b\Leftrightarrow Q_a=-\lc_{ab}\bar Q^b$, so that
$\lc_{ab}$ is a parity operator. The crucial difference between these two types
of vector spherical harmonics is that $\bar Q_a$ is solenoidal, so
\be
\delta^a\bar Q_a=0,~~~\mbox{while}~~~
\delta^aQ_a=-\ell\bra{\ell+1}r^{-1} Q.
\ee
Note also that
\be
\lc_{ab}\delta^a  Q^b=0,~~~\mbox{and}~~~\lc_{ab}\delta^a\bar Q^b=\ell\bra{\ell+1}r^{-1} Q.
\ee
The harmonics are orthogonal: $Q^a\bar Q_a=0$ (for each $\ell$), which implies
that any first-order vector ${\Psi}_a$ can now be written
\be
{\Psi}_a=\sum_{\ell=1}^{\infty} {\Psi}^{(\ell)}\V Q_a^{(\ell)}+\bar
{\Psi}^{(\ell)}\V\bar Q_a^{(\ell)}={\Psi}\V Q_a+\bar {\Psi}\V\bar Q_a.
\ee
Again, we implicitly assume a sum over $\ell$ in the last equality, and the
$\mathsf{V}$ reminds us that ${\Psi}^a$ is a vector expanded in spherical
harmonics.

Similarly we define even and odd tensor spherical harmonics for $\ell\geq2$ as
\ba
Q_{ab}&=& r^2\delta_{\lb a}\delta_{b\rb}Q,~~~\Rightarrow ~~~\hat Q_{ab}=0=\dot
Q_{ab},~~~\delta^2Q_{ab}=\bras{\phi^2-3\E-\ell\bra{\ell+1}r^{-2}}Q_{ab};\\
\bar Q_{ab}&=& r^2\lc_{c\lb a}\delta^c\delta_{b\rb}Q,~~~\Rightarrow ~~~\hat{\bar
Q}_{ab}=0=\dot{\bar Q}_{ab},~~~\delta^2\bar
Q_{ab}=\bras{\phi^2-3\E-\ell\bra{\ell+1}r^{-2}}\bar Q_{ab},
\ea
which are orthogonal: $Q_{ab}\bar Q^{ab}=0$, and are parity inversions of one
another: $Q_{ab}=-\lc_{c\lb a}\bar Q_{b\rb}^{~~c}\Leftrightarrow
\bar Q_{ab}=\lc_{c\lb a} Q_{b\rb}^{~~c}$. Any first-order tensor may be expanded
\be
{\Psi}_{ab}=\sum_{\ell=2}^{\infty} {\Psi}\T^{(\ell)}Q_{ab}^{(\ell)}+\bar
{\Psi}\T^{(\ell)}\bar Q_{ab}^{(\ell)}={\Psi}\T Q_{ab}+\bar {\Psi}\T\bar Q_{ab}.
\ee

We will not write all the equations expanded in spherical harmonics; instead we
list here all the replacements which must be made for scalars, vectors and
tensors. Note that sums over $\ell$ and $m$ are implicit in these equations
below. For brevity, we will sometimes use the aliases
\be
\ll =\ell\bra{\ell+1},~~~~
\lll =\bra{\ell-1}\bra{\ell+2}=\ll-2.
\ee
\be
\begin{array}{rcl|rcl|rcl}
\multicolumn{3}{c}{\mathrm{\textsc{scalar}}}
&\multicolumn{3}{c}{\mathrm{\textsc{vector}}}&\multicolumn{3}{c}{\mathrm{\textsc{tensor}}}\\
\hline
   {\Psi} &=& {\Psi}\S Q                              \vphantom{\;{\;XXX}{XXX}}     &   {\Psi}_a& =&+{\Psi}\V Q_a+\bar {\Psi}\V\bar Q_a                                                   &   {\Psi}_{ab}&=&+{\Psi}\T Q_{ab}+\bar {\Psi}\T\bar Q_{ab}                                                                    \\
   \delta_a {\Psi} &=& r^{-1}{\Psi}\S Q_a\label{SH1}       &   \lc_{ab} {\Psi}^b&=& -\bar {\Psi}\V Q_a+{\Psi}\V\bar Q_a                                         &   \lc_{c\lb a}{\Psi}_{b\rb}^{~~c}&=& -\bar {\Psi}\T Q_{ab}+{\Psi}\T\bar Q_{ab}                                               \\
   \lc_{ab}\delta^b {\Psi} &=& r^{-1}{\Psi}\S \bar Q_a     &   \delta^a{\Psi}_a&=&-\ell\bra{\ell+1}r^{-1}{\Psi}\V Q                                             &   \delta^b {\Psi}_{ab}&=& \:12\lll r^{-1}\bra{-{\Psi}\T Q_{a}+\bar {\Psi}\T\bar Q_{a}}  \\
               &&                                             &   \lc_{ab}\delta^a{\Psi}^b&=&+\ell\bra{\ell+1}r^{-1}\bar {\Psi}\V Q                                 &   \lc_{c\lb d}\delta^d{\Psi}_{a\rb}^{~c}&=&\:12\lll r^{-1}\bra{+\bar {\Psi}\T Q_{a}+ {\Psi}\T\bar Q_{a}}     \\
               &&                                             &   \delta_{\lb a}{\Psi}_{b\rb}&=&r^{-1}\bra{{\Psi}\V Q_{ab}-\bar {\Psi}\V \bar Q_{ab}}              &   &&    \\
               &&                                             &   \lc_{c\lb a}\delta^c {\Psi}_{b\rb}&=& r^{-1}\bra{\bar {\Psi}\V Q_{ab}+ {\Psi}\V \bar Q_{ab}}
               &&&
\end{array}
\ee

\forget{
Scalar:
\ba
{\Psi} &=& {\Psi}\S Q,\\
\delta_a {\Psi} &=& r^{-1}{\Psi}\S Q_a,\label{SH1}\\
\lc_{ab}\delta^b {\Psi} &=& r^{-1}{\Psi}\S \bar Q_a.\label{SH2}
\ea

Vector:
\ba
{\Psi}_a& =&{\Psi}\V Q_a+\bar {\Psi}\V\bar Q_a,\\
\lc_{ab} {\Psi}^b&=& -\bar {\Psi}\V Q_a+{\Psi}\V\bar Q_a,\\
\delta^a{\Psi}_a&=&-\ell\bra{\ell+1}r^{-1}{\Psi}\V Q,\label{SH3}\\
\lc_{ab}\delta^a{\Psi}^b&=&\ell\bra{\ell+1}r^{-1}\bar {\Psi}\V Q,\label{SH4}\\
\delta_{\lb a}{\Psi}_{b\rb}&=&r^{-1}\bra{{\Psi}\V Q_{ab}-\bar {\Psi}\V \bar Q_{ab}},\\
\lc_{c\lb a}\delta^c {\Psi}_{b\rb}&=& r^{-1}\bra{\bar {\Psi}\V Q_{ab}+ {\Psi}\V \bar Q_{ab}}.
\ea

Tensor:
\ba
{\Psi}_{ab}&=&{\Psi}\T Q_{ab}+\bar {\Psi}\T\bar Q_{ab},\\
\lc_{c\lb a}{\Psi}_{b\rb}^{~~c}&=& -\bar {\Psi}\T Q_{ab}+{\Psi}\T\bar Q_{ab},\\
\delta^b {\Psi}_{ab}&=& \bra{1-\:12\ell\bra{\ell+1}}r^{-1}\bra{{\Psi}\T Q_{a}-\bar {\Psi}\T\bar Q_{a}},\label{SH5}\\
\lc_{c\lb d}\delta^d{\Psi}_{a\rb}^{~c}&=&-\bra{1-\:12\ell\bra{\ell+1}}r^{-1}\bra{\bar {\Psi}\T Q_{a}+ {\Psi}\T\bar Q_{a}}.
\ea
}

Each vector and tensor equation produces two harmonics equations for each
$\ell$, one of odd parity and one of even parity, due to the orthogonality of
the vector and tensor harmonics. \forget{The importance of using spherical
harmonics lies in Eqs.~(\ref{SH1}),~(\ref{SH2}),~(\ref{SH3}) and~(\ref{SH4}),
which is where changes occur in the even and odd parity relations; other than
these, spherical harmonics simply re-write the equations (albeit with some sign
changes, and $\ell$'s) without indices.}

We can use these harmonic relations to derive various properties of the
$\delta_a$ derivative. For example, an important relation we use in deriving
the Regge-Wheeler equation in Sec.~\ref{RWsec} is
\be
2\delta_{\lb a}\delta^c\Psi_{b\rb
c}-\delta^2\Psi_{ab}=\bra{\E-\:12\phi^2}\Psi_{ab}.
\ee

\forget{
For brevity, we will sometimes use the aliases
\be
\ll =\ell\bra{\ell+1},~~~~
\lll =\bra{\ell-1}\bra{\ell+2}=\ll-2.
\ee}

\subsection{Odd and even parity perturbations}

After decomposing the equations into their harmonic components, we find that
they split into two independent subsets, which we refer to as even and odd
(parity) perturbations, but are also known as \emph{polar} and \emph{axial}
perturbations. This splitting is analogous to the well known splitting of
scalar, vector and tensor modes in the cosmological situation. The two
independent sets of equations utilise the variables
\ba
\mbox{Odd parity:~}&&\vo\equiv\bra{ \{\bar\E\T,~\H\T,~\bar\Sigma\T,~\bar\zeta\T\},
 ~\{\bar\E\V,~\H\V,~\bar\Sigma\V,~\Omega\V,~\bar\udot\V,~\bar\alpha\V, ~\bar\hatn\V,%~\bar\dotn\V,
 ~\bar X\V,~\bar Y\V,~\bar Z\V\},
 ~\{\H\S,~\Omega\S,~\xi\S\}}
\label{oddvars}\\
\mbox{Even parity:~}&&\ve\equiv\bra{ \{\E\T,~\bar\H\T,~\Sigma\T,~\zeta\T\},
 ~\{\E\V,~\bar\H\V,~\Sigma\V,~\bar\Omega\V,~\udot\V,~\alpha\V, ~\hatn\V,%~\dotn\V,
 ~ X\V,~ Y\V,~ Z\V\},
 ~\{\Sigma\S,~\theta\S\}},\label{evenvars}
\ea
where we have defined the vectors $\vo,\ve$ for later convenience. The
`parity switching' which occurs between the sets of variables (e.g., $\H\T$
appears in the odd parity system) may be seen in the covariant tensor
equations: $\H_{ab}$ always appears alongside a `$\lc_{ab}$' factor relative
to other variables such as $\E_{ab}$; similarly for $\Omega^a$; it's always
seen as $\lc_{ab}\Omega^b$, relative to $\Sigma^a$, say.

Hereafter, we will assume all equations have been decomposed into their
spherical harmonic components, unless stated otherwise, and when we refer to
specific equations that are given in tensor form (such as the evolution
equations above), we will generally assume too that this has been decomposed
implicitly.

\subsection{Time harmonics}

Because the background is static, we can, if we wish, decompose time
derivatives of first order quantities into their Fourier components. This is
simply assuming an $e^{i\omega\tau}$ time dependence for the first order
variables, but we shall make it a bit more precise. Define the time harmonic
functions $T^{(\omega)}$ in the background by
\be
\dot T^{(\omega)}=i\omega T^{(\omega)},~~~\hat
T^{(\omega)}=0=\delta_aT^{(\omega)};~~~\dot\omega=0=\delta_a\omega.
\ee
In the background, this must satisfy
\be
\hat{\dot T}+\udot\dot T=0,
\ee
which implies
\be
\hat\omega=-\udot\omega.
\ee
We may integrate this in the background, in terms of the parameter $x$, or
$r$, to give
\be
\omega=\sigma\coth x=  \sigma\bra{1-\;{2m}{r}}^{-1/2}=\;{2\sigma}{\phi r},
\ee
where $\sigma$ is a  constant. Then any first order variable $\Psi$ (which
will usually be an even or odd parity variable discussed above, but it could
be any first order scalar, vector or tensor) as
\be
\Psi=\sum_\omega \Psi^{(\omega)} T^{(\omega)}=\Psi^{(\omega)} T^{(\omega)}
\ee
where the summation is understood in the last equality (which may be an
integral, and depends on the types of differential equations and their boundary
conditions occurring in the solutions). We can simply replace a dot by
$i\omega$ in the equations, as no confusion should occur.

\forget{
\subsection{Integrability conditions and constraints}

Before we start trying to understand the full set of equations we must make
sure that the equations are consistent with one another. In particular, the
constraint equations~(\ref{conom_anl}) and~(\ref{consig_anl}) must hold when
evolved and propagated; furthermore, each of the propagation and evolution
equations must obey the integrability condition
\be
{\cal C}[\Psi]\equiv\dot{\hat \Psi}-\hat{\dot \Psi}-\udot\dot
\Psi=0,\label{integrab}
\ee
for any first order scalar, vector or tensor $\Psi$. An alternative way of
doing this is to rewrite the evolution equations using the time harmonics, so
that the evolution equations become constraints for the propagation equations.
(The propagation equations then form a first-order linear system of
differential equations supplemented by constraint equations~-- more of this
later.) For now though, we will use Eq.~(\ref{integrab}).

We will present the consistency relations we find in terms of harmonic
functions, because the resulting constraints are a little different [e.g., the
constraint~(\ref{consig_anl}) only applies to the odd-parity equations], and
for practice using the harmonic equations. We use the
constraints~(\ref{conom_anl}) and~(\ref{consig_anl}) to eliminate the variables
$\H\S,~\H\V$ and $\Sigma\S$ where appropriate. \emph{Any equations not
mentioned here are consistent.}

\subsubsection{Odd-parity consistency}

\iffalse
From ${\cal C}[\Omega\S]$ [or from~(\ref{consig_anl})$^\cdot$] we find
\be
\bar Z\V=2\bra{\ell\bra{\ell+1}}^{-1}r\udot\bra{\phi+\udot}\xi\S;\label{barZ}
\ee
propagating this, we find
\be
\bar Y\V=\bra{\ell\bra{\ell+1}}^{-1}r\phi\bra{\phi-2\udot}\xi\S.\label{barY}
\ee
From~(\ref{consig_anl}) and~(\ref{sigmadotfull}), together with~(\ref{Xhat}),
~(\ref{Yhat}) and~(\ref{Zhat}) we find
\be
\bar X\V=-\udot\bar Y\V-\phi\bar Z\V,\label{barX}
\ee
which is just the odd-parity part of Eq.~(\ref{XYZ}).
\fi

%Evolving the odd parity part of Eq.~(\ref{conom_anl}) reveals
%\be
%\bar\E\V=\bra{\ell\bra{\ell+1}r}^{-1}\brac{1+\ell\bra{\ell+1}-\:34r^2\phi^2}\xi\S
%+\:12r^{-1}\bra{\ell+2}\bra{\ell-1}\bar\zeta\T;\label{barEV}
%\ee
%(which propagates)
Evolving the odd parity part of Eq.~(\ref{newcons}) implies
\be
\bar{\dotn}\V=3\bar\Sigma\V;\label{bardotn}
\ee
this now gives us a propagation equation for $\Omega\V$ [or $\bar\Sigma\V$
owing to the coupling between the two; cf., Eq.~(\ref{hatSig_anl})]
\be
\hat\Omega\V-\:13\dot{\bar{\hatn}}\V=\:43
r^{-1}\Omega\S+\:43\bra{\phi-\udot}\bar\Sigma\V
-\bra{\:12\phi+\:73\udot}\Omega\V
+\:23r^{-1}\bra{\ell+2}\bra{\ell-1}\bar\Sigma\T.\label{OmVhat}
\ee
We must also evolve Eq.~(\ref{bardotn}), but this just gives a new evolution
equation for $\bar\dotn\V$,
\be
\dot{\bar\dotn}\V=\:32\udot\bar{\hatn}\V+\:14\phi^{-2}\udot^{-1}
\bras{\ll\bra{\phi^2-4\E}-2\bra{\phi^2+2\udot^2}}\bar X\V -\:32\lll
r^{-1}\bar\zeta\T.
\ee
These two new equations in turn allow us to check ${\cal C}[\Omega\V]$, ${\cal
C}[\bar\Sigma\V]$ and ${\cal C}[\bar\dotn\V]$, all of which give the same new
equation for $\hat{\bar{\hatn}}\V$, which involves a $\ddot{\bar{\hatn}}\V$
term. Evolving the odd parity Eq.~(\ref{Znewcons}) implies an equation for
$\dot{\bar Z}\V$:
\be
\dot{\bar Z}\V=\udot\bra{\phi+\udot}\bras{4\bar\Sigma\V-\Omega\V}.
\ee
${\cal C}[\bar Z\V]$ is now satisfied.

\subsubsection{Even-parity consistency}

Evolving the even parity part of Eq.~(\ref{newcons}) we get
\be
\dotn\V=3\Sigma\V;\label{dotnVcons}
\ee
propagating this gives
\be
\hat{\bar\Omega}\V+\:13\dot{\hatn}\V=r^{-1}\theta\S-\:16\bra{14\udot+5\phi}\bar\Omega\V +
\:13\bra{4\udot-5\phi}\Sigma\V+\:23\bar\H\V+\lll r^{-1}\Sigma\T,\label{OmVbarhat}
\ee
while evolving implies an evolution equation for $\dotn\V$:
\be
\dot{\dotn}\V=\:32\udot\hatn\V-3\E\V+\:32Z\V.
\ee
These new equations allow us to form ${\cal C}[\bar\Omega\V]$, ${\cal
C}[\Sigma\V]$ and ${\cal C}[\dotn\V]$, all of which result in the same
equation, involving $\hat{\hatn}\V$, which involves a $\ddot{\bar{\hatn}}\V$
and a  $\dot\theta\S$ term.

}

\section{The \emph{Regge-Wheeler tensor} and the Regge-Wheeler and Zerilli
equations}
\label{RWsec}

We have presented the full, covariant, gauge-invariant linearised equations for
the propagation of gravitational radiation in a perturbed Schwarzschild black
hole spacetime, and we have discussed the introduction of spherical harmonics,
enabling us to replace $\delta$-derivatives with spherical harmonic indices, as
well as remove the tensorial nature of the equations. We could now introduce a
perturbed metric and calculate all variables in terms of the metric functions,
to show how the standard Regge-Wheeler and Zerilli equations of black hole
perturbation theory may be related to this approach, linking all of our
variables to these Regge-Wheeler and Zerilli functions. We need not bother
however: it is possible to find generalisations of these functions and
correlations directly, an important test of our theory.

\forget{So, before we address the intricacies of solving the equations let's think
about what we can learn from the covariant equations themselves, and how they
relate to the solutions of our harmonic equations and to the well-known
Regge-Wheeler and Zerilli equations that describe the harmonic solutions in
the metric-based approach to black hole perturbations, to which our solutions
must clearly be directly equivalent.} We show in this section that it is
possible to find a gauge- and frame-invariant, transverse-traceless tensor
that satisfies a closed, covariant, gauge-invariant wave equation. We call
this tensor the \emph{Regge-Wheeler tensor}.
%The fact that it is a covariant wave equation
%means, in terms of harmonics, that perturbations of either parity satisfy the
%same equation.
Moreover, we demonstrate that, once decomposed into spherical harmonics, and
with the appropriate radial coordinate, this equation is the Regge-Wheeler
equation for \emph{both parities}, thus unifying both parities in one TT
tensor. We present similar results for the Zerilli equation that normally
describes even parity solutions, but defer the derivation of the \emph{Zerilli
tensor} until later in Sec.~\ref{zertens}.

\subsection{Regge-Wheeler}

We know that gravitational waves
%, represented at least in part by the
%transverse-traceless 1+1+2 tensors $\E_{ab}$ and $\H_{ab}$,
propagate in the perturbed Schwarzschild spacetime, and therefore that the TT
tensors must satisfy (covariant) wave equations of some sort [see
Eqs.~(\ref{Ewave}) and~(\ref{Hwave}) in section~\ref{grav-wave}], as our
solutions must recover the plane wave case when $m=0$. When we investigate
these wave equations for $\E_{ab}$ and $\H_{ab}$, say, we find that they are
not closed: that is, they contain forcing terms from other 1+1+2 tensors, a
feature not present for plane waves. This makes their interpretation and
solution nontrivial. In fact, when we look at the second-order wave equations
obtained from the covariant equations for any of the 1+1+2 tensors we find the
same story. Can we find some combination of the basic tensors that satisfies a
\emph{closed, covariant} wave equation? Indeed we can, and we outline its
derivation here. The formulae are rather formidable, and so we omit a detailed
derivation.

In addition to the obvious 1+1+2 transverse-traceless tensors ($\E_{ab}$,
$\H_{ab}$, $\Sigma_{ab}$ and~$\zeta_{ab}$) it is possible to construct many
TT tensors from $\delta$-derivatives of vectors and scalars, such as
$\delta_{\lb a}X_{b\rb}$, $\delta_{\lb a}a_{b\rb}$, or even $\delta_{\lb
a}\delta_{b\rb}\Omega$, for example. Using the dot and hat equations, along
with the commutators, we can obtain the wave equations satisfied by any of
these tensors, by calculating the wave operator $\ddot
\Psi_{\lb ab\rb}-\hat{\hat \Psi}_{\lb ab\rb}$ for that tensor~$\Psi_{ab}$. \forget{This
will involve other tensor quantities and their $\delta$-derivatives, and we
could, in theory, calculate all possible such equations and systematically
eliminate unwanted terms until we obtain a closed equation. Delightful as this
possibility is, here we will just reveal the answer.}

If we calculate the wave operators for $\zeta_{ab}$ and $\delta_{\lb
a}X_{b\rb}$ we notice that they contain similar terms in $a_a$, $\alpha_a$
and~$\xi$. In fact, if in $\ddot \zeta_{\lb ab\rb}-\hat{\hat \zeta}_{\lb
ab\rb}$ we substitute for  $a_a$ from Eq. (\ref{Xhat}), $\alpha_a$ from
Eq.~(\ref{Xdot}), and for $\xi$ from Eq.~(\ref{barX}) we find, amazingly,
that all variables other than $\delta_{\lb a}X_{b\rb}$ and $\zeta_{ab}$
cancel, leaving the closed wave equation~(\ref{RWtensorwave}) below.

So, we define the dimensionless, gauge-invariant, frame-invariant,
transverse-traceless tensor
\be
%W_{ab}=\zeta_{ab}+\:23\udot^{-1}\phi^{-2}\delta_{\lb a}X_{b\rb},\label{Wab}
W_{ab}=\:12\phi r^2\zeta_{ab}-\:13r^2\E^{-1}\delta_{\lb a}X_{b\rb},\label{Wab}
\ee
which obeys the rather nice wave equation
\be
\ddot W_{\lb ab\rb}-\hat{\hat W}_{\lb ab\rb}-\udot{\hat W}_{\lb ab\rb}
+\phi^2 W_{ab}-\delta^2 W_{ab} =0,\label{RWtensorwave}
\ee
where $\delta^2=\delta^a\delta_a$ is the covariant laplacian on the
sheets (approximate 2-spheres, in this case).

We can expand Eq.~(\ref{RWtensorwave}) into SH: let $W=\{W\T,\bar W\T\}$, then Eq.~(\ref{RWtensorwave})
becomes
\be
%\ddot W-\hat{\hat W}-\bra{\phi+3\udot}\hat
%W+\brac{\;{\ell\bra{\ell+1}}{r^{2}}-\udot\bra{4\phi+\udot}}W=0.\label{RW}
\ddot W-\hat{\hat W}-\udot\hat
W+\brac{\;{\ell\bra{\ell+1}}{r^{2}}+3\E}W=0.\label{RW}
\ee
Note that both the even and odd parity parts of $W_{ab}$ satisfy the same wave
equation~(\ref{RW}).

It turns out that Eq.~(\ref{RW}) is actually the Regge-Wheeler
equation~\cite{RW} in appropriate coordinates,  which we now show. Converting
to the parameter $r$, $\rho\rightarrow r$, and then to the `tortoise'
coordinate of Regge and Wheeler
\be
r_*=r+2m\ln\bra{\;{r}{2m}-1},\label{tortoise}
\ee
and then letting
\be
%\psi=\psi_{RW}=\sqrt{r\bra{r-2m}}W,
\psi=\psi_{RW}=W,
\ee
we find that~(\ref{RW}) becomes:
\be
\bra{\;{d^2}{dr_*^2}+\sigma^2}\psi=V\psi\label{schroed}
\ee
where
\be
V=V_{RW}=\;{\bra{r-2m}}{r^4}\bra{\ell\bra{\ell+1}r-6m},\label{RWpot}
\ee
which is the Regge-Wheeler potential. We will thus refer to $W_{ab}$ as the
Regge-Wheeler tensor.

\subsection{Zerilli}
\label{zersec}

%\subsection{Even parity perturbations and the Zerilli equation}

We have shown that $W_{ab}$ satisfies the Regge-Wheeler equation~(\ref{RW})
regardless of parity. However, it is well known that even parity perturbations
are governed by the Zerilli equation.

For even parity perturbations the variable
\ba
%{\cal Z}=\Sigma\T+\:23r^{-1}\E^{-1}\bar\H\V\label{zer1}
{\cal Z}&=&\:16\phi\bra{\lll-3\E r^2}^{-1}\bra{3 r\Sigma\T+2\E^{-1}\bar\H\V}\\
 &=& \:23 c_3^{-1}\bras{3r\phi\Sigma\T-2\udot^{-1}\bar\H\V}
%\bra{16\E\udot r^4}^{-1}\Sigma\T+i\omega\bras{r\udot
%c_3\bra{3r^2\phi^2-4}}^{-1}Y\V,
\label{zer1}
\ea
is one of two fundamental variables, and it can be shown to satisfy the Zerilli
equation. The wave equation equation for ${\cal Z}$ is
\be
\ddot{\cal Z}-\hat{\hat{\cal Z}}-\udot\hat{\cal
Z}+\:13r^{-2}\brac{\:14c_3+32c_3^{-2}\bra{\ll+1}\lll^2}{\cal Z}=0.
\label{zerilli}
\ee
We have used the abbreviations:
\be
c_j=4\bra{\ll+1}-j r^2\phi^2,
\ee
where we keep the freedom in $j$ for later use.

Making the change to the tortoise coordinate, inserting time harmonic
functions, and changing to the variable
\be
%\psi=\psi_{\cal Z}={\;{\sqrt{r\bra{r-2m}}}{\lll r+6m}}{\cal Z}
\psi=\psi_{\cal Z}={\cal Z}
\ee
we find that~(\ref{zerilli}) is in fact the Zerilli equation,
Eq.~(\ref{schroed}), with
\be
%V=V_{\cal Z}=\;{r-2m}{r^4\brac{\bra{\ell\bra{\ell+1}-1}r+6m}^2}
%\bras{\bra{\ell\bra{\ell+1}-1}^2\bras{r^3\bra{\ell\bra{\ell+1}+7} +24mr^2}
%+36m^2r\bra{\ell\bra{\ell+1}-1} +72m^3}.
V=V_{\cal Z}=\;{r-2m}{r^4\brac{\bra{\ll-1}r+6m}^2}
\bras{\bra{\ll-1}^2\bras{r^3\bra{\ll+7} +24mr^2}
+36m^2r\bra{\ll-1} +72m^3}.\label{zerpot}
\ee

For the even parity perturbations, then, there are two variables which obey
wave equations: the Zerilli variable ${\cal Z}$, and the Regge-Wheeler variable
$W\T$. They are in fact related by
\be
%\bras{4\bra{\ell\bra{\ell+1}+1}-3r^2\phi^2}i\omega W\T =
%-12r^2\E\hat{\cal Z}
%+\;{4}{r^2\phi}\bras{\ell\bra{\ell+1}\bra{\ell-1}\bra{\ell+2} +3r^4\E^2}{\cal
%Z},\label{WofZ}
\hat{\cal Z}=\;{i\omega}{3r^3\phi\udot} W\T +
\;{r^2\phi^2\bras{c_3^2-8\lll \bra{c_3+\ll+4}}+32\ll\lll\bra{\ll+1}}{24r^4\phi^2\udot
c_3}{\cal Z}
\label{WofZ}
\ee
as may be found by considering considering $\dot W\T$, using Eq.~(\ref{Wab}).
Furthermore, it is possible to find an equation of the form
\be
\hat W\T=\bra{\mbox{stuff}}\hat{\cal Z}+%\bra{\mbox{more stuff}}\dot{\cal Z}+
\bra{\mbox{more stuff}}{\cal Z},\label{WofZ2}
\ee
by utilising Eqs.~(\ref{Eadot}),~(\ref{Eabdot}) and~(\ref{Sigmaabdot}), using
the time harmonics throughout, which implies that ${\cal Z}$ may be written as
a function of $W\T$ and its derivatives. This gives
\be
\hat W\T= -\;{r^2\phi^2\bras{c_3^2-8\lll \bra{c_3+\ll+4}}+32\ll\lll\bra{\ll+1}}{24r^4\phi^2\udot
c_3}W\T-\;{9\omega^2r^8\phi^4\udot^2+\ll^2\lll^2}{3i\omega
r^5\phi^3\udot}{\cal Z}\label{ZofW}.
\ee
Thus, \emph{the wave variables are not independent}. Eqs.~(\ref{WofZ})
and~(\ref{ZofW}) are a two dimensional first-order linear system of
differential equations, which could replace the second-order Regge-Wheeler
and Zerilli equations. In fact, it is instructive to rewrite
Eqs.~(\ref{WofZ}) and~(\ref{ZofW}) in matrix form:
\be
{\bra{\begin{array}{c}
   \hat W\T\\
 \hat{\cal Z}
\end{array}}}=\bra{\begin{array}{cc}
   \beta & A\\
   B & -\beta
\end{array}}\bra{\begin{array}{c}
  W\T \\
  {\cal Z}
\end{array}}\label{RWzer}
\ee
where the definitions of $A$, $B$ and~$\beta$ are obvious from above. Notice
that the matrix that couples $\hat W\T$ and $\hat{\cal Z}$ is traceless, and
that the \emph{special quasinormal modes} discussed below correspond to~$A=0$.

This form of writing the Regge-Wheeler and Zerilli equations,
Eq.~(\ref{RWzer}), may be viewed as just a neat way to write two decoupled
second order DE's as two coupled first-order ones; however, it also shows that
%, although it is clearly intrinsic to the physical situation at hand:
the two decoupled second order DEs, which Eq.~(\ref{RWzer}) may replace, must
be representations of the \emph{same} physical situation~\cite{chandra}
(because $W\T$ is a linear combination of ${\cal Z}$ and $\hat{\cal Z}$, and
${\cal Z}$ is a linear combination of $W\T$ and $\hat W\T$). We can use this DE
to find an \emph{odd parity Zerilli variable}, which satisfies the Zerilli
equation~(\ref{zerilli}), quite easily: simply demand that Eq.~(\ref{ZofW})
hold for the odd parity perturbations too, and substitute for $\bar W\T$ from
Eq.~(\ref{Wab}), and then from the propagation equations as appropriate. This
then gives us $\bar{\cal Z}$ as a complicated linear combination of $\bar\E\T$,
$\bar\zeta\T$, $\bar X\V$ and $\bar Y\V$, seemingly unrelated to the even
parity Zerilli variable, Eq.~(\ref{zer1}). This is an illusion, however, and it
is possible to find a form for $\bar{\cal Z}$ very similar to Eq.~(\ref{zer1}),
and consequently a Zerilli tensor, but we defer this until later, in
Sec.~\ref{zertens}.

%In this case, Eq.~(\ref{RWzer}) applies equally well to the odd parity
%perturbations as well, but ${\cal Z}$ has no connection to the odd parity
%variables, other than through $\bar W\T$ and Eq.~(\ref{ZofW}); this would then
%put ${\cal Z}$ on the same footing as other odd parity variables, as we will
%see below, in Sec.~\ref{oddsolssec}.

\subsection{Quasinormal modes}\label{QNMsec}

The Regge-Wheeler and Zerilli equations have been studied in some detail over
the years, and their solution is a complicated business~\cite{nollert}.
The relevant boundary conditions for the two equations are
those that represent a GW perturbation which propagates outwards at infinity
($r\sim r_*\rightarrow\infty$) and inwards to the horizon ($r\rightarrow
2m$,~$r_*\rightarrow-\infty$)~-- that is, there are no GW propagating in from
infinity or out of the horizon. The form of the Regge-Wheeler and Zerilli
variables corresponding to this are
\be
\psi\sim e^{i\sigma r_*}~~~\mbox{as}~~~r_*\rightarrow-\infty~~~\mbox{and}
~~~\psi\sim e^{-i\sigma r_*}~~~\mbox{as}~~~r_*\rightarrow+\infty,\label{BCs}
\ee
where $\psi=W$ or $\psi={\cal Z}$; see, e.g.,~\cite{nollert}. It turns out that
the only solutions to Eq.~(\ref{schroed}), with potentials~(\ref{RWpot})
or~(\ref{zerpot}) with boundary conditions~(\ref{BCs}) require \emph{discrete}
values of the frequency parameter $\sigma$, with $\Im(\sigma)>0$; these are
referred to as \emph{quasinormal frequencies}, and the solutions constructed
from them as \emph{quasinormal modes} (QNMs). Because of the $e^{i\omega \tau}$
time dependence, these decay exponentially in time, which corresponds to energy
radiated to infinity or the horizon as GW. This damping in time is important as
$\psi$ grows exponentially as $r\rightarrow\infty$. In particular, it also
means that the spacetime is not flat at spacelike infinity, but it is flat at
future null infinity (i.e., along a null ray).

The factor in front of the ${\cal Z}$ in Eq.~(\ref{ZofW}) is rather
interesting: evaluate (in terms of $r$ say) to find
\be
9\omega^2r^8\phi^4\udot^2+\ll^2\lll^2=\bra{12m\sigma}^2+\bra{\ll\lll}^2
\ee
which has roots at
\be
2m\sigma=\pm \;i6 \ell\bra{\ell-1}\bra{\ell+1}\bra{\ell+2}\label{specialfreq}
\ee
with the `$+$' root corresponding to the frequency of the `special
(quasi-normal) mode' discussed in~\cite{liu-M,nollert}, which is the only QNM
with $\Re(\sigma)=0$. Since the frequency is purely imaginary the special QNM
is not oscillatory in time but merely decays exponentially.

The potentials for the Regge-Wheeler and Zerilli equations when converted into
Schr\"odinger form, $V_{RW}$ and $V_{\cal Z}$, although appearing very
different functionally, are actually very similar (see, e.g.,~\cite{nollert}),
becoming identical as $\ell\rightarrow\infty$, with peaks lying just beyond the
photon sphere, $r=3m$; as $\ell\rightarrow\infty$ the peaks approach the photon
sphere (so that the solutions $W$ and ${\cal Z}$ become identical for
$\ell\rightarrow\infty$). Thus these one dimensional wave equations represent
gravitational waves scattering off the photon sphere, with the same reflection
and transmission coefficients~\cite{chandra}.
%We can understand the
%scattering if we think of a gravity wave coming from infinity with an impact
%parameter just larger than $3m$; it will start shaking the spacetime near the
%hole, and the location of the photon sphere along with it. Because the gravity
%wave is not localised, `some' of it will pass within the photon sphere and be
%dragged into the horizon, while the rest will escape to infinity; hence the
%reflection and transmission coefficients, when viewed as a one dimensional
%problem.

It is important to note, however, that the Regge-Wheeler and Zerilli equations
allow physical solutions with boundary conditions other than~(\ref{BCs}), and
hence with $\Im(\sigma)\leq0$, but these solutions will represent  GW incoming
partially from infinity. These are not relevant for GW detection, but are of
interest in their own right, as perturbations of bounded regions, say, may
involve such waves. %We will, therefore, often show plots with $\sigma$ real.

\section{Solving the equations}
\label{solution-section}

\forget{
When the 1+3 perturbation formalism is applied in a cosmological setting one
can separate quite nicely the scalar, vector, and tensor modes, and study them
largely independently: in particular, one can find non-trivial pure tensor
modes. This turns out not to be the case for the 1+1+2 black hole perturbations
we are studying, as is easily seen from the equations: retaining only tensor
variables leads to a set of constraints from the scalar and vector propagation
and evolution equations that imply that each tensor is constant (and therefore
zero). This results from the fact that the types of vacuum perturbation are
restricted. We are extremely limited in the variables we can `switch off'~--
once a frame is chosen (that is, once a specific observer congruence~$u^a$ is
selected~-- see Sec.~\ref{evensolssec} below), setting any vector or tensor to
zero will result in no perturbations whatsoever. We can turn off a scalar, with
the result that either the even or odd perturbations are zero, depending on the
choice. From a mathematical viewpoint, this is rather unfortunate, as it means
that the whole system of equations must be solved at once; one cannot look at
certain subsets of the equations, to gain an understanding of the whole. }

%\subsection{Overview}
\label{solsoverview}

Before we discuss the solution of our system of equations, it is worthwhile
giving an overview of the structure of the equations. We have three distinct
types of equations: propagation, evolution, and constraint. The propagation
equations are to be considered the key \emph{differential} equations, while
the evolution and constraints may be considered as auxiliary algebraic
equations. This is because the structure of the background varies in the
radial direction, so the hat-derivative cannot be expanded in harmonic
functions, while the dot- and $\delta$-derivatives are \emph{perturbation}
derivatives, as they do not occur in the background equations, and can thus
be expanded in harmonics.   We can then analyse and solve the large system of
equations using matrix methods. This will provide interesting insight into
the problem of black hole perturbations, since it allows us, at any radial
position from the black hole, to treat the harmonic variables
in~(\ref{oddvars}) and~(\ref{evenvars}) as `coordinates'~-- i.e., as a
particular choice of basis vectors~--  in an abstract, 33-dimensional vector
space~${\cal V}_{33}$: the non-propagation equations then tell us that only
fourteen degrees of freedom are present in the evolution and constraint
equations (six in the odd parity, eight in the even), and these then allow us
to work out how a subset of this this 14-dimensional solution subspace
of~${\cal V}_{33}$ propagates radially (that is, they provide the
differential system for the remaining degrees of freedom, but there are some
undetermined variables), \emph{without us having to explicitly specify a
basis for the 14-dimensional vector subspace of~${\cal V}_{33}$}.

Assume a spherical harmonic decomposition of all the equations. Let
$\mathbf{V}$ denote the 33-dimensional vector (element of~${\cal V}_{33}$)
with the variables~$\vo=$(\ref{oddvars}) and~$\ve=$(\ref{evenvars}) as
elements; arrange the vector thus:
\be
\mathbf{V}=(\mathrm{Odd~variables}~|~\mathrm{Even~variables})=(\vo,\ve).
\ee
Further, introduce the harmonic expansion in time, so that dot derivatives are
everywhere replaced by $i\omega$; the  evolution equations  without a
propagation derivative in them then become a set of 18 algebraic equations.
\forget{Then, three things happen:
\begin{enumerate}
\item the \emph{constraint} equations, involving no dot or hat derivatives,
are unaffected and remain a set of 10 purely algebraic (linear) equations in
the 33 variables;
\item the evolution equations now also become a set of 23 \emph{algebraic}
equations (which each contain~$i\omega$);
\item most of the propagation equations are unaffected, but in those that
also contain a dot derivative this is replaced with~$i\omega$.
\end{enumerate}}
 The propagation equations are then of the form
\be
\hat{\mathbf{V}}_{28}=\mathbf{P V},\label{propgeneral}
\ee
where $\mathbf{V}_{28}$ denotes the vector consisting of the 28 elements
of~$\mathbf{V}$ which have a propagation equation. Note that there is no
propagation equation for $a^a$ (indeed, $a^a$ only acts as a forcing term in
the differential equations~-- it is completely undetermined), and some
variables do not have an individual propagation equation: there are
propagation equations for $\Sigma_a-\lc_{ab}\Omega^b$ and
$\Sigma-\:23\theta$, but not for each of these variables separately (and
these \emph{can't} be generated, as the evolution equations and constraints
contain the same degeneracy).
 $\mathbf{P}$ is the $28\times33$
propagation matrix, which contains $i\omega$ terms subsuming evolution
equations which have hat derivatives in them. \forget{We will discuss its form
in more detail later, but note that since there are no independent propagation
equations for $\Sigma\S$ and $\theta\S$, only the combination
$\Sigma\S-\:23\theta\S$, we cannot give a~$\hat{\theta}\S$ equation.}

The 18 remaining evolution equations take the form, in matrix notation
\be
%\dot{\mathbf{V}}=\mathbf{E V} ~~~\Leftrightarrow~~~\bra{-i\omega\mathbf{I}+
\mathbf{E}\mathbf{V}=\mathbf{0},
\label{Eeq}
\ee
where $\mathbf{E}$ is a $18\times33$ matrix. The 9 constraints are
\be
\mathbf{C V}=\mathbf{0},
\ee
where $\mathbf{C}$ is a $9\times33$ matrix. Recall that the equations
decouple into two sets of equations of opposing parity, which is reflected in
the structure of the matrices~$\mathbf{E,~P,~C}$, which are all in
block-diagonal form:
\be
\left(
\begin{array}{c|c}
  \begin{array}{l}
    \mbox{Odd} \\
    \mbox{parity}
  \end{array} & 0 \\\hline
  0 & \begin{array}{l}
    \mbox{Even} \\
    \mbox{parity}
  \end{array}
\end{array}
\right),
\ee
thus dividing ${\cal V}_{33}$ into two divorced subspaces. (We will denote the
odd parity upper block of the matrices by a subscript $\mathbf{\mathsf{O}}$,
and the even lower block by $\mathbf{\mathsf{E}}$.) This means that we can
treat the odd and even subspaces separately.\forget{, and, moreover, that the
complexities of the even solution (related to the freedom to choose a frame and
the corresponding absence of a $\hat{\theta}\S$ equation) can be ignored when
the odd case is considered, making the odd case rather easier to understand.}
We will find that the odd equations reduce to a third-order system (i.e., a
coupled three-dimensional first-order system of differential equations), and
the even equations to a fourth-order system, both of which do not close: the
odd has two undetermined variables (from the $\Sigma_a-\lc_{ab}\Omega^b$
degeneracy, and lack of an equation determining $a^a$), while the even has
three (from the $\Sigma_a-\lc_{ab}\Omega^b$ and $\Sigma-\:23\theta$
degeneracies, as well as the $a^a$ business), all of which reflect freedom in
choosing our frame.

In principle, we could adjoin the matrix $\mathbf{C}$ to $\mathbf{E}$ giving
a total of 27 \emph{algebraic} relations between the variables, represented
by the $27\times33$ matrix
\be
              \mathbf{F}\equiv \left(\frac{\mathbf{E}}{\mathbf{C}}\right)
\ee
with
\be
              \mathbf{F}\mathbf{V} = \mathbf{0}.
\ee
However, only one constraint equation, Eq.~(\ref{Znewcons}), is \emph{not}
represented in the evolution equations (as there is no equation for $Z^a$).
With this exception the constraints evolve and propagate consistently, so eight
elements of $\mathbf{C}$ do not contain any more information than $\mathbf{E}$
and $\mathbf{P}$. To put this more elegantly, the fact that the constraints
evolve consistently implies that eight rows of $\mathbf{C}$ are just linear
combinations of the rows of the matrix $\mathbf{E}$, so that the rank
of~$\mathbf{F}$ is 19, just one more than that of~$\mathbf{E}$. Let
$\mathbf{F}_{\cal L}$ denote the 19 linearly independent rows of $\mathbf{F}$.
\forget{Let us consider, then, equation~(\ref{Eeq}), which constitutes a set of 22
linear simultaneous equations for the 33 variables. These turn out to be
independent equations (so that the rank of $\mathbf{E}$, or equivalently
$\mathbf{F}$, is~22).} Formally, this means that the solution vector
$\mathbf{V}$ in ${\cal V}_{33}$ must lie in the ($33-19=14$-dimensional)
\emph{null space}, ${\cal N}_{14}$, of~$\mathbf{F}_{\cal L}$. Since all
equations propagate consistently we can think of this solution space, and the
propagation equations acting on it, in an abstract way. We see that the
propagation equations are in a certain sense \emph{independent} of the
particular variables we choose to represent the solution.

To actually obtain solutions, however, we must reintroduce `coordinates'
in~${\cal N}_{14}$. To this end, we use $\mathbf{F}_{\cal L}$ to eliminate 19
variables, leaving just~14 (it being largely a matter of choice exactly
which~14, provided they span~${\cal N}_{14}$); let us denote them by
$\mathbf{v}$. The remaining variables can then all be expressed in terms of
these fourteen `coordinate variables'. We may therefore write
\be
\mathbf{V}=\mathbf{M}\mathbf{v},\label{solform1}
\ee
where $\mathbf{M}$ is a $33\times14$ matrix of the form
\be
\bra{\begin{array}{c|c}
  \begin{array}{lcr}
   \longleftarrow & 6 & \longrightarrow
  \end{array} &
  \begin{array}{lcr}
    \longleftarrow~ & ~8~ & ~\longrightarrow
  \end{array} \\
%  & \\
  & \\
  \mbox{\sc odd} & \mbox{\sc even}\\
%  & \\
  &
\end{array}
}.
\ee
Because of the frame freedom evident in the degenerate variables, we will split
the vector $\mathbf{v}$ into two parts:
$\mathbf{v}=(\mathbf{v}_D,\mathbf{v}_F)$, a `determined' part containing 9
variables which have an individual propagation equation and a `free' part
containing 5 [say, $(\Omega\V,\bar a\V~|~\bar\Omega\V,a\V,\theta\S)$] which do
not. Inserting Eq.~(\ref{solform1}) into the propagation equation,
Eq.~(\ref{propgeneral}), will then finish off the solution, resulting in the
equation
\be
\hat\mathbf{v}_{D}=\mathbf{B}\mathbf{v}_D+\mathbf{Av}_F,\label{solform2}
\ee
where $\mathbf{B}$ is a $9\times9$ matrix and $\mathbf{A}$  is $9\times5$,
giving the solution in the form of a nine-dimensional, first-order
(non-autonomous) dynamical system for the vector $\mathbf{v}_D$, with forcing
terms from the five undetermined variables. The solution~(\ref{solform1}) is
then guaranteed to propagate using~(\ref{solform2}). \forget{Now, we have noted
above that there is one propagation equation `missing' for $\theta\S$ (say), so
one row of $\mathbf{B}$ must therefore contain a $d/d\rho$ term (in the even
parity part of $\mathbf{B}$), no matter how we choose $\mathbf{v}$.  This
reflects our remaining frame freedom (indeed, we can choose a differential
equation for $\theta\S$ to satisfy, if we wish); we will discuss the
intricacies below, but for now we note that there is a frame choice which
allows us to reduce Eq.~(\ref{solform2}) to a 6 dimensional system, with
$\mathbf{B}$ taking the rather pleasant form
\be
\mathbf{B}=\bra{\begin{array}{c|c}
  \mathbf{A} & 0 \\\hline
  0 & \mathbf{A}
\end{array}};\label{solform3}
\ee
that is, the 3-dimensional differential equation is the same for both parities.
We will now discuss the details of finding $\mathbf{v}$, $\mathbf{A}$ and
$\mathbf{M}$.}

From a naive argument based on counting degrees of freedom of our frame
vectors we would expect to be able to eliminate five variables through a
careful use of all frame freedom: the congruence~$u^a$ can be changed by
boosting to any new frame moving with some (first-order) three-velocity,
giving three degrees of freedom; given a $u^a$, we may further make any
first-order change in~$\n^a$ that preserves $u^a\n_a=0$, giving two degrees
of freedom. Indeed, there are five unknowns in Eq~(\ref{solform2}),
corresponding precisely to these five degrees of freedom, which implies that
five equations (rows) of Eq.~(\ref{solform2}) \emph{do not represent true
dynamical degrees of freedom}. We may specify a frame not only by directly
specifying $\mathbf{v}_F$ (the most obvious being $\mathbf{v}_F=0$),
selecting some 9-dimensional subspace of~${\cal N}_{14}$, but by specifying
any variable which can somehow be related to an element of $\mathbf{v}_F$
through either Eq.~(\ref{solform2}) or Eq.~(\ref{solform1}), thus defining
that element of $\mathbf{v}_F$, provided, of course, this results in a
non-zero, self consistent solution. It turns out, in fact, that the only
variables we can't specify in this way are $\E_{ab},
\H_{ab}, W_{ab},
\hat W_{ab}$ (the latter being equivalent for these purposes
to the Zerilli tensor defined later)~-- these are frame invariant. Below we
choose our frame so that 5 elements of $\mathbf{v}_D$ are zero, thus placing
5 rows of Eq.~(\ref{solform2}) into $\mathbf{M}$, and explicitly having our
core dynamical system represent the four true dynamical degrees of freedom~--
a four dimensional subspace of ${\cal N}_{14}$.

\forget{ An alternative, and more sophisticated, view of this frame
freedom has already been touched upon in section~\ref{solsoverview}. The
constraint and evolution equations restrict the solution variables to lie in an
14-dimensional subspace,~${\cal N}_14$, of the full 33-D space~${\cal V}_{33}$
at each radial position. Within this subspace all the equations propagate
consistently. If we now use the frame freedom to further constrain the solution
space, by setting

!!!!!!!!!!!! insert some frame choice here !!!!!!!!!!!!!!!!!!!

, say, or choosing some other conditions on the variables, we can select some
9-dimensional subspace of~${\cal N}_14$.}

To summarise:
\begin{itemize}
%\item there are 33 variables (24 Ricci rotation coefficients + 10
%Weyl tensor components - 1~degree of freedom corresponding to a rotation of the
%two basis vectors in the sheet);
\item[{\tiny{{$^\bullet$}}}] there are 28 propagation equations for 33 variables, which suggests that there are
33-28=5 `frame' degrees of freedom in the choice of the two basis vectors
$u^a$ and~$n^a$;
\item[{\tiny{{$^\bullet$}}}] once we have used the time-independence of the background to
harmonically decompose the evolution equations, these, combined with the
original constraint equations, give rise to a total of 19 linearly
independent algebraic relations (not all of the 27 evolution and constraint
equations~-- represented by $\mathbf{F}$~-- are independent);
\item[{\tiny{{$^\bullet$}}}] using these algebraic relations to eliminate 19 variables leaves
$33-19=14$ variables to be solved for;
\item[{\tiny{{$^\bullet$}}}] since the contraint and evolution equations propagate consistently
(imposing the constraints `commutes' with the hat derivative) we can be sure
that the propagation equations for eliminated variables can be dropped, since
they will follow from the propagation equations for the remaining variables;
\item[{\tiny{{$^\bullet$}}}] we find then that 9 propagation equations (for~$\mathbf{v}_D$)
remain (leaving the $14-9=5$ frame degrees of freedom,~$\mathbf{v}_F$);
\item[{\tiny{{$^\bullet$}}}] finally, since we can choose the 5 frame degrees of freedom
more or less arbitrarily we really only have $9-5=4$ true dynamical
propagation equations, as we would expect (to see this, imagine choosing 5 of
the 9 elements of~$\mathbf{v}_D$ to be anything at all; then the propagation
equations for those variables just fix all five elements of~$\mathbf{v}_F$,
so that the only unknowns that remain are the 4 components of~$\mathbf{v}_D$
for which we have propagation equations.
\end{itemize}

\subsection{Determining the full solution: Finding $\mathbf{v},~\mathbf{M}$ and $\mathbf{B}$}

\subsubsection{Odd}
\label{oddsolssec}

\subsubsubsection{General Frame}

If we don't specify a frame choice, and choose our solution vector as, say,
$\mathbf{v}_{D\odd}=(\bar\Sigma\T,\bar\zeta\T,\bar X\V,\bar\udot\V)$, then
there are two undetermined variables, which we can choose to be
$\mathbf{v}_{F\odd}=(\Omega\V,\bar a\V)$, as witnessed in the propagation
equation for this solution vector:
\be
\hat\mathbf{v}_D=\mathbf{B\gen\odd v}_D+\mathbf{A}\odd\mathbf{v}_F.\label{crap}
\ee
All other variables are linear combinations of elements of $\mathbf{v}_D$,
except $\bar\Sigma\V$, which depends on $\Omega\V$; nothing depends on $\bar
a\V$. Thus, $\mathbf{v}_F$ represents a frame freedom in the odd-parity
variables, and we can specify it at will.

\subsubsubsection{Specific Frame}

Here, we will choose the frame specifically such that $\bar
Y\V=\bar\udot\V=0$. This immediately implies that $\xi\S=\Omega\S=\bar
a\V=\bar X\V=\bar Z\V=\Omega\V=0$, which is a rather decent simplification.
For our reduced basis vector we will choose
\be
\mathbf{v}= \bra{\begin{array}{c}
  \bar W\T \\
  \hat{\bar W}\T
\end{array}};
\ee
i.e., the governing DE will be the Regge-Wheeler equation. The remaining
variables are then given by
\be
\vo=\bra{\begin{array}{c}
 \bar\E\T \\ \H\T \\ \bar\Sigma\T \\ \bar\zeta\T\\
 \bar\E\V \\ \H\V \\ \bar\Sigma\V \\ \Omega\V\\ \bar\udot\V \\ \bar\alpha\V\\
 \bar\hatn\V %\\ \bar\dotn\V
 \\ \bar X\V \\ \bar Y\V \\ \bar Z\V
 \\ \H\S \\ \Omega\S \\ \xi\S
\end{array}}=
\mathbf{M\odd v}=\bra{\begin{array}{cc}
    -C/2\phi^2 r^4   &  -2/\phi r^2     \\
   \bra{-c_{-3}+8r^2\omega^2+16}/4{i\omega\phi r^4} ~    &   ~
   -C/{2i\omega\phi^2r^4}    \\
    1/i\omega r^2    &   2/i\omega\phi r^2     \\
      2/\phi r^2   &   0     \\
      \lll/\phi r^3   &  0      \\
     0& -\lll/i\omega\phi r^3   \\
      -\lll/i\omega\phi r^3    &0      \\
      0&0     \\
      0 & 0\\
       \lll/i\omega\phi r^3 & 0 \\
      0   &     0   \\
      0   &    0    \\
      0   &    0    \\
      0   &    0    \\
      -\ll\lll/i\omega\phi r^4   &    0    \\
      0   &   0     \\
      0   &0
\end{array}}\bra{\begin{array}{c}
  \bar W\T \\
  \hat{\bar W}\T
\end{array}}
\ee
where we have used
\be
C=3\phi^2r^2-4.
\ee

\forget{
We have discussed in section~\ref{solsoverview} that the solutions for our
variables may be put in the form of a solution matrix $\mathbf{M}$ and vector
of basis variables $\mathbf{v}$, together with a first order ode for
$\mathbf{v}$.}

\forget{
\subsubsection{Odd}
\label{oddsolssec}

\forget{
Here we wish to find $\mathbf{M}\odd$, $\mathbf{v}\odd$ and $\mathbf{B}\odd$,
defined in section~\ref{solsoverview}, but first we'll discuss the method of
solution specific to the odd case. We can solve the full system of equations as
follows. First note that we have 16 variables which we must
determine:~(\ref{oddvars}). In the evolution equations, use the time harmonic
decomposition to turn these into `constraints' for the propagation equations
(i.e., the evolution equations propagate consistently, when time harmonics are
used). The propagation equations then form a first-order linear system of
differential equations, one for each variable. The evolution equations are now
linear equations in the 16 variables, and can thus be solved easily. They are
not all independent which we turn to now.

There are evolution equations for each variable; however, of these the
evolution equations for $\bar{\hatn}\V$~(\ref{OmVhat}),
$\bar\E\T$~(\ref{Eabdot}), and $\H\T$~(\ref{Habdot}) are really propagation
equations (for $\Omega\V$, $\H\T$ and $\bar \E\T$ respectively), so these may
not be included in the set. This leaves $16-3=13$ equations in 16 unknowns,
which means that there are 3 unknown functions in the evolution equations,
which must be determined by the propagation equations, and form our solution
vector $\mathbf{v}\odd$. The remaining 13 variables will be determined as
linear combinations of these 3 variables; these linear combinations give
$\mathbf{M}\odd$. We are guaranteed that these solutions will propagate because
the evolution equations, from whence they came, do. As the (actual~-- i.e.,
non-evolution) constraints~(\ref{conom_anl}), (\ref{consig_anl}),
(\ref{barEV}), (\ref{barZ}), (\ref{barY}), (\ref{barX}) and~(\ref{bardotn})
evolve consistently with the evolution equations, they are just linear
combinations of the evolution equations, when time harmonics are used, so they
don't give any more information as regards the full solution.

The problem of the full solution is thus reduced to deciding which variables we
would like to choose as the basis for the full solution. We will choose $\bar
X\V$, $\bar W\T$ and $\hat{\bar W}\T$, for reasons which will become clear. We
can solve for $\bar\zeta\T$ immediately from Eq.~(\ref{Wab})
\be
%\bar\zeta\T=\bar W\T+\:23r^{-1}\udot^{-1}\phi^{-2}\bar X\V.\label{zetabarsol}
\bar\zeta\T=2r^{-2}\phi^{-1}\bar W\T+\:23r^{-1}\udot^{-1}\phi^{-2}\bar X\V.\label{zetabarsol}
\ee
To insert $\hat{\bar W}\T$ into the equations, eliminate $\bar\hatn\V$ from
from~(\ref{xihat}) and~(\ref{zetahat}); or, simply calculate $\hat{\bar W}\T$
from~(\ref{zetabarsol}),
\be
%\bar\E\T=-\hat{\bar W}\T-\phi\bar W\T.\label{ETbarsol}
\bar\E\T=-r^{-2}\phi^{-1}\bras{2\hat{\bar W}\T+\bra{\phi-2\udot}\bar W\T}.\label{ETbarsol}
\ee
This gives the solution for $\bar\E\T$. We can now insert
Eqs.~(\ref{zetabarsol}) and~(\ref{ETbarsol}) into the 13 evolution equations,
and solve for each variable as discussed.}

The problem of the full solution is thus reduced to deciding which variables we
would like to choose as the basis for the full solution. There are evolution
equations for each variable; however, of these the evolution equations for
$\bar{\hatn}\V$~(\ref{OmVhat}), $\bar\E\T$~(\ref{Eabdot}), and
$\H\T$~(\ref{Habdot}) are really propagation equations (for $\Omega\V$, $\H\T$
and $\bar \E\T$ respectively), so these may not be included in the set. This
leaves $16-3=13$ equations in 16 unknowns, which means that there are 3 unknown
functions in the evolution equations, which must be determined by the
propagation equations, and form our solution vector $\mathbf{v}\odd$. We will
choose $\bar X\V$, $\bar W\T$ and $\hat{\bar W}\T$ for this, to keep the
Regge-Wheeler tensor at the core of our solution.

The full solution, in terms of these three variables is~$\mathbf{M}\odd$, which
is found from solving the 13 evolution equations:
\be
\begin{array}{c|c|c|c|c}
    \mbox{\sc odd}  &  \mbox{factor}  &  \bar X\V   &    \bar W\T   &    \hat{\bar W}\T   \\\hline
    \bar\E\T        &            &  0   &    r^{-2}\phi^{-1}\bra{2\udot-\phi}   &    -2r^{-2}\phi^{-1}   \\%\hline
    \H\T            &       i\omega^{-1} & 0    &  r^{-4}\phi^{-1}\bras{\:14c_{-3}-2\bra{2+r^2\omega^2}}    &  r^{-2}\phi^{-1}\bra{\phi-2\udot}     \\%\hline
    \bar\Sigma\T    &      i\omega^{-1}   &  -\:23r^{-1}\phi^{-2}   &  -r^{-2}    &   -2r^{-2}\phi^{-1}    \\%\hline
    \bar\zeta\T     &            &  \:23r^{-1}\udot^{-1}\phi^{-2}   &    2r^{-2}\phi^{-1}   &   0    \\%\hline
    \bar\E\V        &            &  -\phi^{-1}   &  \lll r^{-3}\phi^{-1}     &  0     \\%\hline
    \H\V            &      i\omega^{-1} &  -\udot\phi^{-1}   &  0    &   \lll r^{-3}\phi^{-1}    \\%\hline
    \bar\Sigma\V    &      i\omega^{-1}  &  -\;{\udot\bra{2\udot-\phi}+2\omega^2}{9\phi^2\udot}   &   -\:13\lll r^{-3}\phi^{-1}    &  0     \\%\hline
    \Omega\V        &      i\omega^{-1}      &   -\;{4\udot\bra{2\udot-\phi}+2\omega^2}{9\phi^2\udot}  &    -\:43\lll r^{-3}\phi^{-1}   &    0   \\%\hline
    \bar\hatn\V     &            &   \:29\E^{-2}\bras{\udot\bra{5\udot-7\phi}+2\omega^2}  &    -\:83\lll r^{-3}\E^{-1}   &  0     \\%\hline
    \bar\dotn\V     &    i\omega^{-1} &  -\;{\udot\bra{2\udot-\phi}+2\omega^2}{3\phi^2\udot}   &    -\lll r^{-3}\phi^{-1}   &   0    \\%\hline
    \bar X\V        &            &  1   &    0   &    0   \\%\hline
    \bar Y\V        &            &  \:13\E^{-1}\bra{\phi-2\udot}   &   0    &    0   \\%\hline
    \bar Z\V        &            &   -\:23\phi^{-2}\bra{\phi+\udot}  &    0   &   0    \\%\hline
    \H\S            &      i\omega^{-1}      &    0 &    \ll\lll r^{-4} \phi^{-1}   &    0   \\%\hline
    \Omega\S        &      i\omega^{-1}      &  \:13\ll r^{-1}\phi^{-2}   &   0    &  0     \\%\hline
    \xi\S           &            &   -\:13\ll r^{-1}\udot^{-1}\phi^{-2}  &   0    & 0
\end{array}\label{oddfull}
\ee
where the table gives the coefficients for each variable and a common factor.
In  equations like these, $i\omega$ may be understood as a differential
operator, where every factor of $i\omega$ represents a time derivative on the
function it multiplies; in the cases where there is a common factor of
$i\omega^{-1}$, this should be taken to the left hand side of the equation, if
one wishes to convert the equations back to dot-derivatives.

The full solution~(\ref{oddfull}) is known, then, when the functions $\bar W\T$
and $\bar X\V$ are known; $\bar W\T$ is determined from the Regge-Wheeler
equation~(\ref{RW}), while $X\V$ is determined from the propagation equation
for $\bar X\V$~(\ref{Xhat}), which now becomes
\be
%\hat{\bar X}\V=-\:16\udot^{-1}\bras{\udot\bra{\phi+4\udot}+4\omega^2}\bar X\V
%-2\lll r^{-1}\phi^2\bar W\T.\label{barXVde}
\hat{\bar X}\V=-\:23\bras{r^{-2}\phi^{-1}+\omega^2\udot^{-1}}\bar X\V
-4\lll r^{-3}\phi\bar W\T.\label{barXVde}
\ee
This equation has a driving term which is proportional to $\bar W\T$; given
that $\bar W\T$ satisfies a second order equation, $\bar X\V$ must therefore
satisfy a third order equation (and it can be shown that it cannot satisfy a
second order equation). \forget{Indeed, one can replace $\bar W\T$ by $\bar
X\V$ in Eq.~(\ref{oddfull}), so that each variable is a linear combination of
$\bar X\V$ and its first and second hat-derivatives. (Of course, this is true
of any of the variables $\{\bar X\V,
\bar Y\V, \bar Z\V,
\Omega\S,
\xi\S\}$~-- any of these would do equally well.) However, we choose to write
the solutions in the form above as it is the Regge-Wheeler equation in the
standard metric approach which governs all odd parity perturbations; once $\bar
W\T$ is determined, then $\bar X\V$ is determined up to a constant by
Eq.~(\ref{barXVde}).

For completeness, we note that in terms of the more familiar $r$ parameter,
Eq.~(\ref{barXVde}) takes the form
\be
\;{d\bar X\V}{dr}=-\;{2\sigma^2r^3+m}{3m\bra{r-2m}}\bar X\V-\;{8\lll}{r^4}\bar
W\T.
\ee}

%Moreover, it can be shown that
%\be
%\bar X\V=a_1\hat{\bar W}\T +a_2{\bar W}\T,
%\ee
%although we can't solve the coupled first-order des that $a_1$ and $a_2$ must
%satisfy; this does imply though that only one variable is enough to solve the
%system....???

We have therefore found our odd solution,  given by~(\ref{oddfull}), with the
basis vector for the solution
\be
{\mathbf{v}}\odd=\bra{\begin{array}{c}
  \bar X\V \\
  \bar W\T \\
  \hat{\bar W}\T
\end{array}},
\ee
which satisfies
\be
\hat{\mathbf{v}}\odd=\mathbf{B\odd{v}\odd}\label{fullsol1-odd}
\ee
and the  the $3\times3$ matrix $\mathbf{B}\odd$ is
\be
\mathbf{B}\odd=\bra{\begin{array}{ccc}
  -\:23\bras{r^{-2}\phi^{-1}+\omega^2\udot^{-1}}&
-4\lll r^{-3}\phi & 0 \\
  0 & 0 & 1 \\
  0  & {\ll{r^{-2}}+3\E}-\omega^2& -\udot
\end{array}}.\label{Bodd}
\ee
Meanwhile, our variables may be written as
\be
{\mathbf{V}}\odd={\mathbf{M}}\odd{\mathbf{v}}\odd
\ee
where ${\mathbf{M}}\odd$ is a $15\times3$ matrix corresponding to the
variable~(\ref{oddvars}), with components given by~(\ref{oddfull}).

%It is useful to visualise the odd part of ${\cal N}_8$ we have found as a 3-d
%space with coordinates $\bar X\V$, $\bar W\T$ and $\hat{\bar W}\T$; the zeros
%in~(\ref{oddfull}) tell us the sector in which each variable exists.

In Figs.~\ref{X-W-hatW-l=20-sigma=6.6-r=2.4..10.ps}~--
\ref{X-r-odd-l=2-QNM-sigma=sR=.30105,sI=.47828-r=2..4.ps} we show some aspects
of the solution of the differential equation~(\ref{fullsol1-odd}),
concentrating on the behaviour of $\bar X\V$, as the Regge-Wheeler equation is
well understood. In Fig.\ref{X-W-hatW-l=20-sigma=6.6-r=2.4..10.ps}, we show the
full solution vector $\mathbf{v}\odd$ for variables which do not give rise to a
QNM, which shows the solution curve decaying with increasing distance from the
horizon. The key point is that $\bar W\T$ and $\hat{\bar W}\T$ do not decay,
whereas $\bar X\V$ does, when $\sigma$ is real. In
Fig.~\ref{X-r-odd-l=2..10-sigma=6.6-r=2.4..9.ps} we show $\bar X\V$ as a
function of $r$ for different values of $\ell$, and $\sigma$ real again. This
shows amplitude and wavelength increasing with $\ell$. In
Figs.~\ref{X-r-odd-l=2-QNM-sigma=sR=.37367,sI=.08896,sR=.34671,sI=.27391-r=2..15.ps}
and~\ref{X-r-odd-l=2-QNM-sigma=sR=.30105,sI=.47828-r=2..4.ps} we show $\bar
X\V$ as a function of $r$ for the first three QNM frequencies when
$\ell=2$~\cite{nollert}. For the first fundamental frequency, the solution is
pretty unexciting, but this is not so for the second mode: we see some
interesting wiggles before the solution curve gets down to the business of
growing very fast when $r\gtrsim10$, although this is not shown. It looks
pretty similar to the curve in
Fig.~\ref{X-r-odd-l=2-QNM-sigma=sR=.30105,sI=.47828-r=2..4.ps}, which is the
third frequency, and can be seen to grow very fast while the wavelength
shrinks, as the distance from the horizon is increased.

\begin{figure}[ht]
\includegraphics[width=13cm]{{X-W-hatW-l=20-sigma=6.6-r=2.4..10_LAB.ps}}
\caption{\small Solution curve of the components of the vector $10^{10}\mathbf{v}\odd$
parameterised by $r$ for $\ell=20$ and $\sigma=6.6$, with $m=1$. Boundary
conditions are (arbitrarily) set at ${\mathbf{v}\odd}(r=2.3)=10^{-8}(10,1,1)$.
The plot ends in the  circles at $r=10$ in the centre, after spiraling in from
the outside at $r\sim 2.4$. Oscillations of $\bar X\V$ die off much more
quickly than $\bar W\T$.
\label{X-W-hatW-l=20-sigma=6.6-r=2.4..10.ps}}
\end{figure}

\begin{figure}[ht]
\includegraphics[width=11cm]{{X-r-odd-l=2..10-sigma=6.6-r=2.4..9_LAB.ps}}
\caption{\small Solution curve of the components of  $\bar X\V$
as a function of  $r$ for $\ell=2,\cdots,10$ and $\sigma=6.6$, with $m=1$.
Boundary conditions are (arbitrarily) set at
${\mathbf{v}\odd}(r=2.3)=10^{-8}(10,1,1)$. This shows the amplitude and
wavelength of the oscillations of $\bar X\V$ increasing with $\ell$; $\ell=2$
hugs the $\bar X\V=0$ axis, while the wildest oscillations occur for $\ell=10$.
\label{X-r-odd-l=2..10-sigma=6.6-r=2.4..9.ps}}
\end{figure}

\begin{figure}[ht]
\includegraphics[width=10cm]{{X-r-odd-l=2-QNM-sigma=sR=.37367,sI=.08896,sR=.34671,sI=.27391-r=2..15_LAB.ps}}
\caption{\small Solution curve of the components of (the real part of) $\bar X\V$
as a function of  $r$ for $\ell=2$ and the first two quasinormal mode
frequencies~\cite{NS,nollert} $\sigma=0.37367+0.08896i$ and
$\sigma=0.34671+0.27391i$, with $m=1$. Boundary conditions are (arbitrarily)
set at ${\mathbf{v}\odd}(r=2.003)=10^{-8}(10^{-2},1,1)+i10^{-9}(1,1,1)$. For
the first QNM, nothing much happens~-- one bump before settling to zero; for
the second, we have some rather interesting oscillations. After they become
very gentle around $r\sim10$, they start to oscillate again, with a much longer
wavelength; this continues as $r\rightarrow\infty$, with the oscillations
becoming larger and larger. See Sec.~\ref{QNMsec} for a discussion of QNMs.
\label{X-r-odd-l=2-QNM-sigma=sR=.37367,sI=.08896,sR=.34671,sI=.27391-r=2..15.ps}}
\end{figure}

\begin{figure}[ht]
\includegraphics[width=10cm]{X-r-odd-l=2-QNM-sigma=sR=.30105,sI=.47828-r=2..4_LAB.ps}
\caption{\small Solution curve of the components of (the real part of) $\bar X\V$
as a function of  $r$ for $\ell=2$ for the third quasinormal mode
frequency~\cite{NS,nollert} $\sigma=0.30105+0.47828i$, with $m=1$. Boundary
conditions are as in the previous plot. In contrast with
Fig.~(\ref{X-r-odd-l=2-QNM-sigma=sR=.37367,sI=.08896,sR=.34671,sI=.27391-r=2..15.ps}),
we see the growing oscillations start right away from the horizon, and grow
very fast with $r$: by $r\sim 10$, $\bar X\V\sim10^{20}$!
\label{X-r-odd-l=2-QNM-sigma=sR=.30105,sI=.47828-r=2..4.ps}}
\end{figure}

}

\subsubsection{Even}\label{evensolssec}

\subsubsubsection{General Frame}

If we don't specify a frame choice, and choose our solution vector as, say,
$\mathbf{v}_{D\even}=(\Sigma\T,\zeta\T,\Sigma\V+\bar\Omega\V,a\V, X\V, Y\V)$,
then there are three undetermined variables, which we can choose to be
$\mathbf{v}_{F\even}=(a\V,\bar\Omega\V,\theta\S)$. The underlying propagation
equation is then
\be
\hat\mathbf{v}_D=\mathbf{B\gen\even v}_D+\mathbf{A}\even\mathbf{v}_{F}\label{crap2}
\ee
All other variables are linear combinations of elements of
$\mathbf{v}_{D\even}$, except $\Sigma\S$ which depends on $\theta\S$. Recall
that these freedoms arise from the nature of the propagation and evolution
equations: only the combinations of variables, $\Sigma_a-\lc_{ab}\Omega^b$
and $\Sigma-\:23\theta$, are determined by the propagation and evolution
equations, and not the individual variables themselves (although they appear
in different combinations in other equations); meanwhile, $a^a$ is only
present on the rhs of the propagation equations. Thus, $a\V,\bar\Omega\V$ and
$\theta\S$ represent frame freedom in the even-parity variables, and we can
specify these at will (or other variables which indirectly fix these).

\subsubsubsection{Specific Frame}

To concur with the odd case above, we will choose here a frame in which
$\udot\V=Y\V=Z\V=0$. We choose
\be
\mathbf{v}= \bra{\begin{array}{c}
   W\T \\
  {\cal Z}
\end{array}},
\ee
so that the governing DE will be Eq.~(\ref{RWzer}). Thus, the basis vector in
${\cal V}_{33}$ can easily be converted to the Regge-Wheeler equation, or the
Zerilli equation, depending on ones frame of mind [by substituting for $W\T$ or
${\cal Z}$ from the rhs of Eq.~(\ref{RWzer})]. The other variables are linear
combinations of the elements of this solution basis vector:
\be
\ve=\bra{\begin{array}{c}
\E\T \\ \bar\H\T \\ \Sigma\T \\ \zeta\T\\
 \E\V \\ \bar\H\V \\ \Sigma\V \\ \bar\Omega\V \\ \udot\V \\ \alpha\V\\
 \hatn\V \\
% \dotn\V\\
  X\V \\  Y\V \\  Z\V\ \\
 \Sigma\S \\ \theta\S
\end{array}}=
\mathbf{M\even v}=\bra{\begin{array}{cc}
   2\bra{12\E^2r^4-C\lll}/c_3\phi^2r^4      &   -\bras{12\omega^2\phi^2\E r^6+C\ll\lll}/2i\omega\phi^3r^5     \\
     2i\omega/\phi r^2    &  c_3\bra{c_{-3}-8}/8\phi^2r^3      \\
     4\lll/i\omega c_3r^2    &  -\bras{6\omega^2\E r^4+\ll\lll}/\omega^2\phi r^3      \\
     2/\phi r^2    &  2\ll\lll/i\omega\phi^2r^3      \\
     -\lll/\phi r^3    &  -\ll\lll^2/i\omega\phi^2r^4      \\
    -12 \lll\E/i\omega c_3r    &    -3\lll\udot/2\omega^2 r^2    \\
    \lll/i\omega\phi r^3      &
         -\ll\lll\bras{\lll c_3+6\E^2r^4}/\omega^2\phi^2r^4     \\
      0     &  6\ll\lll\udot^2/\omega^2c_3       \\
      0&0\\
       \lll C/i\omega\phi r^3 c_3   & -\ll\lll C/4\omega^2\phi^2r^4   \\
      0   &    12\ll\lll\udot/i\omega c_3    \\
       0  &   -3\ll\lll\udot/i\omega r^2      \\
       0  &    0    \\
        0 &   0     \\
      0   &   \ll\lll\udot\bras{5C^2-24C\bra{L-1}-32\bra{3L-2}}/2\omega^2\phi^2r^3c_3    \\
       0  &-3\ll\lll\udot^2\bra{C+8}/\omega^2\phi rc_3
\end{array}}\bra{\begin{array}{c}
  W\T \\
  {\cal Z}
\end{array}}.
\ee

\forget{
Although we have chosen a frame for which~$\udot^a=0$, we have not used up all
of our frame freedom in the even case, since we may still choose different
timelike congruences~$u^a$, apparent by the absence of a propagation equation
for $\theta\S$. The constraint and evolution equations restrict the solution
variables to lie in an 8-dimensional subspace,~${\cal N}_8$, of the full 31-D
space~${\cal V}_{31}$ at each radial position.

We have 15 variables to determine~(\ref{evenvars}) and we have evolution
equations for each of these variables except $Z\V$, but of these, the equation
for $\theta\S$~(\ref{Zhat}), is a propagation equation for $Z\V$; evolution
equations for ${\hatn}\V$~(\ref{OmVbarhat}), $\E\T$~(\ref{Eabdot}), and
$\bar\H\T$~(\ref{Habdot}) are really propagation equations (for $\bar\Omega\V$,
$\bar\H\T$ and $\E\T$ respectively). Thus we have $15-1-4=10$ independent
equations for 15 unknowns, leaving a 5-D solution vector $\mathbf{v}\even$,
which we must choose (recall that any choice is fine provided it spans the even
part of ${\cal N}_8$). We can now use the frame freedom to further constrain
the solution space (and the dimension of $\mathbf{v}\even$), by setting
$\theta\S=0$, say, or choosing some other condition on the variables, we select
some \emph{subspace} of~${\cal N}_8$ at each radius.
 We will briefly consider three frame choices, each
of which has certain advantages.
\forget{\begin{enumerate}
\item The general and zero expansion frame;\label{frame1}
\item The `cross-over' frame, in which the underlying DE takes the same form as
the odd parity case [see Eq.~(\ref{solform3})];\label{frame2}
\item The `covariant' frame, where the governing dynamics of both parities may
be given as two \emph{tensorial} DE's.\label{frame3}
\end{enumerate}}

\subsubsubsection{The general and zero expansion frames}

Because there is no independent propagation equation for $\theta\S$, let
${\mathbf{v}}\even\gen=(X\V, Z\V, W\T, {\cal Z})$, where the `$\mathsf{g}$'
stands for `general frame'; then, the 4-dimensional dynamical system in general
is
\be
\hat{\mathbf{v}}\even\gen={\mathbf{B\even\gen v\even\gen}}+i\omega r^{-1}\theta\S\bra{\begin{array}{c}
  0 \\
  1 \\
  0 \\
  0
\end{array}}\label{soleveven-gen}
\ee
where the $4\times4$ matrix $\mathbf{B}\even\gen$ is rather complicated, but a
key point is that the DE's for $W\T$ and ${\cal Z}$ decouple from the rest and
in no way depend on our choice of $\theta\S$. In addition, the variables $\E\T$
and $\bar\H\T$ only depend on $W\T$ and ${\cal Z}$. Hence these variables are
\emph{frame-invariant}. The other crucial point is that the DE for $X\V$
depends on $Z\V$; thus, until $\theta\S$ is given, these variables are unknown.
Therefore, we can equally well choose one of $X\V$ or $Z\V$ as our frame
freedom, as this then effectively fixes $\theta\S$; we exploit this in the
frame choices we discuss below.

We show some solutions of the differential equation~(\ref{soleveven-gen}), when
$\theta=0$ (non-expanding frame) in
Figs.~\ref{Z-W-hatW-l=20-sigma=6.6-r=2.3..9.ps}~--
\ref{Z-r-l=20-sigma=6.6-r=2.3..5-l=2..10.ps}.

%We will not give the solution matrix $\mathbf{M}\even\gen$ in this frame.

\begin{figure}[ht]
\includegraphics[width=11cm]{{Z-W-hatW-l=20-sigma=6.6-r=2.3..9_LAB.ps}}
\caption{\small Solution curve of $Z\V$ and $W\T$ against $r$
for $\ell=20$ and $\sigma=6.6$ and $m=1$, in the $\theta=0$ frame. Boundary
conditions are (arbitrarily) set at
${\mathbf{v}}\gen(r=2.3)=10^{-9}(10,10,1,(i\sigma)^{-1}10)$. This shows
oscillations of $Z\V$ decaying very fast moving away from $r\sim2.3$ (where the
plot starts in the upper left hand corner); in contrast, the amplitude of $W\T$
does not decay. This is reflected in the curve at large $r\lesssim9$ being
oscillations in the $Z\V=0$ plane.
\label{Z-W-hatW-l=20-sigma=6.6-r=2.3..9.ps}}
\end{figure}

\begin{figure}[ht]
\includegraphics[width=11cm]{{Z-X-hatW-l=20-sigma=6.6-r=2.3..9_LAB.ps}}
\caption{\small Solution curve of the previous plot,
Fig.~\ref{Z-W-hatW-l=20-sigma=6.6-r=2.3..9.ps}, but now showing $X\V$ instead
of $W\T$. $X\V$ decays very quickly too.
\label{Z-X-hatW-l=20-sigma=6.6-r=2.3..9.ps}}
\end{figure}

\begin{figure}[ht]
\includegraphics[width=11cm]{{X-Z-r-even-l=2-QNM-sigma=sR=.37367,sI=.08896-r=2..8_LAB.ps}}
\caption{\small $X\V$ and $Z\V$ vs. $r$ for $\ell=2$ and the fundamental
quasinormal frequency $\sigma=0.37367+0.08896i$. Boundary conditions are chosen
${\mathbf{v}\even\gen}(r=2.003)=10^{-8}(10^{-2},1,1,1)+i10^{-9}(1,1,1,1)$, and
the plot starts at $r=2.03$ ($Z\V$ has already undergone one oscillation). Both
$X\V$ and $Z\V$ tend to zero for this frequency.
\label{X-Z-r-even-l=2-QNM-sigma=sR=.37367,sI=.08896-r=2..8.ps}}
\end{figure}

\begin{figure}[ht]
\includegraphics[width=11cm]{{Z-r-l=20-sigma=6.6-r=2.3..5-l=2..10_LAB.ps}}
\caption{\small Solution curve of the previous plots,
Fig.~\ref{Z-W-hatW-l=20-sigma=6.6-r=2.3..9.ps}, but now showing $Z\V$ against
$r$ for different values of $\ell=2,\cdots,10$. For $\ell=2$, at the top of the
plot, there are no oscillations, and $Z\V$ just tends to zero; $\ell$ increases
as we move from curve to curve down the plot until $\ell=10$, where there are
large oscillations, but still decaying to zero as $r$ increases.
\label{Z-r-l=20-sigma=6.6-r=2.3..5-l=2..10.ps}}
\end{figure}

\subsubsubsection{The `cross-over' frame}

We will choose here a `crossover' frame, in which the solutions become similar
in form to the odd parity case, and the differential equation for the full
solution is of the form~(\ref{solform3}). To do this, demand that $X\V$, which
satisfies~(\ref{Xhat}), has a propagation equation of the same form as in the
odd parity case, Eq.~(\ref{barXVde}):
\be
\hat{X}\V=-\:23\bras{r^{-2}\phi^{-1}+\omega^2\udot^{-1}}X\V
-4\lll r^{-3}\phi W\T;\label{XVde-crossover}
\ee
this fixes $Z\V$ and hence $\theta\S$. Indeed, all variables can be written now
in terms of $\{X\V, W\T, {\cal Z}\}\leftrightarrow\{X\V, W\T, {\hat W\T}\}$; we
will use the former, involving ${\cal Z}$ to try to keep the equations simple,
although it makes comparison with the odd case a little more difficult~-- we
can use Eq.~(\ref{ZofW}) to convert.
\be
\begin{array}{c|c|c|c}
    \mbox{\sc even}&       X\V     &   2r^{-2}\phi^{-1}W\T     &  \:12c_3r^{-1}\phi^{-1}  {\cal Z}\\\hline
    \E\T         &     0     &   \;{16\bra{\ll+1}-3r^2\phi^2c_1}{4r^2\phi c_3}     & \;{4\ll\lll-3r^2\phi^2\bras{r^2\omega^2\bra{r^2\phi^2-4}-\ll\lll}}{i\omega r^4\phi^2 c_3}       \vphantom{\;{X\;{X\;{x}{X}}{X}}{X\;{X}{X}}}  \\
    \bar\H\T     &    0     &    i\omega    &   \:14\ll\bra{\phi+\udot}-3\udot    \vphantom{\;{X\;{X\;{x}{X}}{X}}{X\;{X}{X}}}   \\
    \Sigma\T     &    -\:23\bra{i\omega r\phi^2}^{-1}     &    2\lll\phi\bra{i\omega c_3}^{-1}    &   \;{4\ll\lll-3r^2\phi^2\bras{r^2\omega^2\bra{r^2\phi^2-4}+\ll\lll}}{\omega^2r^4\phi^2}     \vphantom{\;{X\;{X\;{x}{X}}{X}}{X\;{X}{X}}}  \\
    \zeta\T      &    \;{8r}{3\phi\bra{r^2\phi^2-4}}     &   1     &   0     \vphantom{\;{X\;{X\;{x}{X}}{X}}{X\;{X}{X}}}  \\
    \E\V         &    -\phi^{-1}     &    -\:12\lll r^{-1}    &   -\:12\ll\lll\bra{i\omega r^3\phi}^{-1}     \vphantom{\;{X\;{X\;{x}{X}}{X}}{X\;{X}{X}}}  \\
    \bar\H\V     &     \;{r^2\phi^2-4}{4i\omega r^2\phi^2}    &    -\;{3\lll\phi\bra{r^2\phi^2-4}}{4i\omega r c_3}    &   \;{3\lll\bra{r^2\phi^2-4}\bras{4r^4\phi^2\omega^2+\ll\bra{3r^2\phi^2-4}}}{8\omega^2 r^5\phi^2c_3}     \vphantom{\;{X\;{X\;{x}{X}}{X}}{X\;{X}{X}}}  \\
    \Sigma\V     & \;{16\bra{r^2\phi^2-1}-r^4\phi^2\bra{3\phi^2+16\omega^2}}{18i\omega r^2\phi^3\bra{r^2\phi^2-4}}        &   \;{\lll\bra{3r^2\phi^2-4}}{6i\omega rc_3}     &   -\;{\ll\lll\bras{r^4\phi^2\bra{9\phi^2+16\omega^2}-8\bra{3r^2\phi^2-2}}}{12\omega^2r^5\phi^2c_3}     \vphantom{\;{X\;{X\;{x}{X}}{X}}{X\;{X}{X}}}  \\
    \bar\Omega\V & \;{2\bras{r^4\phi^2\bra{3\phi^2+4\omega^2}-16\bra{r^2\phi^2-1}}}{9i\omega r^2\phi^3\bra{r^2\phi^2-4}}    &   \;{\lll\bras{r^2\phi^2\bra{25\ll+24-18r^2\phi^2}}+4\ll\bra{\ll+1}}{3i\omega r^3\phi^2 c_3}        & -\;{\ll\lll\bras{r^2\phi^2\bra{12\ll-r^2\bra{9\phi^2+8\omega^2}+8\bra{\ll\lll-2}}}}{6\omega^2\phi^3 r^5 c_3}       \vphantom{\;{X\;{X\;{x}{X}}{X}}{X\;{X}{X}}}  \\
    \hatn\V      &    \;{2r^4\phi^2\bra{33\phi^2+32\omega^2}-16\bra{19r^2\phi^2-10}}{9\phi^2\bra{r^2\phi^2-4}^2}     &    \;{4\lll\bra{\ll-4r^2\phi^2}}{3r\phi\bra{r^2\phi^2-4}}    &    \;{\ll\lll\bra{c_{-3}-16}}{3i\omega r^3\phi^2\bra{r^2\phi^2-4}}    \vphantom{\;{X\;{X\;{x}{X}}{X}}{X\;{X}{X}}}  \\
    \dotn\V      &    \;{16\bra{r^2\phi^2-1}-r^4\phi^2\bra{3\phi^2+16\omega^2}}{6i\omega r^2\phi^3\bra{r^2\phi^2-4}}     &    \;{\lll\bra{3r^2\phi^2-4}}{2i\omega r c_3}    &  \;{\ll\lll\bras{8\bra{3r^2\phi^2-2}-r^4\phi^2\bra{9\phi^2+16\omega^2}}}{4\omega^2r^5\phi^3c_3}      \vphantom{\;{X\;{X\;{x}{X}}{X}}{X\;{X}{X}}}  \\
    X\V          &    1      &    0   &     0   \vphantom{\;{X\;{X\;{x}{X}}{X}}{X\;{X}{X}}}  \\
    Y\V          &    -\:23c_3\phi^{-1}\bra{r^2\phi^2-4}^{-1}     &    0    &   {\ll\lll}\bra{i\omega r^3\phi}^{-1}     \vphantom{\;{X\;{X\;{x}{X}}{X}}{X\;{X}{X}}}  \\
    Z\V          & -\:23\phi^{-2}\bra{\phi+\udot} &  \;{\lll\bras{4\ll\bra{\ll+1}+r^2\phi^2\bra{24r^2\phi^2+31\ll+32}}}{3 r\phi c_3}  & \;{\ll\lll \bras{16\bra{\ll\lll-1}+r^2\phi^2\bra{r^2\bra{32\omega^2+45\phi^2}-48\bra{\ll+1}}}}{{12} i\omega r^{5}\phi^{3} c_3}\vphantom{\;{X\;{X\;{x}{X}}{X}}{X\;{X}{X}}}
%\\    \theta\S     &         &         &        &        \\
%    \Sigma\S     &         &         &        &
\end{array}\label{allevensols}
\ee
Note that $r^2\phi^2-4=-4\phi\udot r^2$. The solutions for $\Sigma\S$ and
$\theta\S$ are not explicitly given as they are huge, but may be obtained
relatively easily from the above equations.

In this crossover frame we have our basis vector
\be
\mathbf{v}\even\cross=\bra{\begin{array}{c}
  X\V \\
  W\T \\
  \hat W\T
\end{array}}
\ee
with solution matrix $\mathbf{M}\even\cross$ given by~(\ref{allevensols}) (plus
the equations for $\theta\S$ and $\Sigma\S$, which were too large to give), and
our 3-dimensional differential equation
\be
\hat{\mathbf{v}}\even\cross=\mathbf{B\odd{v}\even\cross}\label{fullsol1-evencross}
\ee
where $\mathbf{B}\odd$ is given by Eq.~(\ref{Bodd}). This is the frame in which
the full solution takes the form of Eqs.~(\ref{solform2}) and~(\ref{solform3}),
with $\mathbf{A}=\mathbf{B}\odd$.

\subsubsubsection{The `fundamental' frame}

In this crossover frame, we chose $Z\V$ (and hence $\theta\S$, etc.) such that
Eqs.~(\ref{barXVde}) and~(\ref{XVde-crossover}) are satisfied: that is, $\bar
X\V$ and $X\V$ satisfy the same ordinary differential equation \emph{after
decomposition into SH.} However, we cannot write down an equation for $\hat
X^a$ \emph{in tensor form} because of the sign of the $W$-term in
Eqs.~(\ref{barXVde}) and~(\ref{XVde-crossover}); these are of opposite sign if
we form the equations from a single tensor equation. If instead we demand that
$X^a$ satisfy
\be
\ddot X_a-\:32\udot\hat
X_a=\bra{\udot^2-\:14\E}X_a-12\E\bra{\:14\phi^2-\E}\delta^b
W_{ab},\label{X_a_de}
\ee
so that the odd parity part of this equation is given by Eq.~(\ref{barXVde}),
and the even parity part by
\be
\hat{X}\V=-\:23\bras{r^{-2}\phi^{-1}+\omega^2\udot^{-1}}X\V
+4\lll r^{-3}\phi W\T;\label{XVde-crossover2}
\ee
with a sign change on the $W$ term from Eq.~(\ref{XVde-crossover}), then we
must make the following changes to the above equations. Let
\be
{\cal
F}[\Psi]\equiv\bra{\Psi\mathrm{~in~crossover~frame~given~by~Eq.~(\ref{allevensols})}}-\bra{\Psi\mathrm{~
in~this~frame}}
\ee
for any $\Psi$. Then the following changes must be made:
\be
{\cal F}\bras{Z\V}=\;{16}{3}\lll r^{-3}\phi^{-1} W\T =i\omega{\cal
F}\bras{\bar\Omega\V} =\udot{\cal F}\bras{\hatn\V}
\ee
with changes to be made to $\theta\S$ and $\Sigma\S$.

The purpose of choosing this frame lies in the fact that the dynamical
behaviour of the spacetime is given by two covariant, gauge-invariant tensorial
equations: one is a closed wave equation~(\ref{RWtensorwave}), and the other,
Eq.~(\ref{X_a_de}), is a more unusual oscillatory equation, with a forcing term
from the wave equation. Crucially, therefore, this extra part of the solution
doesn't travel on null geodesics.

\forget{
The procedure to find the full solution is as in the odd parity case; that of
viewing the evolution equations as constraints for the propagation equations,
and solving the system of linear equations. We have 15 variables to
determine~(\ref{evenvars}). We have evolution equations for each of these
variables except $Z\V$, but of these, the equation for $\theta\S$~(\ref{Zhat}),
is a propagation equation for $Z\V$; ${\hatn}\V$~(\ref{OmVbarhat}),
$\E\T$~(\ref{Eabdot}), and $\bar\H\T$~(\ref{Habdot}) are really propagation
equations (for $\bar\Omega\V$, $\bar\H\T$ and $\E\T$ respectively), so these
may not be included in the set. Thus we have $15-1-4=10$ independent equations
for 15 unknowns. The true constraints, Eqs.~(\ref{conom_anl}),~(\ref{newcons})
and~(\ref{dotnVcons}), evolve (i.e., are linear combinations of the evolution
equations when time harmonics are used) so give no additional information.

The equation for $\Sigma\S$~(\ref{XYZ}), must be considered as an equation for
$\bra{\Sigma\S-\:23\theta\S}^\cdot$; similarly, there are no independent
propagation equations for $\Sigma\S$ and $\theta\S$, only the combination
$\bra{\Sigma\S-\:23\theta\S}$. What this tells us is that there is a remaining
frame freedom in choosing $u^a$, which we will take as a freedom in the
expansion: the field equations cannot determine $\theta$.

Before we find the full solution then, we must decide what to do with
$\theta\S$. Let's first choose the five variables as the set $\{\theta\S, X\V,
Z\V, \E\T, \bar\H\T\}$, in terms of which we can solve the full system of
evolution equations. We can show that $\bar\H\T$ and $\E\T$ may be written in
terms of ${\cal Z}$ and $W\T$. We can find $\bar\H\T$ by calculating $\hat{\cal
Z}$, given~(\ref{zer1}), and is a linear combination of ${\cal Z}$ and
$\hat{\cal Z}$; similarly we find the solution for $\E\T$ by calculating $\hat
W\T$ from~(\ref{Wab}), which is a linear combination of $\hat{\cal Z},~\hat
W\T$ and $W\T$. Therefore, our five variables may be considered as $\{\theta\S,
X\V, Z\V, W\T, {\cal Z}\}$.

The solution structure is as follows: only $\Sigma\S$ depends on $\theta\S$,
reflecting the fact that $\bra{\Sigma\S-\:23\theta\S}$ is determined by the
field equations, and not $\Sigma\S$ and $\theta\S$ independently. There are
only three variables which depend on $Z\V$: $\hatn\V$, $\bar\Omega\V$
and~$\Sigma\S$. All other variables depend on $\{X\V, W\T, {\cal
Z}\}\leftrightarrow\{X\V, W\T, {\hat W\T}\}$, in parallel with the odd parity
case (but the functional dependence for the solution is different). In
addition, only $\E\T$ and $\bar\H\T$ don't depend on $X\V$. Thus the solution
for all the even parity variables is completely determined when $\{\theta\S,
X\V, Z\V, W\T, {\cal Z}\}$ are know. The solution for $X\V$ comes directly from
Eq.~(\ref{Xhat}), which includes $Z\V$, while the solution for $Z\V$ comes
from~(\ref{Zhat})~-- but this equation has a $\theta\S$ term on the rhs, which
we are free to choose. So, until $\theta\S$ is given, $Z\V$ is undetermined as
is $X\V$ [cf.~(\ref{Xhat})]. But what value for $\theta\S$ is sensible? By
judicious choice we can set any one variable except $\E\T$ and $\bar\H\T$ to
zero (or to whatever value we like), because all variables but these two depend
on $X\V$, and $X\V$ can be chosen at will by choosing $\theta\S$ appropriately.
It is no accident that only $\E\T$ and $\bar\H\T$ are frame invariant: these
are the carriers of gravity waves which cannot be `gauged away' in this manner
due to their non-local nature. One cannot eliminate the effects of gravity
waves by moving appropriately! }

\forget{
\subsection{The significance of the frame freedom}
\label{solution-frame}

Although we have chosen a frame for which~$\udot^a=0$, we have not used up
all of our frame freedom, since we may still choose different timelike
congruences~$u^a$. What is the effect of making particular frame choices,
and can we achieve any great simplification by choosing an especially
auspicious frame?

From a naive argument based on counting degrees of freedom we would expect to
be able to eliminate five variables through a careful use of all frame freedom
(including that in~$\n^a$, which we have fixed): the congruence~$u^a$ can be
changed by boosting to any new frame moving with some (first-order)
three-velocity, giving three degrees of freedom; given a $u^a$, we may further
make any first-order change in~$\n^a$ that preserves $u^a\n_a=0$, giving two
degrees of freedom. Clearly, this by no means provides us with a systematic
method for simplifying the equations, and we should not expect to be able to
eliminate variables willy-nilly, but, as we see in section~\ref{thetaframe} it
can be a useful starting point.

An alternative, and more sophisticated, view of this frame freedom has already
been touched upon in section~\ref{solsoverview}, and will be expanded upon
below. The constraint and evolution equations restrict the solution variables
to lie in an 8-dimensional subspace,~${\cal N}_8$, of the full 31-D
space~${\cal V}_{31}$ at each radial position. Within this subspace all the
equations propagate consistently. If we now use the frame freedom to further
constrain the solution space, by setting $\theta\S=0$, say, or choosing some
other condition on the variables, we select some \emph{subspace} of~${\cal
N}_8$ at each radius. Since we are imposing an external constraint this need
not be consistent with the propagation equations and we can obtain different
systems of equations depending on this frame choice.

\subsubsection{The effect of changes of frame}
\label{thetaframe}

As an example of how we might use the frame freedom to simplify our equations
consider any timelike congruence $u^a$ in the true spacetime, with
expansion~$\theta$. Now consider another timelike congruence $\tilde{u}^a$
whose difference from $u^a$ is first order: that is, make the replacement
\[
     u^a \mapsto \tilde{u}^a = u^a + v e^a \quad \hbox{(to first order in
$v$),}
\]
for some unit vector $e^a$ ($u^a e_a=0$) and some first-order quantity $v$, so
that
\[
      \tilde{h}_{ab} = h_{ab} + v(u^a e^b + e^a u^b) %\quad
                                  %   \hbox{(to first order).}
\]
Then calculate
\[
      \tilde{\theta} = \tilde{h}^{ab} \del_a \tilde{u}_b
                   = \theta + e^a\del_a v + v(e^a\dot{u}_a + h^{ab}\del_a e_b).
\]
There are many possible choices for $e^a$, but if we take $e^a=n^a$, and demand
$\tilde{\theta}=0$, then we get
\[
              \hat{v} + ({\cal A}+\phi)v + \theta = 0,
\]
which is a simple first-order linear equation for the speed~$v$ given the
expansion~$\theta$ of the original congruence~$u^a$. A solution for~$v$ exists
(and could easily be written down), and for this choice of $v$ and $e^a=n^a$
the congruence $\tilde{u}^a$ has zero expansion: we have eliminated one
(troublesome) variable through a judicious choice of frame.
We have not yet exhausted the
frame freedom (we could have made a different choice for $e^a$, and still
achieved zero expansion), so it is possible to eliminate further 1+1+2
variables, but we will satisfy ourselves with this.

Before introducing the mathematically simple `crossover' frame, we will discuss
the general case, with $\theta\S$ unspecified.

\subsubsection{The general frame}\label{generalframe}

As mentioned above, we may choose our five dependent variables as $\{\theta\S,
X\V, Z\V, W\T, {\cal Z}\}$, which are the components of our solution vector.
However, $\theta\S$ only appears in the solution to $\Sigma\S$, and the
propagation equation for $Z\V$. Because there is no independent propagation
equation for $\theta\S$, let ${\mathbf{v}}\even\gen=(X\V, Z\V, W\T, {\cal Z})$,
where the `$\mathsf{g}$' stands for `general frame'; then, the 4-dimensional
dynamical system in general is
\be
\hat{\mathbf{v}}\even\gen={\mathbf{B\even\gen v\even\gen}}+i\omega r^{-1}\theta\S\bra{\begin{array}{c}
  0 \\
  1 \\
  0 \\
  0
\end{array}}\label{soleveven-gen}
\ee
where the $4\times4$ matrix $\mathbf{B}\even\gen$ is given by
\ba
{\mathbf{B}}\even\gen=&&\left(
    \begin{array}{cc}
      \:16\bra{5\phi+2\udot-4\udot^{-1}\omega^2} &       \:32\phi^2 \\
-\;{16r^4\phi^2\omega^2\bra{3r^2\phi^2+4}+2\bra{r^2\phi^2-4}\bra{3r^2\phi^2c_{15/2}-8}}{36r^4\phi^4\bra{r^2\phi^2-4}} &  -\:52\phi-3\udot          \\
          0        &      0                                                                                                \\
            0      &       0
    \end{array}\right.~~~\cdots\nonumber\\
&& \cdots~~~\left.   \begin{array}{cc}
         \;{\lll \bras{r^2\phi^2\bra{4c_3-\ll}-4\ll\bra{\ll+1}}}{r^{5}\phi c_3}          &   \;{\ll\lll\brac{r^2\phi^2\bras{48\bra{\ll+1}+r^2\bra{32\omega^2-45\phi^2}}-16\bra{\ll\lll-1}}}{i\omega r^6\phi^2}          \\
                \;{2\lll\bra{3r^2\phi^2+4}\bra{\ll-c_3}}{3r^5\phi^2c_3}         &    -\;{\ll\lll\bras{\ll r^2\phi^2\bra{r^2\phi^2+4}+\bra{4-3r^2\phi^2}\bra{\:13 r^4\phi^2\omega^2+\:{7}{16}r^4\phi^4+r^2\phi^2-\:13}}}{i\omega r^8\phi^5}         \\
              -\;{r^2\phi^2\bras{c_3^2-8\lll \bra{c_3+\ll+4}}+32\ll\lll\bra{\ll+1}}{24r^4\phi^2\udot c_3}     &   -\;{9\omega^2r^8\phi^4\udot^2+\ll^2\lll^2}{3i\omega r^5\phi^3\udot}          \\
             \;{i\omega}{3r^3\phi\udot}      &    \;{r^2\phi^2\bras{c_3^2-8\lll \bra{c_3+\ll+4}}+32\ll\lll\bra{\ll+1}}{24r^4\phi^2\udot c_3}         \
    \end{array}\right).\label{Bgeneral}
\ea
This is the differential equation underlying the even parity perturbations, as
every variable is proportional to $\mathbf{v}\even\gen$, in this gauge. Note,
however, that the DE's that ${\cal Z}$ and $W\T$ satisfy do not depend in any
way on our choice of $\theta\S$, reflected in Eq.~(\ref{soleveven-gen}), and in
the lower left $2\times 2$ matrix in Eq.~(\ref{Bgeneral}) being zero. In
addition, the variables $\E\T$ and $\bar\H\T$ only depend on $W\T$ and ${\cal
Z}$. Hence these variables are \emph{frame-invariant}.

We show some solutions of the differential equation~(\ref{soleveven-gen}), when
$\theta=0$ (non-expanding frame) in
Figs.~\ref{Z-W-hatW-l=20-sigma=6.6-r=2.3..9.ps}~--
\ref{Z-r-l=20-sigma=6.6-r=2.3..5-l=2..10.ps}.

We will not give the solution matrix $\mathbf{M}\even\gen$ in this frame.

\begin{figure}[ht]
\includegraphics[width=11cm]{{Z-W-hatW-l=20-sigma=6.6-r=2.3..9_LAB.ps}}
\caption{\small Solution curve of $Z\V$ and $W\T$ against $r$
for $\ell=20$ and $\sigma=6.6$ and $m=1$, in the $\theta=0$ frame. Boundary
conditions are (arbitrarily) set at
${\mathbf{v}}\gen(r=2.3)=10^{-9}(10,10,1,(i\sigma)^{-1}10)$. This shows
oscillations of $Z\V$ decaying very fast moving away from $r\sim2.3$ (where the
plot starts in the upper left hand corner); in contrast, the amplitude of $W\T$
does not decay. This is reflected in the curve at large $r\lesssim9$ being
oscillations in the $Z\V=0$ plane.
\label{Z-W-hatW-l=20-sigma=6.6-r=2.3..9.ps}}
\end{figure}

\begin{figure}[ht]
\includegraphics[width=11cm]{{Z-X-hatW-l=20-sigma=6.6-r=2.3..9_LAB.ps}}
\caption{\small Solution curve of the previous plot,
Fig.~\ref{Z-W-hatW-l=20-sigma=6.6-r=2.3..9.ps}, but now showing $X\V$ instead
of $W\T$. $X\V$ decays very quickly too.
\label{Z-X-hatW-l=20-sigma=6.6-r=2.3..9.ps}}
\end{figure}

\begin{figure}[ht]
\includegraphics[width=11cm]{{X-Z-r-even-l=2-QNM-sigma=sR=.37367,sI=.08896-r=2..8_LAB.ps}}
\caption{\small $X\V$ and $Z\V$ vs. $r$ for $\ell=2$ and the fundamental
quasinormal frequency $\sigma=0.37367+0.08896i$. Boundary conditions are chosen
${\mathbf{v}\even\gen}(r=2.003)=10^{-8}(10^{-2},1,1,1)+i10^{-9}(1,1,1,1)$, and
the plot starts at $r=2.03$ ($Z\V$ has already undergone one oscillation). Both
$X\V$ and $Z\V$ tend to zero for this frequency.
\label{X-Z-r-even-l=2-QNM-sigma=sR=.37367,sI=.08896-r=2..8.ps}}
\end{figure}

\begin{figure}[ht]
\includegraphics[width=11cm]{{Z-r-l=20-sigma=6.6-r=2.3..5-l=2..10_LAB.ps}}
\caption{\small Solution curve of the previous plots,
Fig.~\ref{Z-W-hatW-l=20-sigma=6.6-r=2.3..9.ps}, but now showing $Z\V$ against
$r$ for different values of $\ell=2,\cdots,10$. For $\ell=2$, at the top of the
plot, there are no oscillations, and $Z\V$ just tends to zero; $\ell$ increases
as we move from curve to curve down the plot until $\ell=10$, where there are
large oscillations, but still decaying to zero as $r$ increases.
\label{Z-r-l=20-sigma=6.6-r=2.3..5-l=2..10.ps}}
\end{figure}

}

%Note, however, that the DE's that ${\cal Z}$ and $W\T$ satisfy do not depend in
%any way on our choice of $\theta\S$, reflected in Eq.~(\ref{soleveven-gen}),
%and in the lower left $2\times 2$ matrix in Eq.~(\ref{Bgeneral}) being zero. In
%addition, the variables $\E\T$ and $\bar\H\T$ only depend on $W\T$ and ${\cal
%Z}$. Hence these variables are \emph{frame-invariant}.

\forget{
\subsubsection{Crossover frame}

We will choose here a `crossover' frame, in which the solutions become similar
in form to the odd parity case, and the differential equation for the full
solution is of the form~(\ref{solform3}). To do this, demand that $X\V$, which
satisfies~(\ref{Xhat}), has a propagation equation of the same form as in the
odd parity case, Eq.~(\ref{barXVde}):
\be
\hat{X}\V=-\:23\bras{r^{-2}\phi^{-1}+\omega^2\udot^{-1}}X\V
-4\lll r^{-3}\phi W\T;\label{XVde-crossover}
\ee
this fixes $Z\V$ and hence $\theta\S$. Indeed, all variables can be written now
in terms of $\{X\V, W\T, {\cal Z}\}\leftrightarrow\{X\V, W\T, {\hat W\T}\}$; we
will use the former, involving ${\cal Z}$ to try to keep the equations simple,
although it makes comparison with the odd case a little more difficult~-- we
can use Eq.~(\ref{ZofW}) to convert.
\be
\begin{array}{c|c|c|c}
    \mbox{\sc even}&       X\V     &   2r^{-2}\phi^{-1}W\T     &  \:12c_3r^{-1}\phi^{-1}  {\cal Z}\\\hline
    \E\T         &     0     &   \;{16\bra{\ll+1}-3r^2\phi^2c_1}{4r^2\phi c_3}     & \;{4\ll\lll-3r^2\phi^2\bras{r^2\omega^2\bra{r^2\phi^2-4}-\ll\lll}}{i\omega r^4\phi^2 c_3}       \vphantom{\;{X\;{X\;{x}{X}}{X}}{X\;{X}{X}}}  \\
    \bar\H\T     &    0     &    i\omega    &   \:14\ll\bra{\phi+\udot}-3\udot    \vphantom{\;{X\;{X\;{x}{X}}{X}}{X\;{X}{X}}}   \\
    \Sigma\T     &    -\:23\bra{i\omega r\phi^2}^{-1}     &    2\lll\phi\bra{i\omega c_3}^{-1}    &   \;{4\ll\lll-3r^2\phi^2\bras{r^2\omega^2\bra{r^2\phi^2-4}+\ll\lll}}{\omega^2r^4\phi^2}     \vphantom{\;{X\;{X\;{x}{X}}{X}}{X\;{X}{X}}}  \\
    \zeta\T      &    \;{8r}{3\phi\bra{r^2\phi^2-4}}     &   1     &   0     \vphantom{\;{X\;{X\;{x}{X}}{X}}{X\;{X}{X}}}  \\
    \E\V         &    -\phi^{-1}     &    -\:12\lll r^{-1}    &   -\:12\ll\lll\bra{i\omega r^3\phi}^{-1}     \vphantom{\;{X\;{X\;{x}{X}}{X}}{X\;{X}{X}}}  \\
    \bar\H\V     &     \;{r^2\phi^2-4}{4i\omega r^2\phi^2}    &    -\;{3\lll\phi\bra{r^2\phi^2-4}}{4i\omega r c_3}    &   \;{3\lll\bra{r^2\phi^2-4}\bras{4r^4\phi^2\omega^2+\ll\bra{3r^2\phi^2-4}}}{8\omega^2 r^5\phi^2c_3}     \vphantom{\;{X\;{X\;{x}{X}}{X}}{X\;{X}{X}}}  \\
    \Sigma\V     & \;{16\bra{r^2\phi^2-1}-r^4\phi^2\bra{3\phi^2+16\omega^2}}{18i\omega r^2\phi^3\bra{r^2\phi^2-4}}        &   \;{\lll\bra{3r^2\phi^2-4}}{6i\omega rc_3}     &   -\;{\ll\lll\bras{r^4\phi^2\bra{9\phi^2+16\omega^2}-8\bra{3r^2\phi^2-2}}}{12\omega^2r^5\phi^2c_3}     \vphantom{\;{X\;{X\;{x}{X}}{X}}{X\;{X}{X}}}  \\
    \bar\Omega\V & \;{2\bras{r^4\phi^2\bra{3\phi^2+4\omega^2}-16\bra{r^2\phi^2-1}}}{9i\omega r^2\phi^3\bra{r^2\phi^2-4}}    &   \;{\lll\bras{r^2\phi^2\bra{25\ll+24-18r^2\phi^2}}+4\ll\bra{\ll+1}}{3i\omega r^3\phi^2 c_3}        & -\;{\ll\lll\bras{r^2\phi^2\bra{12\ll-r^2\bra{9\phi^2+8\omega^2}+8\bra{\ll\lll-2}}}}{6\omega^2\phi^3 r^5 c_3}       \vphantom{\;{X\;{X\;{x}{X}}{X}}{X\;{X}{X}}}  \\
    \hatn\V      &    \;{2r^4\phi^2\bra{33\phi^2+32\omega^2}-16\bra{19r^2\phi^2-10}}{9\phi^2\bra{r^2\phi^2-4}^2}     &    \;{4\lll\bra{\ll-4r^2\phi^2}}{3r\phi\bra{r^2\phi^2-4}}    &    \;{\ll\lll\bra{c_{-3}-16}}{3i\omega r^3\phi^2\bra{r^2\phi^2-4}}    \vphantom{\;{X\;{X\;{x}{X}}{X}}{X\;{X}{X}}}  \\
    \dotn\V      &    \;{16\bra{r^2\phi^2-1}-r^4\phi^2\bra{3\phi^2+16\omega^2}}{6i\omega r^2\phi^3\bra{r^2\phi^2-4}}     &    \;{\lll\bra{3r^2\phi^2-4}}{2i\omega r c_3}    &  \;{\ll\lll\bras{8\bra{3r^2\phi^2-2}-r^4\phi^2\bra{9\phi^2+16\omega^2}}}{4\omega^2r^5\phi^3c_3}      \vphantom{\;{X\;{X\;{x}{X}}{X}}{X\;{X}{X}}}  \\
    X\V          &    1      &    0   &     0   \vphantom{\;{X\;{X\;{x}{X}}{X}}{X\;{X}{X}}}  \\
    Y\V          &    -\:23c_3\phi^{-1}\bra{r^2\phi^2-4}^{-1}     &    0    &   {\ll\lll}\bra{i\omega r^3\phi}^{-1}     \vphantom{\;{X\;{X\;{x}{X}}{X}}{X\;{X}{X}}}  \\
    Z\V          & -\:23\phi^{-2}\bra{\phi+\udot} &  \;{\lll\bras{4\ll\bra{\ll+1}+r^2\phi^2\bra{24r^2\phi^2+31\ll+32}}}{3 r\phi c_3}  & \;{\ll\lll \bras{16\bra{\ll\lll-1}+r^2\phi^2\bra{r^2\bra{32\omega^2+45\phi^2}-48\bra{\ll+1}}}}{{12} i\omega r^{5}\phi^{3} c_3}\vphantom{\;{X\;{X\;{x}{X}}{X}}{X\;{X}{X}}}
%\\    \theta\S     &         &         &        &        \\
%    \Sigma\S     &         &         &        &
\end{array}\label{allevensols}
\ee
Note that
\be
r^2\phi^2-4=-4\phi\udot r^2.
\ee
The solutions for $\Sigma\S$ and $\theta\S$ are not explicitly given as they
are huge, but may be obtained relatively easily from the above equations.

In this crossover frame we have our basis vector
\be
\mathbf{v}\even\cross=\bra{\begin{array}{c}
  X\V \\
  W\T \\
  \hat W\T
\end{array}}
\ee
with solution matrix $\mathbf{M}\even\cross$ given by~(\ref{allevensols}) (plus
the equations for $\theta\S$ and $\Sigma\S$, which were too large to give), and
our 3-dimensional differential equation
\be
\hat{\mathbf{v}}\even\cross=\mathbf{B\odd{v}\even\cross}\label{fullsol1-evencross}
\ee
where $\mathbf{B}\odd$ is given by Eq.~(\ref{Bodd}). This is the frame in which
the full solution takes the form of Eqs.~(\ref{solform2}) and~(\ref{solform3}),
with $\mathbf{A}=\mathbf{B}\odd$. }

\forget{
In this crossover frame, we have chosen $Z\V$ (and hence $\theta\S$, etc.) such
that Eqs.~(\ref{barXVde}) and~(\ref{XVde-crossover}) are satisfied: that is,
$\bar X\V$ and $X\V$ satisfy the same ordinary differential equation
\emph{after decomposition into SH.} However, we cannot write down an equation
for $\hat X^a$ \emph{in tensor form} because of the sign of the $W$-term in
Eqs.~(\ref{barXVde}) and~(\ref{XVde-crossover}); these are of opposite sign if
we form the equations from a single tensor equation; this arises because this
term is from a $\delta^a W_{ab}$ term, which gives a sign change between
parities when decomposed into SH, by Eq.~(\ref{SH5}). To be clear, then, if we
demand that $X^a$ satisfy
\be
\ddot X_a-\:32\udot\hat
X_a=\bra{\udot^2-\:14\E}X_a-12\E\bra{\:14\phi^2-\E}\delta^b W_{ab},
\ee
so that the odd parity part of this equation is given by Eq.~(\ref{barXVde}),
and the even parity part by
\be
\hat{X}\V=-\:23\bras{r^{-2}\phi^{-1}+\omega^2\udot^{-1}}X\V
+4\lll r^{-3}\phi W\T;\label{XVde-crossover2}
\ee
with a sign change on the $W$ term from Eq.~(\ref{XVde-crossover}), then we
must make the following changes to the above equations. Let
\be
{\cal
F}[\Psi]\equiv\bra{\Psi\mathrm{~in~crossover~frame~given~by~Eq.~(\ref{allevensols})}}-\bra{\Psi\mathrm{~
in~this~frame}}
\ee
for any $\Psi$. Then the following changes must be made:
\ba
{\cal F}\bras{Z\V}&=&\;{16}{3}\lll r^{-3}\phi^{-1} W\T\\
&=&i\omega{\cal F}\bras{\bar\Omega\V}\\
&=&\udot{\cal F}\bras{\hatn\V}
\ea
with changes to be made to $\theta\S$ and $\Sigma\S$. }

}

\forget{
\subsection{Solution structure}

In the odd parity case, our solution is relatively simple because everything is
fixed by our choice of frame $\udot^a=0$, but in the even case this is not the
case, as we still had our freedom in $u^a$ to fix (which we didn't simply fix
at the beginning, as we did with $\n^a$, to illustrate the freedom). This may
be thought of as a \emph{first-order} rotation of the 4-vector $u^a$, or simply
as our specific choice of observers in the model. (Note that a \emph{zeroth
order} rotation of $u^a$ will lose gauge invariance.) We have given details of
the solution in our crossover frame, which is chosen \emph{specifically} to
simplify the solution form of the equations, in particular the DE for $X\V$.
However, as a \emph{physical} frame it's a bit unlikely, as may be seen from an
equation like $\theta\S\propto\mathbf{v}\even\cross$, which gives quite a
complicated wavy expansion;  this would be an unusual choice for real observers
to make, since at least part of this motion must come from a force applied by
them (due to the nature of the non-geodesic observers in the background).
Therefore, we also gave the form of the solution in general, which can be
easily specialised to the case of non-expanding observers (and hence easy to
set up physically), which is the case for all the plots
of~Eq.(\ref{soleveven-gen}). Both cases are gauge invariant, because a specific
frame choice has been made. }

\subsection{Gravity waves}
\label{grav-wave}

The plane wave solutions are given by setting $\udot=\phi=0$, with $\n^a$
lying in the direction of propagation (although, if we keep $\phi=2/r$~--
i.e., set~$m=0$~-- then it represents the same solution, but $\n^a$ is a
radial direction with an arbitrary centre). In this case only the four TT
tensors are non-zero, and $\E_{ab}$ and $\H_{ab}$ represent the curvature of
plane gravity waves: dynamical tidal forces orthogonal to the direction of
propagation. In the BH case they therefore may be thought of as representing
the same part of the GW, although with distortion from the BH
itself~\cite{PT}.
\forget{We have made quite a fuss writing all variables in terms of the usual
Regge-Wheeler and Zerilli variables, because these satisfy the standard
equations of black hole perturbation theory, and are the only variables (which
we have found) which satisfy wave equations that are decoupled from all other
variables. However, it is the transverse traceless tensors $\E_{ab}$ and
$\H_{ab}$ which carry the gravity waves traveling along $\n^a$~\cite{PT}.} This
is seen in the  wave equations which these tensors satisfy which no longer
close, and have forcing terms from other quantities (in a general frame):
\ba
\ddot\E_{\lb ab\rb}-\hat{\hat \E}_{\lb ab\rb}-\delta^2\E_{ab}
-\bra{\phi+5\udot}\hat\E_{\lb ab\rb} -\bra{\phi^2+8\udot^2 -5\E
}\E_{ab}&=&2\bra{\phi-2\udot}\delta_{\lb a}\E_{b\rb}
+3\E\bra{\phi-2\udot}\zeta_{ab},\label{Ewave}\\
%\ddot\H_{\lb ab\rb}-\hat{\hat \H}_{\lb ab\rb}-\bra{\phi+5\udot}\hat\H_{\lb ab\rb}
%+2\udot\bra{\phi-2\udot}\H_{ab}&=& \hbox{(source terms for $\H_{ab}$)}.
\ddot\H_{\lb ab\rb}-\hat{\hat \H}_{\lb ab\rb}-\delta^2\H_{ab}-{\udot}\hat\H_{\lb ab\rb}
-\bra{\:14\phi^2-2\udot^2+\:72\E}\H_{ab}&=&\bra{\:72\phi-\udot}\delta_{\lb
a}\H_{b\rb}
 -3\phi\E\lc_{c\lb
a}\Sigma_{b\rb}^{~~c}.
%\nonumber\\&-&\:32\E\lc_{c\lb
%a}\delta^c\Sigma_{b\rb}+\:12\E\lc_{c\lb a}\delta^c\dotn_{b\rb}.
\label{Hwave}
\ea
Thus, the principle forcing term for $\E_{ab}$ comes from the \emph{shear of
$n^a$}, while $\H_{ab}$ is principally forced by the \emph{shear of $u^a$}.
\forget{Recall that  our definition of $W_{ab}$, Eq.~(\ref{Wab}), involves the shear of
$\n^a$, together with an electric Weyl contribution, while our Zerilli
variable, Eq.~(\ref{zer1}), is made up of a magnetic Weyl part, and a
contribution from the shear of $u^a$.}

These TT tensors decouple from all the other variables, but not as wave
equations, so it is worthwhile giving the differential equations which these
satisfy. Let $\mathbf{\cal W}=(\mathbf{\cal W}\odd,\mathbf{\cal W}\even)$ where
\be
\mathbf{\cal W}\odd=\bra{
\begin{array}{c}
  \bar\E\T \\
  \H\T
\end{array}},
~~~\mbox{and}~~~
\mathbf{\cal W}\even=\bra{
\begin{array}{c}
  \E\T \\
  \bar\H\T
\end{array}}.
\ee
Then $\mathbf{\cal W}$ satisfies the 4-dimensional first order DE
\be
\hat{\mathbf{\cal {W}}}=\chi^{-1}\bra{\begin{array}{cc}
  \mathbf{\Upsilon}\odd & 0 \\
  0 & \mathbf{\Upsilon}\even
\end{array}}\mathbf{{\cal W}}\label{GWeqn}
\ee
where
\ba
\chi&=&r^2\phi\brac{r^2\phi^2\bras{8\ll-r^2\bra{\phi^2+256\omega^2}}-16},\\
\mathbf{\Upsilon}\odd&=&\bra{
    \begin{array}{cc}
      -B_2-3r^4\phi^4\bra{r^2\phi^2+4} & B_1 \\
      -B_1-4\ll\lll\bra{i\omega}^{-1}r^2\phi^3 & B_2 \
    \end{array}},\\
\mathbf{\Upsilon}\even&=&\bra{
    \begin{array}{cc}
      -B_2-3r^4\phi^4\bra{r^2\phi^2+4}+480r^4\phi^2\omega^2 & -4\bra{B_1-240i\omega^3r^6\phi^3} \\
      -B_1 & -B_2 \
    \end{array}},~~~\mbox{and}\\
B_1&=&i\omega
r^2\phi\bras{-r^4\phi^2\bra{9\phi^2+16\omega^2}+8r^2\phi^2\bra{2\ll+1}-16},\\
B_2&=& -\:32r^6\phi^6
+6r^4\phi^2\bras{\phi^2\bra{\ll-1}+64\omega^2}+8r^2\phi^2\bra{\ll-1}+32.
\ea
Pretty untidy, but the structure is simple. Note that although decoupling
these equations will result in second order ODE's for each of the four
variables, these are not true wave equations, because of the $\omega$'s in
$\mathbf{\Upsilon}$; powers of $\omega$ up to 3 are found in
Eq.~(\ref{GWeqn}), which correspond to third order dot-derivatives.

%We show a
%couple of representative plots, in Figs. for the odd system. The even case
%looks virtually identical (except the spirals go in the opposite direction for
%the conditions given).

In fact, although the Regge-Wheeler tensor, as we have written it, involves
$\zeta_{ab}$ which is a kinematical term, it may be written from \emph{purely
Weyl} contributions, using e.g., Eqs.~(\ref{Sigmaabdot}) and~(\ref{Eabdot}),
or just Eq.~(\ref{Habdot}). Therefore, the entire solution may be related
relatively simply to the Weyl curvature. Indeed, the variables
$\E_{ab},\H_{ab},W_{ab}$ are all frame invariant so the relations between
these given for the odd and even cases apply regardless of frame; hence,
$W_{ab}$ and $\hat W_{ab}$ may be given as linear combinations of~$\E_{ab}$
and~$\H_{ab}$, though not in tensorial form.

\subsection{The Zerilli tensor}
\label{zertens}

\forget{
We have discussed at the end of the section on the Zerilli variable,
Sec.~\ref{zersec}, how it is possible to find an odd parity Zerilli variable
as a linear combination of $\bar\E\T$, $\bar\zeta\T$ and $\xi\S$, which was
seemingly unrelated to the even parity form of the Zerilli variable. However,
we may utilise our solution for the odd parity variables
[Eq.~(\ref{solform1}), although the details of this `general frame' solution
are not given here] in order to write $\bar{\cal Z}$ as a linear combination
of $\bar\Sigma\T$, $\bar Y\V$ and $\H\S$, thus suggesting a connection with
the even parity form, Eq.~(\ref{zer1}):
\ba
\bras{\:19\ll^2\lll^2\omega^{-2}r^{-8}\phi^{-4}\udot^{-2}+1}\bar{\cal Z}&=&
\bra{16\E\udot r^4}^{-1}\Sigma\T+i\omega\bras{r\udot
\bra{3r^2\phi^2-4}^2}^{-1}Y\V\nonumber\\
&&-\;{3\phi^2r^2\bra{\phi^2r^2-4\ll}+16\bra{\ll+1}}
{8\phi^3\udot^3r^6c_3\bra{3\phi^2r^2-4}}\H\S .
\ea
As yet, the similarity with the even case is not quite clear, although nearly
there. If we note that the factor multiplying $\bar{\cal Z}$ on the left hand
side is actually a constant (in fact, it is just the special quasi-normal
mode factor, discussed earlier), then we can \emph{re-define} our definition
of $\bar{\cal Z}$, without affecting the differential equation it obeys. In
addition, the factor multiplying $\bar Y\V$ doesn't quite correspond to that
of the even analogue; therefore we use the fact that $\bar Y\V\propto\xi\S$
to adjust for this, to allow this desired correspondence.
%We may then
%construct the Zerilli tensor as follows: define the operator
%\be
%\Delta\Psi_{ab}=2\delta_{\lb a}\delta^c\Psi_{b\rb c}=\delta^2\Psi_{ab}+\phi\bra{2\udot-\:12\phi}\Psi_{ab}=-\lll r^{-2}\Psi_{ab}
%\ee
%where the last equality has an implicit sum over $\ell$ as usual.
Therefore, we define the gauge- and frame-invariant Zerilli tensor ${\cal
Z}_{ab}$ by the \emph{differential} equation:
\ba
16\bra{3\phi^2r^2-4}\E^2r^2\bras{\bra{4\phi^2r^2-9}-r^2\delta^2}{\cal
Z}_{ab}&=&
-\phi\bra{3\phi^2r^2-4}\bras{\bra{4\phi^2r^2-9}-r^2\delta^2}\Sigma_{ab}\nonumber\\
&+&16\phi^2r^4\delta_{\lb a}\dot Y_{b\rb}-32\phi^2r^4\lc_{c\lb
a}\delta^c\delta_{b\rb}\dot\xi\nonumber\\
&+&\;{2}{3\E}\bras{\bra{3\phi^4r^4+1}\lc_{c\lb
a}\delta^c\delta_{b\rb}\H+4r^2\bra{3\phi^2r^2-4}\lc_{c\lb
a}\delta^c\delta_{b\rb}\delta^2\H}
 ,\label{zerillitensor}
\ea
which is related to our odd and even Zerilli variables by
\ba
\bar{\cal Z}\T&=&\bras{\:19\ll^2\lll^2\omega^{-2}r^{-8}\phi^{-4}\udot^{-2}+1}\bar{\cal Z},\\
{\cal Z}\T&=&{\cal Z},
\ea
and, of course, both $\bar{\cal Z}\T$ and ${\cal Z}\T$ satisfy the Zerilli
equation~(\ref{zerilli}). The dot-derivatives in Eq~(\ref{zerillitensor}) can
of course be replaced by the appropriate evolution equations, leading to an
even more depraved equation for ${\cal Z}_{ab}$.

%Our new definition does suffer from the drawback that it is not defined at the
%special QNM frequency, because the new factor involved vanishes for this value
%of $\sigma$. The origin of this factor arises from Eq.~(\ref{ZofW}), and is not
%put in by hand; it does appear to be entwined in the Zerilli tensor.

%Indeed, the entire contribution to black hole perturbation theory of the
%Zerilli tensor may just come down to this special QNM frequency; it does not
%seem to have much other purpose.

The complexity of the definition of this Zerilli tensor~(\ref{zerillitensor})
is somewhat startling, especially given the simplicity of the analogous
Regge-Wheeler tensor. Indeed, the necessity of defining ${\cal Z}_{ab}$ via a
differential equation really does not inspire confidence in its possible
fundamental nature; were it not for the harmonic splitting we are permitted
to do due to the nature of the background, we would not be able to actually
write down a closed wave equation for ${\cal Z}_{ab}$~-- the closed tensorial
equation it obeys is
\ba
&&\bra{4\phi^2r^2-9-r^2\delta^2}\bras{\ddot{\cal Z}_{ab}-\hat{\hat{\cal
Z}}_{ab}
    -\udot\hat{\cal Z}_{ab}}\nonumber\\&&~~+\:13r^{-2}\brac{
    \:14 \left( 4{\phi}^{2}{r}^{2}-8-4{r}^{2}{\delta}^{2} \right) ^{3
}+32 \left( \:74{\phi}^{2}{r}^{2}-2-{r}^{2}{\delta}^{2} \right)
 \left( \:74{\phi}^{2}{r}^{2}-5-{r}^{2}{\delta}^{2} \right) ^{2}
}{\cal Z}_{ab}=0.
\ea
%If we try to convert
%Eq.~(\ref{zerilli}) into tensorial form, we run into trouble because of the
%$c_3^{-2}$ factor which appears~-- multiplying the equation through by this
%factor would give $\delta^4$ terms operating on each term; hence, it would
%not be a true tensorial wave equation.
Hence, only the spherical harmonic amplitudes of ${\cal Z}_{ab}$, $\bar{\cal
Z}\T$ and ${\cal Z}\T$, obey a true wave equation.}

We have discussed at the end of the section on the Zerilli variable,
Sec.~\ref{zersec}, how it is possible to find an odd parity Zerilli variable
as a linear combination of $\bar\E\T$, $\bar\zeta\T$ and $\xi\S$, which was
seemingly unrelated to the even parity form of the Zerilli variable. However,
we may utilise our solution for the odd parity variables
[Eq.~(\ref{solform1}), although the details of this `general frame' solution
are not given here] in order to write $\bar{\cal Z}$ as a linear combination
of $\bar\Sigma\T$, $\H\V$ and $\H\S$, thus suggesting a connection with the
even parity form, Eq.~(\ref{zer1}):
\be
\bras{\:19\ll^2\lll^2\omega^{-2}r^{-8}\phi^{-4}\udot^{-2}+1}\bar{\cal Z}=
\:23
c_3^{-1}\bras{3r\phi\bar\Sigma\T-2\udot^{-1}\H\V}-\:19r^{-3}\phi^{-2}\udot^{-2}\H\S.
\ee
As yet, the similarity with the even case is not quite clear, although nearly
there. If we note that the factor multiplying $\bar{\cal Z}$ on the left hand
side is actually a constant (in fact, it is just the special quasi-normal mode
factor, discussed earlier), then we can \emph{re-define} our definition of
$\bar{\cal Z}$, without affecting the differential equation it obeys.
%We may then
%construct the Zerilli tensor as follows: define the operator
%\be
%\Delta\Psi_{ab}=2\delta_{\lb a}\delta^c\Psi_{b\rb c}=\delta^2\Psi_{ab}+\phi\bra{2\udot-\:12\phi}\Psi_{ab}=-\lll r^{-2}\Psi_{ab}
%\ee
%where the last equality has an implicit sum over $\ell$ as usual.
Therefore, we define the Zerilli tensor ${\cal Z}_{ab}$ by the
\emph{differential} equation:
\be
\bras{2-r^2\bra{2\E+\delta^2}}{\cal
Z}_{ab}=\:12r\phi\Sigma_{ab}-\:13r\udot^{-1}\lc_{c\lb a}\delta^c\H_{b\rb}
-\:19r^{-1}\E^{-2}\bras{2-r^2\bra{2\E+\delta^2}}\lc_{c\lb
a}\delta^c\delta_{b\rb}\H,\label{zerillitensor}
\ee
\iffalse
\ba
&&r\bras{4+r^2\bra{\phi^2-12\E-4\delta^2}}\bra{\phi^2-3\E
-\delta^2}^2\bra{\E-\:12\phi^2+\delta^2}{\cal Z}_{ab}
-9r^3\phi^4\udot^2\bra{\phi^2-12\E-4\delta^2}\ddot{\cal
Z}_{\lb ab\rb}=\nonumber\\
&&~~~~~~-18r^2\phi^5\udot^2\ddot\Sigma_{\lb ab\rb} +12r^2\phi^4\udot\lc_{c\lb
a}\delta^c\ddot\H_{b\rb} +\phi^2\bras{4+r^2\bra{\phi^2-12\E-\delta^2}}\lc_{c\lb
a}\delta^c\delta_{b\rb}\ddot\H,\label{zerillitensor}
\ea
\fi
which is related to our odd and even Zerilli variables by
\ba
\bar{\cal Z}\T&=&\bras{\:19\ll^2\lll^2\omega^{-2}r^{-8}\phi^{-4}\udot^{-2}+1}\bar{\cal Z},\\
{\cal Z}\T&=&{\cal Z},
\ea
and, of course, both $\bar{\cal Z}\T$ and ${\cal Z}\T$ satisfy the Zerilli
equation~(\ref{zerilli}).

%Our new definition does suffer from the drawback that it is not defined at the
%special QNM frequency, because the new factor involved vanishes for this value
%of $\sigma$. The origin of this factor arises from Eq.~(\ref{ZofW}), and is not
%put in by hand; it does appear to be entwined in the Zerilli tensor.

%Indeed, the entire contribution to black hole perturbation theory of the
%Zerilli tensor may just come down to this special QNM frequency; it does not
%seem to have much other purpose.

The complexity of the definition of this Zerilli tensor~(\ref{zerillitensor})
is somewhat startling, especially given the simplicity of the analogous
Regge-Wheeler tensor. Indeed, the necessity of defining ${\cal Z}_{ab}$ via a
differential equation really does not inspire confidence in its possible
fundamental nature; were it not for the harmonic splitting we are permitted
to do due to the nature of the background, we would not be able to actually
write down a closed wave equation for ${\cal Z}_{ab}$~-- the closed tensorial
equation it obeys is
\ba
&&\bra{4\phi^2r^2-9-r^2\delta^2}\bras{\ddot{\cal Z}_{ab}-\hat{\hat{\cal
Z}}_{ab}
    -\udot\hat{\cal Z}_{ab}}\nonumber\\&&~~+\:{16}{3}r^{-2}\brac{
     \left( {\phi}^{2}{r}^{2}-2-{r}^{2}{\delta}^{2} \right) ^{3
}+2 \left( \:74{\phi}^{2}{r}^{2}-2-{r}^{2}{\delta}^{2} \right)
 \left( \:74{\phi}^{2}{r}^{2}-5-{r}^{2}{\delta}^{2} \right) ^{2}
}{\cal Z}_{ab}=0.
\ea
%If we try to convert
%Eq.~(\ref{zerilli}) into tensorial form, we run into trouble because of the
%$c_3^{-2}$ factor which appears~-- multiplying the equation through by this
%factor would give $\delta^4$ terms operating on each term; hence, it would
%not be a true tensorial wave equation.
Hence, only the spherical harmonic amplitudes of ${\cal Z}_{ab}$, $\bar{\cal
Z}\T$ and ${\cal Z}\T$, obey a true wave equation.

\forget{
The complexity of the definition of this Zerilli tensor~(\ref{zerillitensor})
is somewhat startling, especially given the simplicity of the analogous
Regge-Wheeler tensor. Indeed, the necessity of defining ${\cal Z}_{ab}$ via a
differential equation really does not inspire confidence in its possible
fundamental nature; were it not for the harmonic splitting we are permitted
to do due to the nature of the background, we would not be able to actually
write down a closed wave equation for ${\cal Z}_{ab}$. If we try to convert
Eq.~(\ref{zerilli}) into tensorial form, we run into trouble because of the
$c_3^{-2}$ factor which appears~-- multiplying the equation through by this
factor would give $\delta^4$ terms operating on each term; hence, it would
not be a true tensorial wave equation. Only the spherical harmonic amplitudes
of ${\cal Z}_{ab}$, $\bar{\cal Z}\T$ and ${\cal Z}\T$, obey a true wave
equation.}

\forget{
\subsection{Near and far: where is $X^a$ important?}\label{ass}

We have found that the vector $X^a$ is important in describing GW emanating
from a black hole. Although it is determined by the Regge-Wheeler tensor
$W_{ab}$ up to a constant by Eq.~(\ref{barXVde})~-- strictly speaking, the
harmonic moments of $X^a$ are determined up to a constant by the harmonic
moments of $W_{ab}$~-- it is a driving variable in its own right for a full
solution to GW's from a perturbed black hole, because of the nature of the
solution~(\ref{fullsol1-odd}). It is important, therefore, to try to understand
the behaviour of $X^a$, as well as its implications for the other variables; in
particular, the asymptotic regimes (as $r\rightarrow 2m$ and
$r\rightarrow\infty$) are interesting. Solutions of the Regge-Wheeler equations
are well understood, so we will not investigate these further here.

[From now on we will be rather sloppy in distinguishing $X\V$ and $\bar X\V$,
and $W\T$ and $\bar W\T$, as the odd and even parts obey the same
equations~(\ref{fullsol1-odd}) in the crossover frame; we will often use just
$X$ and $W$ to mean both parity cases. We can't do this for the other
variables, of course, nor in the general frame.]

\subsubsection{Far field: what on Earth does it all mean?}

First off, we'll have a look at the region far from the black hole, as
$r\rightarrow\infty$. The leading behaviour of the background variables as
$r\rightarrow\infty$ is
\ba
\phi &\rightarrow & \;{2}{r}+{\cal O}\bra{\;{1}{r^2}},\nonumber\\
\udot &\rightarrow & \;{m}{r^2}+{\cal O}\bra{\;{1}{r^3}},\nonumber\\
\omega&\rightarrow&\sigma\brac{1+{\cal O}\bra{\;{1}{r}}},\nonumber\\
\;{d}{d\rho}&\rightarrow& \brac{1+{\cal O}\bra{\;{1}{r}}}\;{d}{dr}.\label{farfield}
\ea
We have used the symbol `$\rightarrow$' to denote the leading behaviour. Recall
that
\be
\E =-\;{2m}{r^3}~~~~(\mbox{always}).
\ee

The far field behaviour of the Regge-Wheeler tensor and the Zerilli variable is
\be
W,~{\cal Z}\rightarrow{e^{\pm i\sigma r}}~~~ \mbox{as}~~~
r\rightarrow\infty;\label{Wfarfield}
\ee
where the $\pm$ refers to incoming and outgoing radiation at infinity,
respectively. However, solutions to the Regge-Wheeler equation~(\ref{RW}) and
Zerilli equation~(\ref{zerilli}) with this boundary condition (with a
corresponding one at the horizon~-- see below) imply that $\Im(\sigma)>0$,
which implies a wave of infinite amplitude being transmitted to infinity. This
is the rather complicated issue of quasi-normal modes; the essential point,
however, is that there is an $e^{i\sigma t}$ time dependence, so these waves of
infinite amplitude at $r=\infty$ (and $\tau=-\infty$) are exponentially damped
in time, and thus represent energy radiated from the black hole from an initial
perturbation. We refer to~\cite{nollert} for further details. What this means
for us is that we can't give an absolute `leading behaviour' expansion for each
variable, because we can't do so for $W$ and hence $X$; instead we can give the
leading behaviour of the coefficients~(\ref{oddfull}) and~(\ref{allevensols}),
to understand which variables are important in the far field.

\paragraph{Odd variables and the crossover frame}

For $r$ sufficiently large, and when $W$ remains finite, the equations for
$X\V$ and $\bar X\V$ decouple from the Regge-Wheeler part to become
\be
\hat X\V\rightarrow \;{d{X\V}}{dr}=-\;{2\sigma^2 r^2}{3m}X\V+ {\cal O}\bra{r}X\V
\label{XVseriesde}
\ee
(and the same for $\bar X\V$); so, to leading order in $r$
\be
\bar X\V,~ X\V\rightarrow\exp\brac{-\;{2\sigma^2}{9m}r^3}.\label{Xfalloff}
\ee
Whenever $\Re(\sigma)\neq\pm\Im(\sigma)$, this falls off extremely fast. In the
special case where $\Re(\sigma)=\pm\Im(\sigma)$, we look for the first term in
Eq.~(\ref{XVseriesde}) without a $\sigma^2$ term [which requires a higher order
expansion of~(\ref{farfield})], which is $-X\V/3r$; hence, in this case
$X\V\rightarrow r^{-1/3}$. Note that the coupling in~(\ref{XVseriesde}) with
$W\T$ happens at order $1/r^4$.

At large $r$, then, whenever $W$ is finite,  the Regge-Wheeler tensor behaves
like
\be
W_{ab}\rightarrow r\zeta_{ab}
\ee
with the Coulomb part, the Weyl curvature contribution, decaying, bound to the
source. At the QNM frequencies, however, the behaviour of $X^a$ is less clear.

The full asymptotic solution, may be deduced from~(\ref{oddfull}) in the odd
case, and~(\ref{allevensols}) in the even crossover frame. The leading
$r$-dependence of each odd variable is
\ba
&&
\begin{array}{c|c|c|c|c|c|c|c|c|c}
  \mbox{\sc odd}  &  \bar\E\T &  \H\T                         & \bar\Sigma\T                  &\bar\zeta\T       &r^{-1}\bar\E\V       & r^{-1}\H\V                  & r^{-1}\bar\Sigma\V            & r^{-1}\Omega\V   &    \\\hline
  \bar{X}\V       &  0        &  0                            & (6i\sigma)^{-1} \,\,r         & (6m)^{-1}\,\,r^3 & -1/2                & (m/2i\sigma) \,\,r^{-2}     & -(i\sigma /18m) \,\,r^3       & -(i\sigma /18m) \,\,r^3 \vphantom{\;{\;XX}{XX}} &  \\\hline
  \bar W\T        &  -r^{-2}  &  -(i\sigma) \,\, r^{-1}       & \bra{i\sigma}^{-1}\,\,  r^{-2}& r^{-1}           &  (\lll/2)\,\,r^{-3} &0                            &  (\lll/6i\sigma) \hfil r^{-3} & (2\lll/3i\sigma) \,\,r^{-3}&\\\hline
  \hat{\bar W}\T  &  -r^{-1}  &-\bra{i\sigma}^{-1}\,\,  r^{-2}& \bra{i\sigma}^{-1}\,\, r^{-1} & 0                &   0                 &  -(\lll/2i\sigma)\,\, r^{-3}&  0                            &
  0&
\end{array}~~~\cdots\nonumber\\
&&~~~~~~~\cdots~~~
\begin{array}{c|c|c|c|c|c|c|c|}
 & r^{-1}\bar\hatn\V     & r^{-1}\bar\dotn\V         & r^{-1}\bar Y\V  & r^{-1}\bar Z\V   & r^{-2}\H\S                     & r^{-2}\Omega\S              & r^{-2} \xi\S           \\\hline
 & (\sigma^2/9m^2)\,\,r^5& -(i\sigma/6m) \,\, r^3    & -(3m)^{-1}\,\,r &  -1/3            &  0                             & -(\ll/12i\sigma)\,\, r^{-1} &  -(\ll/12m) \,\, r   \\\hline
 & 4\lll/3m\,\,r^{-1}    & (\lll/2i\sigma)\,\, r^{-3}&  0              &  0               &  -(\ll\lll/2i\sigma)\,\, r^{-5}&  0                          &   0              \\\hline
 & 0                     &   0                       &  0              &  0               &  0                             &  0                          &0
\end{array}
\ea
while for the even variables in the crossover frame we find
\ba
&&
\begin{array}{c|c|c|c|c|c|c}
  \mbox{\sc even}  & \E\T                                    & \bar\H\T                                  & \Sigma\T                                & \zeta\T              & r^{-1}\E\V                            &                              \\\hline
   X\V             &  0                                      &  0                                        & -\bra{6i\sigma}^{-1}\,r                 & -\bra{6m}^{-1}\,r^3  & -\bra{1/2}                           &   \\\hline
   W\T             &  -\bra{\kappa\sigma^2 m\ll\lll}\,r^{-1} & \bra{12\kappa i\sigma^3 m^2}\, r^{-1}     &  -\bra{\kappa i\sigma m\ll\lll}\,r^{-1} &  r^{-1}              & -\bra{6\kappa\sigma^2 m^2}\,r^{-3}  &   \\\hline
   \hat W\T        &  -\bra{12\kappa\sigma^2 m^2}\, r^{-1}   & -\bra{\kappa i\sigma m \ll\lll}\, r^{-1}  &  -\bra{12\kappa i\sigma m^2}\, r^{-1}   &  0                   &  \bra{\kappa m\ll\lll^2/2}\, r^{-3} &
\end{array}~~~\cdots\nonumber\\
&&~~~\cdots~~~
\begin{array}{c|c|c|c|c|c|c}
   & r^{-1}\bar\H\V                              & r^{-1}\Sigma\V                          & r^{-1}\bar\Omega\V                     & r^{-1}\hatn\V                             & r^{-1}\dotn\V                         & \\\hline
   & -\bra{m/i\sigma}\,r^{-2}                    & -\bra{i\sigma/18m}\,r^3                 & \bra{i\sigma/18m}\,r^3                 & \bra{\sigma^2/9m^2}\,r^5                  & -\bra{i\sigma/6m}\,r^3                & \\\hline
   & \bra{\kappa i\sigma m\ll\lll^2/2}\,r^{-3}   & {\kappa i\sigma\ll^2\lll^2/72}\,r^{-1}  & \kappa i\sigma\ll^2\lll^2/72\,r^{-1}   & 4\lll/3m-\kappa\sigma^2 m\ll\lll \,r^{-1} & \kappa i\sigma\ll^2\lll^2/12\,r^{-1}  & \\\hline
   & \bra{6\kappa i\sigma m^2\lll}\,r^{-3}       & \kappa i\sigma\ll\lll/3 \,r^{-1}        &  \kappa i\sigma\ll\lll/3\,r^{-1}       & \kappa\ll^2\lll^2/12   \,r^{-1}           & \kappa i\sigma\ll\lll  \,r^{-1}       &                      %\\
\end{array}~~~\cdots\nonumber\\
&&~~~~~~~~~~~~~~~~~~~\cdots~~~
\begin{array}{c|c|c|c|c|}
  & r^{-1}Y\V                              &  r^{-1}Z\V                             & r^{-2}\theta\S                             &   r^{-2}\Sigma\S                                 \\\hline
  & \bra{\lll/6m}\,r                       &  -\bra{1/3}                            & \bra{\ll/6i\sigma} \,r^{-1}                & -\bra{\ll/18i\sigma}\,r^{-1}                    \\\hline
  & -\bra{\kappa\ll^2\lll^3/12}\,r^{-3}    &  -\kappa\sigma^2\ll^2\lll^2/18\,r^{-1} & -\bra{2\kappa i\sigma^3 m\ll\lll/3}\,r^{-1}&  -\bra{4\kappa i\sigma^3 m\ll\lll/9}\,r^{-1}    \\\hline
  & -\bra{\kappa m\ll\lll^2}\,r^{-3}       &  -2\kappa\sigma^2 m\ll\lll/3\,r^{-1}   & \bra{\kappa i\sigma\ll^2\lll^2/18}\,r^{-1} & \bra{\kappa i\sigma\ll^2\lll^2/27}\,r^{-1}
\end{array}
\ea
We have used the abbreviation
\be
\kappa=12\bras{\bra{12\sigma m}^2+\ll^2\lll^2}^{-1};
\ee
recall that this blows up for the `special frequency', Eq.~(\ref{specialfreq}).

In the above tables we have multiplied the vector harmonic components by $1/r$
and the scalar by $1/r^2$, so that the factors for each variable are
appropriately normalised with one another (so, e.g., consider the even parity
vector $v_a=\delta_a \psi$; converting to SH, we have $v\V=r^{-1}\psi\S$;
similarly for tensors). This way we can see which terms dominate at large $r$.

\paragraph{The general frame}

In the general frame, the Regge-Wheeler equation is still satisfied, as it is
frame invariant, as is the Zerilli equation, and they both variables have a
${e^{i\sigma r}}$ behaviour. However, now we find that when $W\T$ and ${\cal
Z}$ remain finite
\be
Z\V\sim rX\V\rightarrow r^{-7/2}\exp\brac{-\;{2\sigma^2}{9m}r^3}
\ee
which falls off even faster than before.

\subsubsection{Approaching the horizon}

We'll look here at the behaviour of Eq.~(\ref{fullsol1-odd}) near the horizon,
$r\rightarrow2m_+$. First we define the lapse function as our radial parameter
\be
\alpha\equiv\hat r= \:12r\phi=\sqrt{1-\;{2m}{r}}
\ee
so that $r\rightarrow2m_+\Leftrightarrow\alpha\rightarrow0_+$. The leading
behaviour of the background variables near the horizon is:
\ba
\phi&\rightarrow&\;{1}{m}\alpha+{\cal O}\bra{\alpha^3}\\
\udot&\rightarrow&\;{1}{4m}\alpha^{-1}-\;{1}{2m}\alpha+{\cal O}\bra{\alpha^3}\\
\E&\rightarrow&-\;{1}{4m^2}+\;{3}{4m^2}\alpha^2+{\cal O}\bra{\alpha^4}\\
\omega&\rightarrow&\sigma\alpha^{-1}+{\cal O}\bra{\alpha^4}\\
\;{d}{d\rho}&\rightarrow&\brac{\;{1}{4m}-\;{1}{2m}\alpha^2+{\cal
O}\bra{\alpha^4}}\;{d}{d\alpha}.
\ea

The leading behaviour of the Regge-Wheeler tensor and Zerilli variable is
\be
W,~{\cal Z}\rightarrow
e^{\pm4mi\sigma\ln\alpha}~~~\mbox{as}~~~\alpha\rightarrow 0
\ee
[recall that we are not using the more familiar tortoise coordinate $r_*$,
Eq.~(\ref{tortoise}), so we do not have the same functional behaviour as
Eq.~(\ref{Wfarfield})]. $W$ and ${\cal Z}$ will remain finite provided the $-$
sign is taken for $\Im(\sigma)>0$, and vice versa, of if $\sigma$ is real. For
a more accurate limit further from the horizon, the sine and cosine functions
may be replaced with $\ell$-dependent Bessel functions.

The general frame is not particularly interesting here, so we will just discuss
the crossover solution.

\paragraph{Odd variables and the crossover frame}

Whenever $W$ is finite at the horizon, the equation for $X$ decouples (coupling
comes at order $\alpha W$), and we find
\be
X\rightarrow e^{-\:23\bra{1+16\sigma^2m^2}\ln\alpha},
\ee
which is not very well behaved, unless
$|\Im(\sigma)|>\bra{4m}^{-2}+|\Re(\sigma)|$.

We can give the leading behaviour of each variable, in terms of $\alpha$:
\ba
&&
\begin{array}{c|c|c|c|c|c|c|c}
  \mbox{\sc odd}  &  \bar\E\T                      &  \H\T                                & \bar\Sigma\T                     &\bar\zeta\T                 &r^{-1}\bar\E\V                & r^{-1}\H\V                          &    \\\hline
  \bar{X}\V       &  0                             &  0                                   & \bra{m/3i\sigma} \,\alpha^{-1}   & 4m^2/3\,\alpha^{-1}        & -\bra{1/2}\,\alpha^{-1}      & \bra{8i\sigma m}^{-1} \,\alpha^{-1} &  \\\hline
  \bar W\T        &  \bra{8m^2}^{-1}\,\alpha^{-2}  &  -\bra{i\sigma/2m} \, \alpha^{-2}    & \bra{4i\sigma m^2}^{-1}\,\alpha  & \bra{2m}^{-1}\,\alpha^{-1} &\bra{\lll/16m^3}\,\alpha^{-1} &0                                    &\\\hline
  \hat{\bar W}\T  &  -\bra{2m}^{-1}\,\alpha^{-1}   &\bra{8i\sigma m^2}^{-1}\,  \alpha^{-1}& \bra{2i\sigma m}^{-1}            & 0                          &   0                          &  -\bra{\lll/16i\sigma m^3}          &
\end{array}~~~\cdots\nonumber\\
&&\cdots~~
\begin{array}{c|c|c|c|c}
    & r^{-1}\bar\Sigma\V                                   & r^{-1}\Omega\V                                                            & r^{-1}\bar\hatn\V                            &   \\\hline
    & \bras{\bra{1+16\sigma^2m^2}/36 i\sigma}\,\alpha^{-2} &\bras{\bra{1+4\sigma^2m^2}/9i\sigma}\,\alpha^{-2} \vphantom{\;{X\;xx}{xx}}  & \bras{\bra{5+32\sigma^2m^2}m/9}\,\alpha^{-2}&  \\\hline
    &  \bra{\lll/48i\sigma m^3}                            & \bra{\lll/12i\sigma m^3}                                                  & 2\lll/3m^2                                   &   \\\hline
    &  0                                                   &   0                                                                       & 0                                            &
\end{array}~~\cdots\nonumber\\
&&\cdots~~
\begin{array}{c|c|c|c|c|c|c|c|c|}
 & r^{-1}\bar\dotn\V                                   &  r^{-1}\bar Y\V     & r^{-1}\bar Z\V      & r^{-2}\H\S                     & r^{-2}\Omega\S                   & r^{-2} \xi\S           \\\hline
 & \bras{\bra{1+16\sigma^2m^2}/12i\sigma}\,\alpha^{-2}&   (1/3)\,\alpha^{-1} &  -1/12\,\alpha^{-3} &  0                             & -(\ll/24i\sigma m)\, \alpha^{-1} &  -(\ll/6) \,\alpha^{-1}  \\\hline
 & (\lll/16i\sigma m^3)                                &   0                 &  0                  &  -(\ll\lll/64i\sigma m^5)      &  0                               &   0              \\\hline
 &   0                                                 &   0                 &  0                  &  0                             &  0                               &0
\end{array}
\ea

\ba
&&
\begin{array}{c|c|c|c|c}
  \mbox{\sc even}  & \E\T                                                    & \bar\H\T                                                              & \Sigma\T                                                                    &                              \\\hline
   X\V             &  0                                                      &  0                                                                    & -\bra{m/3i\sigma}\,\alpha^{-1}                                               &   \\\hline
   W\T             &-\bras{\kappa\sigma^2\bra{\ll\lll-3}/2}\,\alpha^{-2}     & \bras{\kappa i\sigma\bra{\ll\lll+48\sigma^2m^2}/8m}\, \alpha^{-2}     &  \bras{\kappa\ll\lll\bra{\ll\lll+48\sigma^2m^2}/96i\sigma m^2}\,\alpha^{-1}  &   \\\hline
   \hat W\T        &-\bras{\kappa\bra{\ll\lll+48\sigma^2m^2}/8m}\,\alpha^{-1}& -\bras{\kappa i\sigma \bra{\ll\lll-3}/2}\, \alpha^{-1}                &  \bras{\kappa\bra{\ll\lll+48\sigma^2m^2}/i\sigma m }                          &
\end{array}~~~\cdots\nonumber\\
&&\cdots~~~
\begin{array}{c|c|c|c|c}
   & \zeta\T                     & r^{-1}\E\V                                                        & r^{-1}\bar\H\V                                                                                  & \\\hline
   & -\bra{4m^2/3}\,\alpha^{-1}  & -\bra{1/2} \,\alpha^{-1}                                          & -\bra{8i\sigma m}^{-1}\,\alpha^{-1}                                                             & \\\hline
   &  \bra{2m}^{-1}\,\alpha^{-1} & \bras{\kappa\lll\bra{\ll^2\lll-48\sigma^2m^2}/64m^3}\,\alpha^{-1} & \bras{\kappa \ll\lll^2\bra{\ll-16\sigma^2m^2}/i\sigma\bra{4m}^4}\,\alpha^{-1}                    & \\\hline
   &  0                          &  \bras{\kappa \ll\lll\bra{\ll+1}/16m^2}                           & \bras{3\kappa\lll\bra{\ll-16\sigma^2m^2}/ i\sigma\bra{4m}^3 }                                    &                      %\\
\end{array}~~~\cdots\nonumber\\
&&\cdots~~~
\begin{array}{c|c|c|c}
& r^{-1}\Sigma\V = r^{-1}\dotn\V/3                                                 & r^{-1}\bar\Omega\V                                                                              &                             \\\hline
& \bras{\bra{1+16\sigma^2m^2}/36i\sigma}\,\alpha^{-2}                              & -\bras{\bra{1+4\sigma^2m^2}9i\sigma}\,\alpha^{-2}                                              &                   \\\hline
& -\bras{\kappa\ll^2\lll^2\bra{1+16\sigma^2m^2}/18i\sigma\bra{4m}^3}\,\alpha^{-2}  & \brac{\kappa\ll\lll\bras{\ll\lll+8\bra{9-\ll\lll}\sigma^2m^2}/9i\sigma\bra{4m}^3}\,\alpha^{-2}&     \\\hline
& -\bras{\kappa\ll\lll\bra{1+16\sigma^2m^2}/72i\sigma\bra{4m}^2}\,\alpha^{-1}      &  -\bras{\kappa\ll\lll\bra{\ll\lll-2+16\sigma^2m^2}/6i\sigma\bra{4m}^2}\,\alpha^{-1}&
\end{array}~~~\cdots\nonumber\\
&&\cdots~~~
\begin{array}{c|c|c|c}
& r^{-1}\hatn\V                                                             & r^{-1}Y\V                                                       &                              \\\hline
& \bras{m\bra{5+32\sigma^2m^2}/9}\,\alpha^{-2}                              & \bras{\bra{\ll+1}/3}\,\alpha^{-1}                              &                    \\\hline
& -\bras{\kappa\ll\lll\bra{\ll\lll+48\sigma^2m^2}/6\bra{4m}^2}\,\alpha^{-2} & -\bras{2\kappa\ll^2\lll^2\bra{\ll+1}/3\bra{4m}^3}\,\alpha^{-1} &     \\\hline
& \bras{\kappa\ll\lll\bra{\ll\lll-3}/24m}\,\alpha^{-1}                      & -\bras{\kappa\ll\lll\bra{\ll+1}/8m^2}&
\end{array}~~~\cdots\nonumber\\
&&\cdots~~~
\begin{array}{c|c|c|c|c|c|c|c|c|}
&  r^{-1}Z\V                                                                                   & r^{-2}\theta\S                             &   r^{-2}\Sigma\S                                 \\\hline
&  -\bra{1/12}\,\alpha^{-3}                                                                    & -\bra{\ll/12i\sigma m} \,\alpha^{-1}       & \bra{\ll/4i\sigma m} \,\alpha^{-1}                    \\\hline
&  -\bras{\ll\lll/12(2m)^3}\brac{\kappa\ll\lll\bra{32m^2\sigma^2-\ll\lll-1}/12-1}\,\alpha^{-3} & \ast\,\alpha^{-3}                          & \ast\,\alpha^{-3}      \\\hline
&  -\kappa\ll\lll\bra{32m^2\sigma^2+\ll\lll-1}/24\bra{2m}^2\,\alpha^{-2}                       & \ast\,\alpha^{-2}                          & \ast\,\alpha^{-2}
\end{array}
\ea
A `$\ast$' was used in the last two entries as the expressions are rather
large. }

\subsection{Discussion}

%\subsection{What are the fundamental variables?}

\forget{
We have seen that in both the odd and even parity regimes the Regge-Wheeler
tensor, defined by Eq.~(\ref{Wab}), and the vector $X^a$ are sufficient to
characterise the full problem of perturbations of Schwarzschild black holes.
All variables, irrespective of parity, depend on $W_{ab}$ and its derivatives,
and $X^a$, although often in quite a complicated manner. Is this surprising?
Well, yes and no, really. Firstly, we should not be too surprised as we know
from the metric approach the the full solution is determined by the
Regge-Wheeler equation for odd parity perturbations, and the Zerilli equation
for even. We find that this is true here [using Eq.~(\ref{WofZ}) in the even
solution~(\ref{allevensols}) to convert to ${\cal Z}$] too; but what of the
extra variable $X^a$? The metric approach does not require this extra variable
to determine the full solution, so why do we get it here? By choosing observers
in the spacetime, we actually have more to determine than in the metric
approach: if we put observers in the spacetime then, from a coordinate point of
view, we have three extra functions to determine in our velocity field
$u^a$\forget{
\be
u^\mu=\bra{1-\;{2m}{r}}^{-1/2}\delta^\mu_{~t}+\epsilon
\bra{f_1\delta^\mu_{~t}+f_2\delta^\mu_{~r}+f_3\delta^\mu_{~\vartheta}+f_4\delta^\mu_{~\varphi}},
\ee
(where $f_\mu=f_\mu(t,r,\vartheta,\varphi)$, in Schwarzschild coordinates,
are four new functions;} (one of the total four is given by the normalisation
condition $u^au_a=-1$) over and above the usual metric functions. The same
argument applies to $\n^a$ too, giving two extra functions (again the
conditions $u^a\n_a=0$ and $\n^a\n_a=1$ eliminate two of a potential four).
This is the origin of our extra degree of freedom in our solutions. Because
the spacetime is vacuum, this detail is not required to understand the
solution; the metric suffices. If the non-vacuum case is to be considered,
however, this sort of thing will be important, as one needs observers before
physical quantities like energy density can be defined; in addition, we can't
define things like the electric and magnetic parts of the Weyl tensor without
them~-- these extra functions would enter into the definition of the electric
Weyl tensor here, because it is non-zero in the background. }

We have seen in our approach that we may define a TT tensor which satisfies
the Regge-Wheeler equation, irrespective of parity, and we have shown that
the Zerilli equation may be derived from this tensor in the even case. The
transformation equations between the two, Eqs.~(\ref{WofZ}) and~(\ref{ZofW}),
then allowed us to derive an odd parity Zerilli variable, and hence a Zerilli
tensor. This transformation between parities is made explicit in our approach
due to the unification properties of $W_{ab}$ and ${\cal Z}_{ab}$. To
contrast with the metric approach, the choice of metric functions which makes
this similarity between the parities explicit is a physically unmotivated
expression, which is a complicated linear combination of the metric
perturbation functions (see, e.g., Eq.~(154) in~\cite{chandra}). Of our two
fundamental tensors, it is fairly clear that the Regge-Wheeler tensor is the
most appealing, for two reasons: it is defined in a clear and simple way, and
it obeys a \emph{covariant} wave equation. Recall that it's the harmonic
amplitudes of ${\cal Z}_{ab}$ which obey wave equations, and not ${\cal
Z}_{ab}$ itself.

In the frame we have chosen where $\udot^a=Y^a=Z^a=0$ we just have $W_{ab}$
governing both the odd and even parity perturbations, which obeys the covariant
wave equation~(\ref{RWtensorwave}).\forget{
\[
\ddot W_{ab}-\hat{\hat W}_{ab}-\udot{\hat W}_{ab}
+\phi^2 W_{ab}-\delta^2 W_{ab} =0.
\]}
The Regge-Wheeler equation describes the dynamically free gravitational field
which propagates at the speed of light. It is given here in its  fully
covariant, gauge- and frame-invariant form. To understand what $W_{ab}$
actually embodies,  recall its definition which applies in any frame,
Eq.~(\ref{Wab}).\forget{:
\[
W_{ab}=\underbrace{\:12\phi
r^2\zeta_{ab}}_{{\mathsf{Shear~of}}~\n^a}-\underbrace{\:13r^2\E^{-1}\delta_{\lb
a}X_{b\rb}}_{\mathsf{Electric~Weyl}}.
\]}
The first term, $\zeta_{ab}$, is just the shear distortion of our sheet
(vibrating 2-`spheres') as we move radially along $\n^a$. The second term is a
little more complicated. Recall that in an exact spherically symmetric
spacetime, $\E$ is the tidal force measured by our static observers $u^a$ in
the direction $\n^a$. There seems to be no reason to change this interpretation
in the real perturbed spacetime - here it will just undergo fluctuations, but
we can still call it the radial tidal force. `Radial' loses its meaning in a
perturbed spacetime, but if we chose a frame in which $n^a\propto\uudot^a$
(i.e., $\udot^a=0$) an observer could always determine this direction because
it would lie precisely in the direction of the external force that they must
apply. Spatial fluctuations in the radial tidal forces are characterised by $
\bra{\sdel_a\E}/{\E} $ which is the \emph{comoving fractional gradient of the tidal
force}; projecting this onto the sheet gives us the gauge invariant
(first-order) part of this 3-vector: $ \bra{\delta_a\E}/{\E}={X_a}/{\E}, $
which tells us the \emph{fractional gradient of the radial tidal forces over a
sheet}~-- i.e., how the radial tidal forces change from point to point on our
vibrating 2-`spheres'. The distortion of this, $
\delta_{\lb a}\bras{\bra{\delta_{b\rb}\E}/{\E}},
$
appearing in $W_{ab}$ is the \emph{shearing distortion of the radial tidal
force gradient}. The Regge-Wheeler tensor is thus of a \emph{shearing} form; it
is this tensor which describes GW around a black hole, through the covariant
form of the Regge-Wheeler equation~(\ref{RWtensorwave}).
\forget{It does not carry their energy, however; this is done by the transverse
traceless tensors $\E_{ab}$ and $\H_{ab}$~\cite{PT}.}

\forget{
This brings us to the following correspondence between variables:
\[
\begin{array}{ccccc}
  \textrm{Electric Weyl} & \leftrightarrow & \textrm{Regge-Wheeler} & \leftrightarrow & \textrm{TT shear of $n^a$, $\zeta_{ab}$} \\
  \textrm{Magnetic Weyl} & \leftrightarrow & \textrm{Zerilli} & \leftrightarrow & \textrm{TT shear of $u^a$, $\Sigma_{ab}$}
\end{array}
\]
as seen, for example, in the wave equations for $\E_{ab}$ and $\H_{ab}$,
Eqs.~(\ref{Ewave}) and~(\ref{Hwave}). Although the Zerilli variable is simply
related to the even parity part of the Regge-Wheeler tensor, via
Eq.~(\ref{ZofW}), it does illustrate the origin of the Zerilli equation, and
the magnetic Weyl curvature it governs. These are not independent perturbations
of course~--  we can't have non-zero $\E_{ab}$ without $\H_{ab}$~-- it does
illustrate the fact that the kinematics of our two frame vectors are enough to
determine the whole spacetime geometry, in particular the propagation of
gravity waves. To be precise, if we look at our solutions~(\ref{oddfull})
and~(\ref{allevensols}), we can see quite easily our solution vector in ${\cal
V}_{31}$, in the crossover frame, could be
\be
\mathbf{v}=\bra{\bar\dotn\V,\bar\zeta\T,\bar\Sigma\T ~|~ \dotn\V,\zeta\T,\Sigma\T}.
\ee
Thus if we measure the covariant kinematical objects  $\zeta_{ab}$,
$\Sigma_{ab}$ and $\dotn^a$, we can determine all the properties of the
spacetime, including the Regge-Wheeler and Zerilli variables.

}
\forget{
We may use $W_{ab}$ to \emph{covariantly characterise} the odd and even parity
perturbations:
\be
\begin{array}{rcc}
\mbox{\sc odd~parity:}&~~~\Leftrightarrow~~~&\delta^a\delta^bW_{ab}=\mbox{`}\div\div
W\mbox{'}=0\\
\mbox{\sc even~parity:}&~~~\Leftrightarrow~~~&\lc^{ac}\delta_c\delta^bW_{ab}=\mbox{`}\curl\div
W\mbox{'}=0
\end{array}
\ee
But, in fact, we can do this more easily: for example, $\xi=0$ characterises
the even perturbations, while $\delta_a X^a=\delta_aZ^a=0$ or $\Sigma=\theta=0$
characterises the odd perturbations. This may be contrasted with the usual
covariant characterisation of cosmological perturbations into scalar, vector
and tensor modes.

As we have emphasised, we have concentrated on $W_{ab}$, as it appears the most
natural tensor in the spacetime. But is $X^a$ a `natural' vector in the same
sense? It would appear not, at least not to the same degree. Our frame freedom
in the even case ensured that we could have chosen any variable except $\E\T$
or $\bar\H\T$ to be our solution parameter with a crossover equation similar to
Eq.~(\ref{barXVde}). It is really the \emph{interrelation} between all the
vectors, tensors and scalars that determines the physical situation, rather
than viewing one or two as `driving' variables, the way one can in, e.g., the
cosmological situation. There, for example, the shear does drive gravitational
radiation, as it acts a source term in the wave equation for the electric Weyl
tensor. The true driving variable here may be considered as $W_{ab}$, with
everything else being determined from it; but is this just a mathematical
nicety, or is it more than this? We believe there is more to it. The two
variables which make up $W_{ab}$ have physical clear interpretations, and the
wave equation for $W_{ab}$ is unaffected by any frame choice we may make. In
addition, there is no other (simple) way to make the tensor $W_{ab}$.}

%\section{Discussion}

\section{Conclusions}

We have presented a new perturbation formalism for dealing with systems with
spherical symmetry in the background, and we have applied this to the simplest
of such systems, the Schwarzschild black hole. Our 1+1+2 splitting allowed the
Schwarzschild solution to be given in covariant form. We then demonstrated that
we can derive the usual equations governing perturbations of the spacetime,
namely the Regge-Wheeler equation~(\ref{RW}), and the Zerilli
equation~(\ref{zerilli}), while discussing in detail our new method. We have
also shown that there exist Regge-Wheeler and Zerilli \emph{tensors} which
unify the odd and even parity perturbations; indeed, the Regge-Wheeler tensor
was shown to satisfy a closed \emph{covariant} wave equation which governs the
dynamics of the whole problem. This sets the basis for future studies of more
general astrophysical systems, which only require an appropriate change of the
background, for which the possible applications are myriad.

The method has several important aspects. The first is the covariant spacetime
splitting itself. In general, the two threading vectors $u^a$ and $n^a$ may be
chosen arbitrarily, defining the sheet  on which the vectors and tensors exist
(strictly speaking, in general the sheet is not a true surface, but a
collection of tangent planes). This makes the approach a halfway house between
the 1+3 approach and the orthonormal tetrad approach, and provides a completion
of the covariant formalism. (Recall that a unique tetrad cannot be defined in
isotropic or locally rotationally symmetric spacetimes~\cite{HvE,henk}, so this
is the ideal compromise between the two in such cases.) For systems with
spherical symmetry in the background, we have seen that when  $\n^a$ is radial
in the background,  the perturbed spacetime becomes a tractable problem,
because all vectors and tensors become first order, allowing decomposition with
suitable harmonic functions~-- spherical harmonics, in this case. Time
harmonics are also introduced to simplify the solution, allowed when the
background is static, but these are not strictly necessary, as the
dot-derivative is a scalar operator, and can be dealt with by standard
techniques. \forget{(We may regain the dot-derivative solution to this problem
by replacing $i\omega\rightarrow d/d\tau$ in all the equations, as already
discussed.)} So far this merely writes the equations in an alternative form. At
this point finding the solution is relatively trivial: our approach simply
requires solving a linear system of algebraic equations, \emph{and this is all
there is to it}. The underlying dynamics are then \emph{automatically} given by
the small system of differential equations that remain~-- wave equations, if
desired, then may be derived (where they exist) by differentiating this
equation.

An important physical aspect of our approach is that it uses a set of
covariantly defined quantities with genuine physical significance, which makes
it clear which objects are crucial for the detection and measurement of
gravitational waves. Put simply, GW detectors essentially measure gravitational
tidal forces; that is, they are sensitive to the dynamical Weyl curvature,
encoded in the electric Weyl tensor, $E_{ab}$, and this dynamical Weyl
curvature forces the working parts of any GW detector through the right hand
side of the geodesic deviation equation. We have shown that there is a
gauge-invariant TT tensor that describes GWs of either parity, and is closely
related to the variation of the radial tidal force. Thus it is clear that our
formalism is dealing with real, physically measurable, objects from the start.
Indeed, we have discussed how a subset of four of all thirty-three variables is
sufficient to determine the full dynamics of the spacetime.

\iffalse
It is worthwhile discussing here the implications of having a static
background: it would be possible to introduce only the vector field $n^a$,
ignoring $u^a$ completely, perform a 1+3 splitting, where the 3-spaces would be
`spherical tubes'~-- 3-cylinders~-- with the tube part in the time direction,
on which all vectors and tensors vanish. By analogy with the FLRW case, where
everything happens in the `time' direction, here everything happens in the
radial direction. The harmonic functions which split the 3-vectors and tensors
would be a mixture of spherical harmonics and the time harmonics, but in
somewhat disguised form. While there are probably benefits to adopting such an
approach for the situation discussed here~-- not least that the equations would
be considerably less in number~-- it is unlikely that the physical picture
would emerge so easily (for example, the dot-derivative has an important
physical meaning, as do the spherical harmonic functions). In addition,
\fi

There are many possible extensions of the work we have presented. The obvious
extension is to perturbations of static stars, but we envisage that our
method will be widely applicable to many other astrophysical situations, such
as systems with a \emph{dynamical} background~-- e.g., collapsing stars, type
Ia supernovae, etc.~-- where the scalar background equations have two
(non-tensorial) derivatives in them. In the static background case, where we
can introduce time-harmonic functions in the perturbed equations, solving the
equations is a simple problem of solving a linear system of equations, and
one is left with a first order non-autonomous system of ordinary differential
equations, whose dimension is small compared to the original system, plus
linear relations among the remaining variables in terms of the basis vector
of the dynamical system. All the physics of the spacetime is contained in
this small dynamical system. With a dynamical background, it may not be as
simple as this, but we do not envisage it being much more difficult to find
the full solution; one may have to be careful choosing one's observers
(perhaps comoving with the matter in the case of a collapsing star). In any
event, there is much that can be achieved. In any situation where there is a
preferred spacelike vector field present, the covariant 1+1+2 sheet formalism
should provide new insight.

%, and the final solution will most likely be in the form of two
%dynamical systems; but no problem.

%RELATE TO PN.

\acknowledgments

We would very much like to thank Peter Dunsby, George Ellis and Roy Maartens
for lots of continuous advice and encouragement on this project. We would
also like to thank Gerold Betschart, Bonita de Swardt, Kostas Kokkotas,
Nazeem Mustapha and Jorge Pullin for useful discussions and/or comments.
%We
%would also like to thank our anonymous referees for inspiring significant
%streamlining of the manuscript.
CAC was funded by NRF (South Africa), and RKB was funded by EPSRC (UK).


\begin{thebibliography}{25}
\bibitem{HvE} G. F.~R.  Ellis and H. van Elst, in M. Lachieze-Rey (ed.),
{\em Theoretical and Observational Cosmology}, NATO Science Series, Kluwer
Academic Publishers~(1998) {\em gr-qc/9812046v4}

\bibitem{BE} Ellis, G.F.R. and Bruni, M. Phys Rev. D {\bf 40} 1804 (1989)

\bibitem{CL} Challinor, A.D. and Lasenby, A.N. Phys. Rev. D {\bf 58} 023001
(1998); Astrophys. J. {\bf 513} 1 (1999)

\bibitem{MGE} Maartens, R., Gebbie, T. and Ellis, G.F.R. Phys. Rev. D {\bf 59}
083506 (1999)

\bibitem{henk}   van Elst, H. and Ellis, G. F. R., Class. Quantum Grav. {\bf 13}
 1099 (1996);
  van Elst, H., Ph.D. thesis, University of London (1996)

\bibitem{LIGO}Barish, B. and Weiss, R., { Phys. Today} {\bf 52},
44-50 (1999); B Willke et al., Class. Quantum Grav. {\bf 19}, 1377 (2002); M
Ando et al., Class. Quantum Grav. {\bf 19}, 1409 (2002); F Acernese et al.,
Class. Quantum Grav. {\bf 19}, 1421 (2002).

\bibitem{chandra} Chandrasekhar, S. The Mathematical Theory of Black Holes.
Oxford, (1983)

\bibitem{RW} Regge, T. and Wheeler, J.A. Phys. Rev. {\bf 108} 1063 (1957)

\bibitem{zerilli} Zerilli, F.J. Phys. Rev. Lett. {\bf 24} 737 (1970)

\bibitem{GS} Gerlach, U. H.  and Sengupta, U. K., Phys. Rev. D, {\bf 19} 2268 (1979);
\emph{ibid.} {\bf 22} 1300 (1980)

\bibitem{MGG} Gundlach, C. and Mart\'\i n-Garc\'\i a, J.M. {\it gr-qc/9906068}
and {\it gr-qc/0012056}

\bibitem{ST} Sarbach, O. and Tiglio, M.: Phys. Rev. D {\bf 64} 084016 (2001)

\bibitem{J} Jezierski, J.: Gen. Rel. Grav. {\bf 31} 1855  (1999) {\it
gr-qc/9801068}

\bibitem{zafiris} Zafiris, E., J. Math. Phys. {\bf 38} 5854 (1997)

\bibitem{TM2} Mason, D.P. and Tsamparlis, M.,  J. Math. Phys. {\bf 26} 2881
(1985)

\bibitem{greenberg} Greenberg, P.J., J. Math. Anal. Applic. {\bf 30} 128 (1970)

\bibitem{TM1} Tsamparlis, M. and Mason, D.P., J. Math. Phys. {\bf 24} 1577
(1983)

\bibitem{BS} Bruni, M., Matarrese, S., Mollerach, S. and Sonego, S., Class.
Quantum Grav. {\bf 14} 2585 (1997); Bruni, M. and Sonego, S.,
 Class. Quantum Grav.,  {\bf 16} L29 (1999)

\bibitem{SW} Stewart, J.M. and Walker, M. Proc. R. Soc. London A {\bf 431} 49
(1974)

\bibitem{thorne} Thorne, K.S., Rev. Mod. Phys. {\bf 52} 299 (1980)

\bibitem{nollert} Nollert, HP. Class. Quantum Grav. {\bf 16} R159 (1999);
Kokkotas, K.D. and Schmidt, B.G., Living Rev. Relativity, {\bf 2} 2 (1999),
{\bf\texttt{http//www.livingreviews.org/Articles/Volume2/1999-2kokkotas/}}

\bibitem{liu-M} Liu, H. and Mashhoon, Class. Quantum Grav. {\bf 13} 233 (1996)

\bibitem{NS} Nollert, HP. and Schmidt, B.G., Phys. Rev. D {\bf 45} 2617 (1992)

\bibitem{PT} Price, R.H. and Thorne, K.S., Phys. Rev. D {\bf 33} 915 (1986)


\end{thebibliography}
\end{document}